\documentclass[structabstract]{aa}
\usepackage[varg]{txfonts}
\usepackage{natbib}
\usepackage{graphicx}
\usepackage{tabularx}
\usepackage{nicefrac}
\usepackage{units}
\usepackage{longtable}
\usepackage{longtable,lscape}

\usepackage{subfig}
\usepackage{float}

\bibpunct{(}{)}{;}{a}{}{,}

\newcommand{\flux} {$\rm{\int{T_{mb}dV}}$} 
\newcommand{\Tp}   {$\rm{T_{peak}}$}

\newcommand{\Vlsr} {$\rm{V_{lsr}}$}

\newcommand{\kms}  {$\rm{km\,s^{-1}}$}

\newcommand{\Eupk} {E${_{\rm up}}/k$}
\newcommand{\Eup} {E${_{\rm up}}$}

\newcommand{\K}    {$\rm{K}$}

\newcommand{\Kkms} {$\rm{K\,km\,s^{-1}}$}

\newcommand{\GHz}  {$\rm{GHz}$}

\begin{document}
%
%
\title{The HIFI spectral survey of AFGL~2591 (CHESS).\\ II. Summary of the survey.
\thanks{\textit{Herschel} is an ESA space observatory with science instruments provided by European-led Principal Investigator consortia and with important participation from NASA.}
}

\authorrunning{M. Ka{\'z}mierczak et al.}
\titlerunning{AFGL 2591 -- HIFI/CHESS survey}
       
\author{M. Ka{\'z}mierczak-Barthel \inst{\ref{inst1}}
\and F.F.S. van der Tak \inst{\ref{inst1},\ref{inst2}}
\and F.P. Helmich \inst{\ref{inst1},\ref{inst2}}
\and L. Chavarr\'{\i}a \inst{\ref{inst3}}
\and K.-S. Wang \inst{\ref{inst4},\ref{inst5}}
\and C. Ceccarelli \inst{\ref{inst6}}
}

\institute{SRON Netherlands Institute for Space Research, Landleven 12, 9747 AD Groningen, The Netherlands\\
     \email{maja.kazmierczak@gmail.com} \label{inst1}
\and Kapteyn Astronomical Institute, University of Groningen, PO Box 800, 9700 AV, Groningen, The Netherlands \label{inst2}
\and Universidad de Chile, Camino del Observatorio 1515, Las Condes, Santiago, Chile \label{inst3}
\and Leiden Observatory, Leiden University, PO Box 9513, 2300 RA, Leiden, The Netherlands \label{inst4}
\and Institute of Astronomy and Astrophysics, Academia Sinica, Taipei, Taiwan \label{inst5}
\and UJF-Grenoble 1/CNRS-INSU, Institut de Plan\'etologie et d'Astrophysique de Grenoble (IPAG) UMR 5274, Grenoble, France \label{inst6}
}

\date{Received / Accepted }

%
%
\abstract%
{}
   {This paper presents the richness of submillimeter spectral features in the high-mass star forming region AFGL~2591. 
}
   {As part of the CHESS (Chemical Herschel Survey of Star Forming Regions) Key Programme, AFGL~2591 was observed by the \textit{Herschel}/HIFI instrument. The spectral survey covered a frequency range from 480 up to 1240\,GHz as well as single lines from 1267 to 1901\,GHz (i.e. CO, HCl, NH$_3$, OH and [CII]). Rotational and population diagram methods were used to calculate column densities, excitation temperatures and the emission extents of the observed molecules associated with AFGL~2591. 
The analysis was supplemented with several lines from ground-based JCMT spectra.
   }
   {From the HIFI spectral survey analysis a total of 32 species were identified (including isotopologues). In spite of the fact that lines are mostly quite weak (\flux $\sim$ few \Kkms), 268~emission and 16~absorption lines were found (excluding blends). 
Molecular column densities range from 6 $\times\, 10^{11}$ to 1 $\times\, 10^{19}$~cm$^{-2}$ and excitation temperatures range from 19 to 175\,K. 
One can distinguish cold (e.g. HCN, H$_2$S, NH$_3$ with temperatures below 70~K) and warm species (e.g. CH$_3$OH, SO$_2$) in the protostellar envelope.
    }
    {}

%
%


\maketitle{}

\label{firstpage}

%
%
\section{Introduction} \label{Introduction}
%
%
Massive stars play a major role in the evolution of galaxies.
From their birth in dense molecular clouds to their death as a supernova explosion, massive stars interact heavily with their surroundings by emitting strong stellar winds and by creating heavy elements \citep{Zinnecker2007}.
They influence the formation of nearby low-mass stars and planets \citep{Bally2005} as well the physical, chemical and morphological structure of galaxies \citep[e.g.,][]{Kennicutt2012}.
Although, massive stars are an important component of galaxies, their formation processes are still unclear. 
It is difficult to observe high-mass star forming regions because of high dust extinction, their large distances and rapid evolution \citep{2014arXiv1402.0919T}. 

High-mass star forming regions are quite rare, so each observational effort is very helpful in solving their puzzle.
One of the goals of the \textit{Herschel Space Observatory} \citep{Pilbratt2010} was to improve our understanding of the high-mass star formation processes. 
Among the several Key Projects devoted to those studies, we focus here on the \textit{Herschel} Key Program CHESS \citep[Chemical Herschel Survey of Star Forming Regions,][]{Ceccarelli2010}. 
The aim of this project is to study the chemical composition of dense regions of the interstellar medium, to understand the chemical evolution of star forming regions and the differences between regions with different masses/luminosities.
The target sources of CHESS are 
the pre-stellar cores I16293E and L1544, 
the outflow shock spot L1157-B1, 
the low-mass protostar IRAS16293-2422,
the intermediate-mass protostar OMC2-FIR 4, 
the intermediate luminosity hot cores NGC 6334I and AFGL 2591
and the high luminosity hot core W51e1/e2. 
Almost the entire spectral range of the HIFI instrument, i.e. 480 to 1910\,\GHz, has been used for the observation of the above listed objects.
In this paper we will focus on the source AFGL 2591.

Spectral surveys cover simultaneously a large variety of molecular and atomic lines.
In this way they offer the possibility to probe cold and warm gas and the fundamental processes which occur in star forming regions. 
Especially, \textit{Herschel's} large frequency range allowed to cover molecular lines from very different energy levels, from light to heavier molecules and therefore study species thoroughly.

AFGL~2591 is one of the CHESS sources.
It is a relatively isolated high-mass protostellar object with a bipolar molecular outflow \citep{vanderTak1999}. 
A massive sub-Keplerian disk has been proposed to exist around source AFGL~2591--VLA 3 \citep{Wang2012}.
AFGL~2591 is located in the Cygnus X region, \mbox{(\textit{l, b}) = 78.$\degr$9, 0.$\degr$71.}
Based on VLBI parallax measurements of 22\,GHz water maser, \citet{Rygl2012} have estimated recently the distance\footnote{The previous distance estimates were uncertain, with values between 1 and 2\,kpc \citep[e.g.][]{vanderTak1999,vanderTak2000}, thus luminosity at 1\,kpc L = 2 $\times\,10^4$\,L$_{\sun}$.}
towards AFGL~2591 of 3.33 $\pm$ 0.11\,kpc, hence, the corresponding luminosity is \mbox{L = 2 $\times\, 10^5$\,L$_{\sun}$} \citep{Sanna2012}.
For a detailed source description see \citet{vanderWiel2013} (hereafter Paper~I) and references therein.

The richness of the detected lines in AFGL~2591 from the \mbox{HIFI/CHESS} spectral survey gives us the opportunity to gain detailed insights into its chemical and physical structure. 
Results from the spectral survey are going to be presented in a series of papers. 
The first one focused on highly excited linear rotor molecules \citep{vanderWiel2013}.
In the present work the entire HIFI spectral survey of AFGL~2591 is presented. 

\citet{vanderWiel2013} studied linear rotor molecules (CO, HCO$^{+},$ CS, HCN, HNC) in the high-mass protostellar envelope. 
This work was based on the \textit{Herschel}/HIFI data together with observations from the ground-based telescopes, JCMT and IRAM~30m.
The line profiles of the observed emissions consist of two components, a narrow one which corresponds to the envelope and a broad component from the outflow. The same nomenclature is used in the present paper.  

This paper starts with the description of the observations and the data reduction of \textit{Herschel} and JCMT spectra (Sect.~2). In~Sect.~3 the general summary of the HIFI/CHESS spectral survey of AFGL~2591 is given. Here, all of the observed species from that survey are presented together with emission and absorption lines analysis.
Discussions and conclusions are given in Sects.~5 and 6, respectively.
Appendix A gives a table with all detected transitions and plots of their line profiles.

%
%
\section{Observations and data reduction} \label{Observations}
%
%
%
%
\subsection{480--1850\,GHz {\textit{Herschel}}/HIFI data} \label{Obs-herschel}
%
%
Observations of AFGL~2591 ($\alpha_{2000}$~$=$~20$^h$29$^m$24$^s$.9, $\delta_{2000}$~$=$~+40${\degr}$11$\arcmin$21$\arcsec$) were obtained with the Heterodyne Instrument for the Far-Infrared \citep[HIFI,][]{Graauw2010} onboard the ESA \textit{Herschel Space Observatory} as a part of the HIFI/CHESS Guaranteed Time Key Programme\footnote{Data are available from:\\ www-laog.obs.ujf-grenoble.fr/heberges/hs3f}. 

A full spectral survey of AFGL~2591 of HIFI bands \mbox{1a~--~5a} (480~--~1240~\GHz, 18.4\,h of observing time) was obtained. Nine additional selected frequencies were observed in \mbox{3.5\,h} of observing time.
The corresponding bands are: 5b (lines: HCl, CO), 6a~(CO), 6b (CO), 7a (NH$_3$, CO) and 7b (CO, OH, [CII]). 

Despite being the second in a series of papers based on HIFI/CHESS data of AFGL~2591 and detailed description of its data reduction process in Paper~I (Van der Wiel et al. 2013), basic information is recalled here as well.

The spectral scan observations were carried out using the dual beam switch (DBS) mode, with the Wide Band Spectrometer (WBS) with a resolution \mbox{of~1.1\,MHz,} corresponding to 0.66 \kms\,at 500~GHz and 0.18 \kms\,at 1850\,GHz.
The~single frequency settings were obtained in the dual beam switch mode as well, with the fast chop and stability optimization options selected.
Table \ref{table:hifi-bands} gives information about the covered frequency range, beam size, noise level, and integration time.

\begin{table}
\setlength{\extrarowheight}{0.7pt}
\caption{Overview over all HIFI bands for the observations of AFGL~2591 (bands: 1a -- 5a are spectral scans, 5b -- 7b single frequency settings).}
\label{table:hifi-bands}
\centering
\begin{tabular}{ccccc}
\hline\hline
Band & Freq. range & Beam size & rms & Obs. time\\
 & [GHz] & [''] & [K] & [s]\\
\hline
1a & 483--558 & 41 & 0.030 & 4591\\
1b & 555--636 & 36 & 0.029 & 4643\\
2a & 631--722 & 31 & 0.026 & 9833\\
2b & 717--800 & 28 & 0.067 & 6407\\
3a & 800--859 & 26 & 0.039 & 4893\\
3b & 858--960 & 23 & 0.067 & 8578\\
4a & 950--1060& 21 & 0.157 & 9137\\
4b &1051--1120& 20 & 0.144 & 6300\\
5a &1110--1240& 18 & 0.147 & 11931\\
5b &1266--1270& 17 & 0.149 & 1380\\
5b &1251--1255& 17 & 0.149 & 2255\\
6a &1496--1499& 14 & 0.117 & 1440\\
6b &1611--1614& 13 & 0.106 & 1392\\ 
7a &1726--1729& 12 & 0.092 & 1575\\
7a &1762--1764& 12 & 0.095 & 1423\\  
7b &1840--1843& 12 & 0.092 & 1711\\
7b &1900--1903& 12 & 0.120 & 1452\\ 
\hline
\end{tabular}
\end{table}

\begin{figure}
\centering
\includegraphics[width=0.5\textwidth]{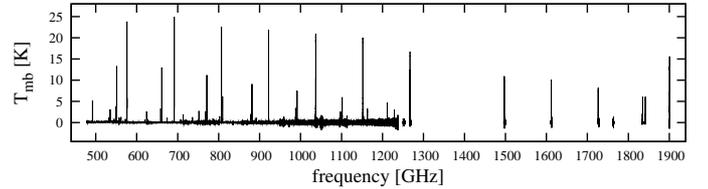}
\caption{Complete baseline-subtracted spectrum. The strongest lines belong to CO and its isotopologues, while at 1901\,GHz [CII] is seen.}
\label{figure:spectrum}
\end{figure}

AFGL~2591 data were completely reduced with the HIPE\footnote{HIPE is a joint development by the Herschel Science Ground Segment Consortium, consisting of ESA, the NASA Herschel Science Center, and the HIFI, PACS and SPIRE consortia.}--\textit{Herschel} Interactive Processing Environment \citep{Ott2010}, version~8.1, using scripts written by the CHESS data reduction team \citep{Kama2013}.
After pipelining, the quality of each spectrum was checked and spectral regions with spurious features (spurs) were flagged. Next, the correction for standing waves was made and a baseline was subtracted (polynomial of order~$\sim$3). 
The final single sideband spectrum is presented in Fig.~\ref{figure:spectrum}.

Strong lines are known to create \textit{ghost} features in the sideband deconvolution process \citep{Comito2002}. To check the importance of this effect on our data, the above steps were repeated with strong lines (especially CO transitions) masked out in the same way as spurs.
The term \textit{strong lines} refers to features of $T_A^*>$1\,K in band 1a to $>$8\,K in band 5a, depending on the amount of lines and the noise level in a given band.
Following the outlined data reduction procedure, two single sideband spectra for bands 1a~--~5a were obtained.
The first set of spectra was used to analyse strong lines (e.g. CO and its isotopologues, HCO$^{+}$). The second one, for line measurements of weak features, i.e.~which were not masked as strong lines (e.g. SO, CH$_3$OH). 

%
%
\subsection{330--373\,GHz JCMT data} \label{Obs-jcmt}
%
%
The excitation analysis of several molecules was complemented by ground-based observations from the James Clerk Maxwell Telescope (JCMT)\footnote{The James Clerk Maxwell Telescope is operated by the Joint Astronomy Centre on behalf of the Science and Technology Facilities Council of the United Kingdom, the Netherlands Organisation for Scientific Research, and the National Research Council of Canada.}. These data are part of the JCMT Spectral Legacy Survey \citep[SLS,][]{Plume2007}. 
The observations were taken with the 16-element Heterodyne Array Receiver Programme B (HARP-B) and the
Auto-Correlation Spectral Imaging System (ACSIS) correlator \citep{Dent2000,Smith2008,Buckle2009}.

The JCMT survey of AFGL~2591 covers the frequency range of \mbox{330 -- 373\,\GHz\,}with a spectral resolution of 1\,MHz \mbox{($\sim$ 0.8\, \kms).} 
The beam size of the JCMT at these frequencies is \mbox{14~--~15$''$,}
the image size is 2$'$.
Detailed information about the data reduction and analysis can be found in \citep{vanderWiel2011}.

%
%
\section{The HIFI spectral survey of AFGL 2591} \label{section-survey}
%
%
%
\subsection{Detections and line profiles} \label{subsection-detections}
%
%
%
\begin{figure*}
\centering
\includegraphics[width=18cm]{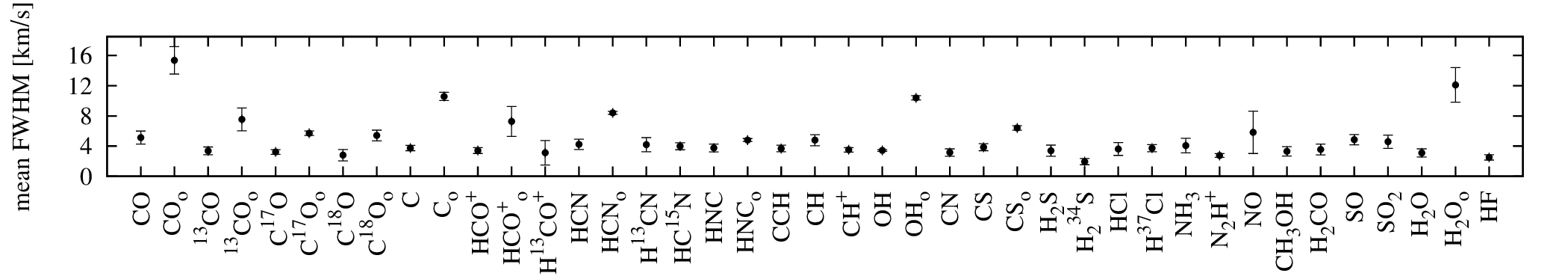}\\
\includegraphics[width=18cm]{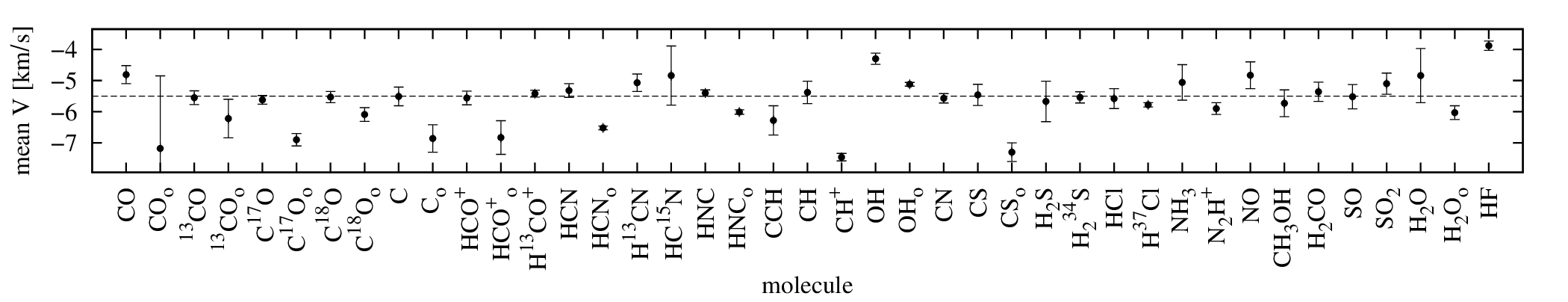}
\caption{The average values of the line widths (top panel) and the average values of the central velocity (bottom panel) from Gaussian fits for the observed emission lines of different molecules (''o'' represents the outflow component). 
The emission lines of the envelope component are centered at $-5.5$\,\kms, as shown by a dashed line at the bottom panel.}
\label{figure:obs}
\end{figure*}

From the \textit{Herschel}/HIFI spectral survey, a total of 32 species (including isotopologues) were identified, resulting in 252 emission and 16 absorption lines (218 different transitions). 
Blended features were excluded from the analysis.
\textit{Herschel} surveys toward different sources revealed many spectral features which are not possible to be identified at this moment \citep[e.g.][]{Wang2011A}. However, no~unidentified lines were found in our spectra.

For the line identification the JPL \citep{Pickett1998} and CDMS \citep{Muller2001,Muller2005} databases were used. 
Line analysis was made with the CASSIS software\footnote{CASSIS (http://cassis.cesr.fr) has been developed by IRAP- UPS/CNRS.}.
The presence of possible transitions resulting from an upper energy level \Eup\,of less than 500~K was checked.
Generally, detected lines have \Eup $<$ 400\,K, except for the high--J CO transitions, which have \Eup\,up to 752\,K. 

All the detected lines in the HIFI survey are presented in Table \ref{table:measurement}, the entire spectrum is shown in Fig.~\ref{figure:spectrum} and corresponding line profiles can be found in Figures~\ref{figure:profiles1}. 
For~the sake of completeness all of the observed lines together with their profiles and measurements are presented in Table~\ref{table:measurement}, including the datasets of Paper~I \citep{vanderWiel2013} and the complementary JCMT data.
 
Although the analysis of our survey revealed no new molecular species, some of our observed species have not been seen toward AFGL 2591 before. 
HIFI with its broad spectral range gave the opportunity to observe for the first time in AFGL 2591 transitions of: HF \citep{Emprechtinger2012}, OH$^+$, CH, CH$^+$ \citep{Bruderer2010b} or C$^+$ and HCl (this work).

Within the object AFGL~2591, CH$_3$OH, SO$_2$ and SO show the highest number of detected transitions (54, 26 and 18 lines, respectively) among its identified species, followed by H$_2$CO as well as CO and its isotopologues.
In the cases of the other molecules, at most a few lines were observed. 
The strongest transitions originate from CO and its isotopologues, HCO$^+$, H$_2$O and OH. 
In comparison, the remaining detected lines are relatively weak due to fluxes below 1~\Kkms.

The line measurements were done in the same way as described in Paper~I.
A Gaussian profile was fitted to each line, using the Levenberg-Marquardt fitter in the line analysis module of CASSIS. 
For most lines, a single Gaussian profile gave a good fit to the profile.
However, in cases of CO, $^{13}$CO, C$^{18}$O, CI, [CII], HCO$^+$, OH and H$_2$O double Gaussian profiles were needed to fit sufficiently narrow and broad line components.
The measured parameters from Gaussian fits of the emission lines (central velocity and full width at half maximum) are plotted in Fig.~\ref{figure:obs} (together with the complementary JCMT data) as an average value for each molecule. 

The narrow and single line components are centered at \mbox{$-$5.5 $\pm$ 0.5\,\kms,} \citep[as derived before by][]{vanderTak1999} and originate from the protostellar envelope. Their line widths are of the order of 3.7~$\pm$~0.9\,\kms. 
Whereas, the broader line components \mbox{(10.9 $\pm$ 4.2\,\kms)} are caused by the outflows and are centered at \mbox{$-$6.3 $\pm$ 0.7\,\kms.}
It was shown in Paper I that the outflow gas is not significantly different from that in the envelope, considering gas density, gas temperature, as well as the chemical balance of CO and HCO$^+$.

%
\subsection{Absorption line analysis} \label{subsection-absorptions}
%
%

\begin{table}
\setlength{\extrarowheight}{0.7pt}
\caption{Fit results for absorptions.}
\label{table:abs}
\centering
\begin{tabular}{lccc}
\hline\hline
Molecule & \Vlsr$^a$ & $\Delta$V$^a$ & N \\
 & [\kms] &[\kms]& [cm$^{-2}$] \\
\hline
CCH    & 0.82& 2.58& 3.3$\pm$1.0 $\times$ 10$^{17}$\\ 
CH$^b$ & 0.21& 2.19& 3.1$\pm$0.9 $\times$ 10$^{13}$\\
CH$^+$ & 4.17& 12.42& 1.1$\pm$0.4 $\times$ 10$^{14}$\\ %
H$_2$S & 0.22& 0.97& 3.5$\pm$0.9 $\times$ 10$^{12}$\\
NH$_3$ & 0.00& 1.30& 1.8$\pm$0.8 $\times$ 10$^{12}$\\
H$_2$O$^b$& -0.50& 2.43& 1.5$\pm$0.6 $\times$ 10$^{13}$\\
OH$^+$$^b$& 3.65& 9.13& 3.0$\pm$1.0 $\times$ 10$^{13}$\\ %
HF     & -0.05& 2.31& 5.2$\pm$1.3 $\times$ 10$^{12}$\\
HF     & -3.88& 2.50& 5.5$\pm$1.4 $\times$ 10$^{12}$\\ %
\hline
H$_2$O &-11.98&13.75& 2.3$\pm$0.6 $\times$ 10$^{13}$\\ 
CH$^+$ &-16.90&9.24& 6.8$\pm$1.3 $\times$ 10$^{13}$\\
HF     &-12.58& 8.81& 1.8$\pm$0.6 $\times$ 10$^{13}$\\
\hline
\end{tabular}
\begin{list}{}{}
\item[$^{\mathrm{a}}$] Errors of \Vlsr and $\Delta$V are listed in Table~\ref{table:measurement}.
\item[$^{\mathrm{b}}$] The average of a few lines from the same velocity component: 2~lines of CH, 2 lines of H$_2$O and 3 lines of OH$^+$.
\end{list}
\end{table}

There are only a few absorption features observed toward AFGL~2591. A foreground cloud at V$_{\mbox{\scriptsize lsr}}$~$\sim$~0\,\kms\,has been detected before by e.g.~\citet{Bruderer2010b,Emprechtinger2012,vanderWiel2013}.
In the CHESS/HIFI dataset we found 16 absorption lines; all measurements are listed together with emissions in Table \ref{table:measurement} and their lines profiles are presented in Figures~\ref{figure:profiles1}. 
Mostly they are red-shifted and associated with the foreground cloud at V$_{\mbox{\scriptsize lsr}}$~$\sim$~0\,\kms.
Three broad, blue-shifted absorptions belong to the outflow lobe.

We derived the molecular column densities using the following relations:
\begin{eqnarray}
\rm{N_{tot}} =  \rm{N_{l}}  \frac{\rm{Q(T_{ex})}}{\rm{g_l}} exp\left({\frac{E_l}{kT_{ex}}}\right) \ [cm^{-2}]
\end{eqnarray}
\begin{eqnarray}
\rm{N_{l}} =  \frac{8\pi \nu^3}{c^3} \frac{\rm{g_l}}{\rm{g_u} \rm{A_{ul}}}   \int \tau \cdot dv  \ [cm^{-2}]
\end{eqnarray}

where $\rm{Q(T_{ex})}$ is the partition function computed at the excitation temperature $\rm{T_{ex}}$,
$\nu$ is the frequency of the observed transition with the Einstein A-coefficient $\rm{A_{ul}}$ and the statistical weights of the lower $\rm{g_l}$ and upper levels $\rm{g_u}$; c is the speed of light and k is the Boltzmann constant.
The line opacity $\tau$ was calculated from the measured brightness temperature $\rm{T_{mb}}$ and the temperature of the background continuum in a single side band $\rm{T_{c}}$, using the relation\, $\tau = -ln(\frac{\rm{T_{mb}}}{\rm{T_{c}}})$.

$\rm{N_{tot}}$ and $\rm{N_{l}}$ are the total column density and the column density in the lower state of transition, respectively. The $\rm{N_{l}}$ may be the same as total column density for the ground state lines, when the excitation temperature is very low \mbox{($\rm{T_{ex}}$ $\sim$ 2.73\,K).} Thus for the ground state transitions we applied the Eq.~2 to calculate column densities. For the absorptions that arise from the excited states we used the Eq.~1 and assumed the excitation temperature of 10\,K as it was derived for the foreground cloud in Paper~I \citep[see Table 4.][]{vanderWiel2013}.

The tentative absorption lines from a foreground cloud at V$_{\mbox{\scriptsize lsr}}$~$\sim$~0\,\kms\,were observed of \mbox{CCH~(7$_{7}$-6$_{6}$} at 611.265\,GHz), CH~(2~transitions: \mbox{$3/2_{2+}-1/2_{1-}$} at 532.724 and \mbox{$3/2_{2-}-1/2_{1+}$} at 536.761\,GHz), \mbox{CH$^+$ ($1-0$} at 835.138\,GHz), \mbox{H$_2$S~($2_{12}-1_{01}$} at~736.034\,GHz), \mbox{NH$_3$~($1_{0}-0_{0}$} at 572.498\,GHz), H$_2$O~(2 transitions: \mbox{$1_{10}-1_{01}$} at 556.936 and \mbox{$1_{11}-0_{00}$} at 1113.343\,GHz), OH$^+$~(3~transitions: \mbox{$J$=$2$--$1, F$=$3/2$--$1/2$} at 971.805, \mbox{$J$=$1$--$1, F$=$3/2$--$1/2$} at 1033.004 and \mbox{$J$=$1$--$1, F$=$3/2$--$3/2$} at 1033.119\,GHz) and \mbox{HF~($1-0$} at 1232.476\,GHz). The estimated column densities for above species are listed in the upper part of Table \ref{table:abs}.  

Three broad absorptions are associated with the outflow (centered at $\sim-$\,13.8\,\kms): H$_2$O, CH$^+$ and HF. 
Their column densities are presented in the lower part of Table \ref{table:abs}.

\citet{Bruderer2010b}, using HIFI, analysed hydrides toward AFGL~2591. Our column density results are in good agreement, within the errors, with their measurements: 
\mbox{3.1 $\times$ 10$^{13}$} and \mbox{2.6 $\times$ 10$^{13}$ cm$^{-2}$} for CH, 
\mbox{6.8 $\times$ 10$^{13}$} and \mbox{1.8 $\times$ 10$^{14}$ cm$^{-2}$} for CH$^+$ outflow component, 
\mbox{1.1 $\times$ 10$^{14}$} and \mbox{1.2 $\times$ 10$^{14}$ cm$^{-2}$} for CH$^+$, and \mbox{3.0 $\times$ 10$^{13}$} and \mbox{6.1 $\times$ 10$^{13}$ cm$^{-2}$} for OH$^+$, our results and from \citep{Bruderer2010b} respectively. \citet{Bruderer2010b} in their spectra found also lines of NH and H$_2$O$^+$. 
These two species are not seen in our dataset, because of a slightly lower quality of spectral scans (\citet{Bruderer2010b} have observations from the single frequency settings).

Recently, based on \textit{Herschel} data, \citep{Barlow2013} detected in the Crab Nebula emission lines of\, $^{36}$ArH$^+$. Absorptions of this ion are also seen toward sources from HEXOS (Herschel Observations of EXtra-Ordinary Sources) and PRISMAS (PRobing InterStellar Molecules with Absorption line Studies) \textit{Herschel} Key Programs \citep{Schilke2014arXiv14037902S}. The $^{36}$ArH$^+$ J=1-0 transition at 617.525~GHz is not seen in our spectra. The upper limit of a~column density is \mbox{7.7 $\times$ 10$^{12}$ cm$^{-2}$} for the width of an absorption line of 1~\kms.

%
\subsection{Emission line analysis} \label{subsection-diagrams}
%
%
\begin{table*}
\setlength{\extrarowheight}{4.6pt}
\caption{Parameters estimated from rotational and population diagrams methods (column densities, excitation temperatures and emission extents) based on HIFI and JCMT data. 
}
\label{table:rot-diag}
\centering
\begin{tabular}{lcc|ll|lll|ccc}      
\hline\hline 
 Molecule & V & FWHM&\multicolumn{2}{c|}{Rotational}& \multicolumn{3}{c|}{Population} & $\tau$  &\Eup\,range& No of trans.\\
\hline
 &[\kms] &[\kms]&  N$_{\mbox{\scriptsize col}}$[cm$^{-2}$] & T$_{\mbox{\scriptsize ex}}$[K] & N$_{\mbox{\scriptsize col}}$[cm$^{-2}$] & T$_{\mbox{\scriptsize ex}}$[K]& size[$''$]&   & [K] &\\
\hline  
CO        &-4.8$^{\pm 0.3}$&5.1$^{\pm 0.9}$& 6.0 $\times$ 10$^{16}$ $_{-0.5}^{+0.5}$ & 162$_{-9}^{+10}$& 1.2 $\times$ 10$^{19}$ $_{-0.6}^{+0.4}$&62$_{-6}^{+4}$& 17$_{-7}^{+3}$&0.1--144& 33--752&12\\
          &-7.2$^{\pm 2.3}$&15.4$^{\pm 1.8}$& 6.0 $\times$ 10$^{16}$ $_{-1.0}^{+1.0}$ & 89$_{-9}^{+10}$& 8.0 $\times$ 10$^{18}$ $_{-0.4}^{+0.4}$&42$_{-2}^{+2}$& 17$_{-1}^{+1}$&0.01--34&33--752&12\\ 
HCO$^+$   &-5.6$^{\pm 0.2}$&3.4$^{\pm 0.4}$& 2.3 $\times$ 10$^{13}$ $_{-1.0}^{+1.0}$ &  35$_{-3}^{+3}$& 1.0 $\times$ 10$^{14}$ $_{-1.0}^{+1.0}$&43$_{-3}^{+2}$& 11$_{-1}^{+1}$&0.08--2.02&43--283&7\\
          &-6.8$^{\pm 0.6}$&7.3$^{\pm 2.0}$& 2.2 $\times$ 10$^{13}$ $_{-1.6}^{+4.2}$&  23$_{-15}^{+15}$& 2.0 $\times$ 10$^{15}$ $_{-1.6}^{+3.3}$ & 19$_{-2}^{+9}$& 9.7$_{-2.4}^{+0.8}$&0.7--35&43--154&4\\                     
HCN       &-5.3$^{\pm 0.2}$&4.2$^{\pm 0.7}$& 4.5 $\times$ 10$^{13}$ $_{-0.5}^{+0.7}$& 31$_{-10}^{+14}$& 1.1 $\times$ 10$^{15}$ $_{-0.7}^{+0.8}$&35$_{-1}^{+1}$& 7.7$_{-0.2}^{+0.2}$&0.2--4.4&43--234&6\\
HNC\tablefootmark{a}&-5.4$^{\pm 0.1}$&3.8$^{\pm 0.5}$&4.8 $\times$ 10$^{12}$ $_{-0.5}^{+6.3}$ &  43$_{-7}^{+9}$& & & & & 44--122&3\\
CCH       &-6.3$^{\pm 0.5}$&3.7$^{\pm 0.5}$&  2.2 $\times$ 10$^{14}$ $_{-0.2}^{+0.3}$ &  22$_{-3}^{+3}$& 1.1 $\times$ 10$^{16}$ $_{-0.5}^{+1.1}$& 25$_{-5}^{+6}$&5.3$_{-0.6}^{+0.8}$&0.3--7.1&42--151&4\\
CN\tablefootmark{a} &-5.6$^{\pm 0.2}$&3.2$^{\pm 0.5}$&        9.7 $\times$ 10$^{13}$ $_{-0.7}^{+0.8}$ &  22$_{-1}^{+1}$& 1.3 $\times$ 10$^{14}$ $_{-0.3}^{+0.3}$& 26$_{-3}^{+4}$&23$_{-4}^{+6}$&0.01--0.18&33--114&3\\
CS        &-5.5$^{\pm 0.4}$&3.9$^{\pm 0.5}$& 7.4 $\times$ 10$^{13}$ $_{-0.4}^{+0.5}$ &26$_{-10}^{+12}$& 4.9 $\times$ 10$^{13}$ $_{-0.6}^{+14.6}$&61$_{-19}^{+7}$& 14$_{-9}^{+4}$&0.01--0.09&66--282&7\\
H$_2$S  &-5.7$^{\pm 0.7}$&3.4$^{\pm 0.8}$&1.1 $\times$ 10$^{13}$ $_{-0.7}^{+1.1}$  &  56$_{-14}^{+25}$& 4.9 $\times$ 10$^{14}$ $_{-0.6}^{+0.9}$& 26$_{-2}^{+3}$&8.9$_{-0.7}^{+0.6}$&0.01--5.6&55--350&5\\
NH$_3$&-5.1$^{\pm 0.6}$&4.1$^{\pm 1.0}$& 2.8 $\times$ 10$^{13}$ $_{-0.2}^{+0.3}$ &  67$_{-5}^{+6}$& 4.8 $\times$ 10$^{13}$ $_{-2.4}^{+11}$& 28$_{-6}^{+3}$&9.6$_{-0.6}^{+1.8}$&0.1--1.6&28--170&5\\
N$_2$H$^+$\tablefootmark{a} &-5.9$^{\pm 0.2}$&2.8$^{\pm 0.3}$& 5.6 $\times$ 10$^{11}$ $_{-1.6}^{+0.1}$ &  19$_{-3}^{+13}$& & & & & 45--125&3\\
NO\tablefootmark{a} &-4.8$^{\pm 0.5}$&5.8$^{\pm 2.8}$&        7.2 $\times$ 10$^{15}$ $_{-0.5}^{+0.7}$ & 25$_{-10}^{+14}$& 1.7$\times$ 10$^{16}$ $_{-0.5}^{+3.6}$&54$_{-14}^{+9}$& 12$_{-6}^{+15}$&0.015--0.021& 36--115&2\\
CH$_3$OH&-5.7$^{\pm 0.5}$&3.3$^{\pm 0.6}$& 1.8 $\times$ 10$^{14}$ $_{-0.7}^{+1.0}$ & 209$_{-62}^{+171}$& 1.5 $\times$ 10$^{17}$ $_{-0.3}^{+0.4}$& 108$_{-7}^{+10}$&1.5$_{-0.1}^{+0.1}$&0.6--10.4& 25--352&49\\
H$_2$CO &-5.4$^{\pm 0.3}$&3.6$^{\pm 0.7}$& 2.0 $\times$ 10$^{13}$ $_{-0.9}^{+1.0}$ & 34$_{-6}^{+12}$& 9.9 $\times$ 10$^{13}$ $_{-0.1}^{+0.1}$& 41$_{-2}^{+2}$&7.3$_{-0.1}^{+0.1}$&0.02--0.61&32--263&14\\
SO   &-5.5$^{\pm 0.4}$&4.9$^{\pm 0.7}$& 1.5 $\times$ 10$^{14}$ $_{-0.2}^{+0.2}$ & 53$_{-20}^{+23}$& 1.9 $\times$ 10$^{16}$ $_{-0.3}^{+0.4}$&64$_{-4}^{+2}$&2.7$_{-0.2}^{+0.3}$&0.1--6.1&26--405&22\\
SO$_2$    &-5.1$^{\pm 0.4}$&4.6$^{\pm 0.9}$& 3.0 $\times$ 10$^{14}$ $_{-1.7}^{+2.0}$ & 92$_{-43}^{+77}$& 5.4 $\times$ 10$^{17}$ $_{-0.6}^{+0.7}$ & 175$_{-4}^{+5}$&0.9$_{-0.1}^{+0.1}$&0.5--8.9&31-354&47\\
H$_2$O&-4.8$^{\pm 0.9}$&3.1$^{\pm 0.6}$& 3.5 $\times$ 10$^{13}$ $_{-1.4}^{+3.0}$ & 63$_{-15}^{+17}$& 2.4 $\times$ 10$^{15}$ $_{-0.5}^{+0.3}$ & 38$_{-1}^{+1}$&9.1$_{-0.6}^{+0.5}$&0.4--104&53--305&8\\
      &-6.0$^{\pm 0.2}$&12.1$^{\pm 2.3}$&  5.5 $\times$ 10$^{13}$ $_{-1.9}^{+3.1}$ & 43$_{-20}^{+35}$& 1.0 $\times$ 10$^{16}$ $_{-0.4}^{+0.4}$ & 31$_{-3}^{+3}$&4.9$_{-0.4}^{+0.3}$&0.2--5.7&101--305&6\\
\hline 
\end{tabular}
\tablefoot{
The second column shows the values of the central velocity of the observed lines. In case of two values, first one corresponds to the envelope component while the second one to the outflow.\\
The last 3 columns 
show first the range of optical depth $\tau$ for observed lines, second the \Eup~range which is covered by observed features and third the number of lines from different energy levels used for the analysis; 
i.e., we observed 4 lines of NO, but they originate only from 2 different energy levels.\\
Population diagram method was used when at least 
4 lines of a given molecule were observed, thus providing no values for HNC and N$_2$H$^+$.
\\
\tablefoottext{a}{Indicates higher uncertainty of measurements because e.g. only 3 different levels were observed.}
}
\end{table*}

\begin{figure*}
\centering
\includegraphics[width=15cm]{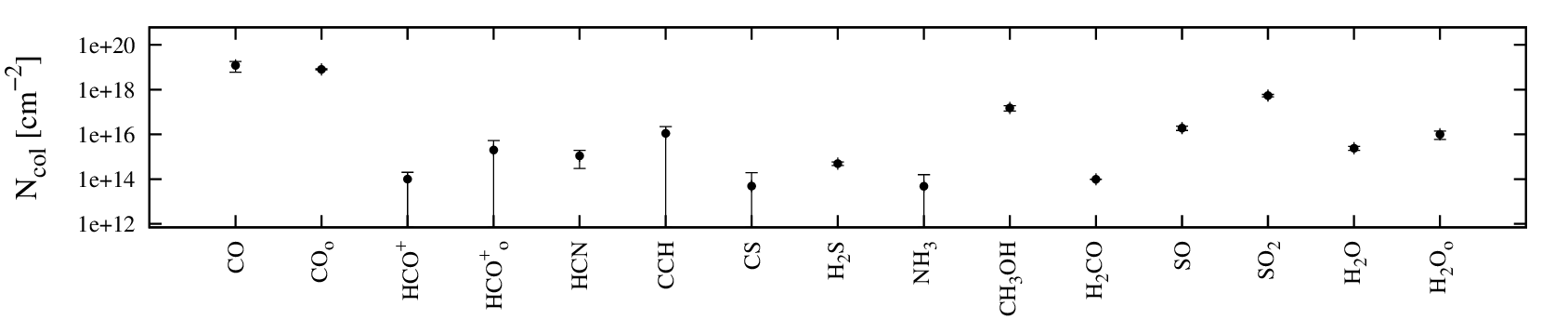}
\includegraphics[width=15cm]{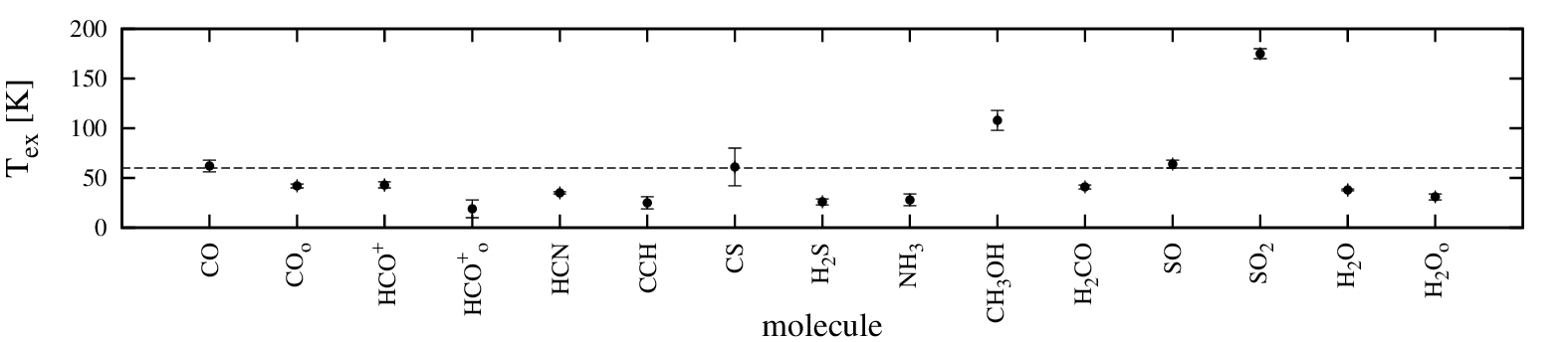}
\caption{Column densities and excitation temperatures estimated from the population diagrams, without the uncertain measurements, ''o'' sign represents the outflow component. }
\label{figure:tkin-survey}
\end{figure*}

\begin{figure*}
\includegraphics[width=3.6cm]{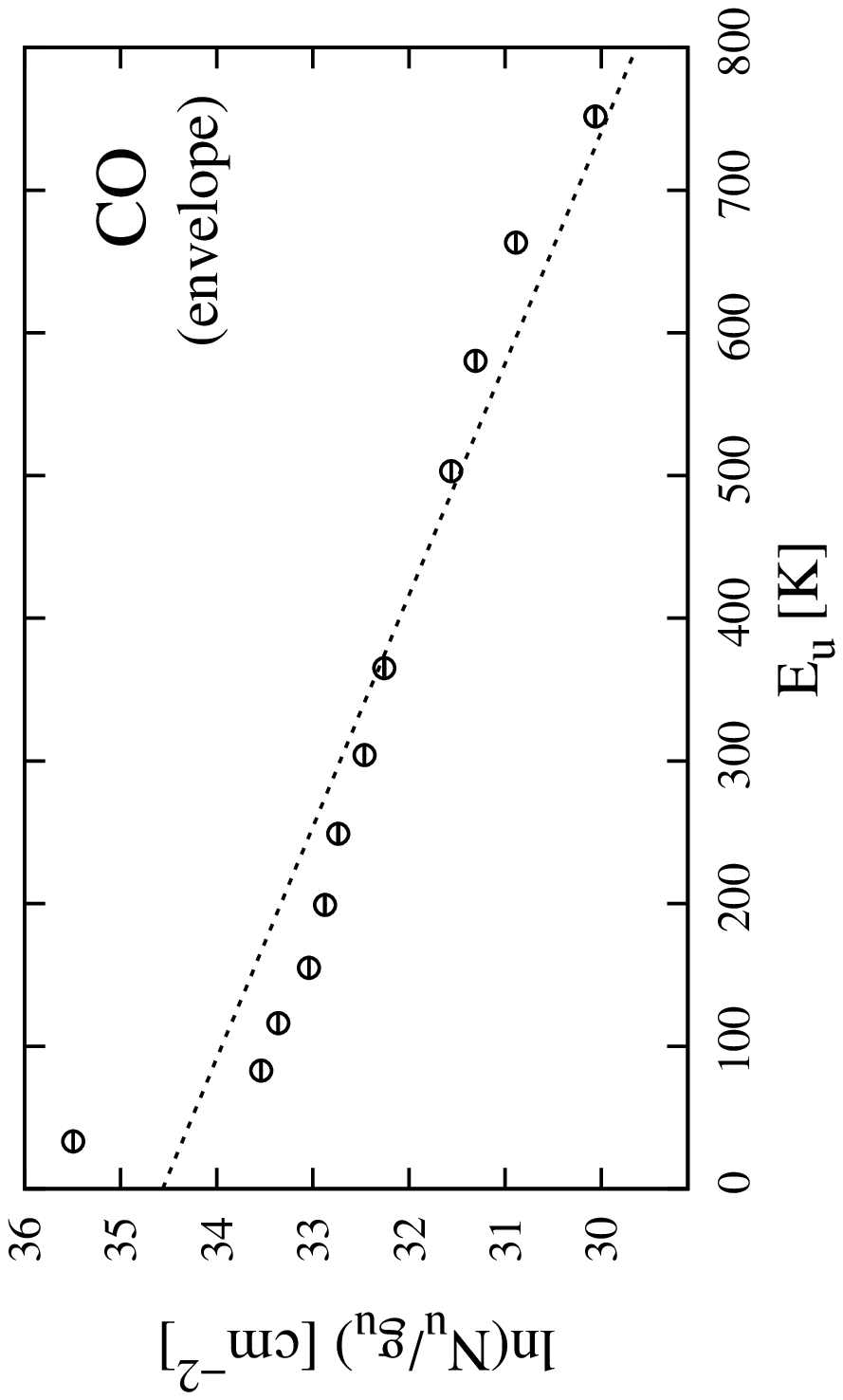}
\includegraphics[width=3.6cm]{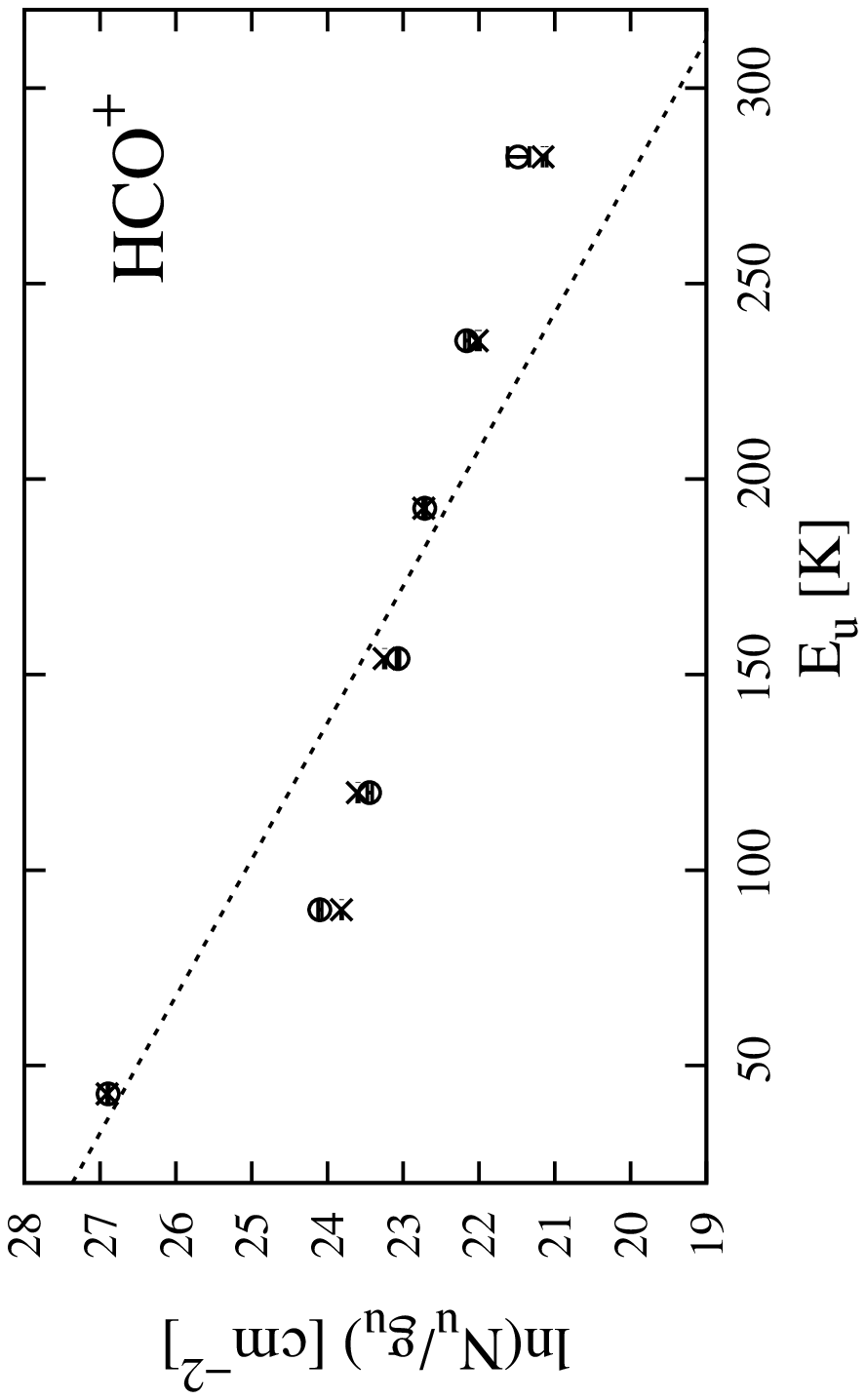}
\includegraphics[width=3.6cm]{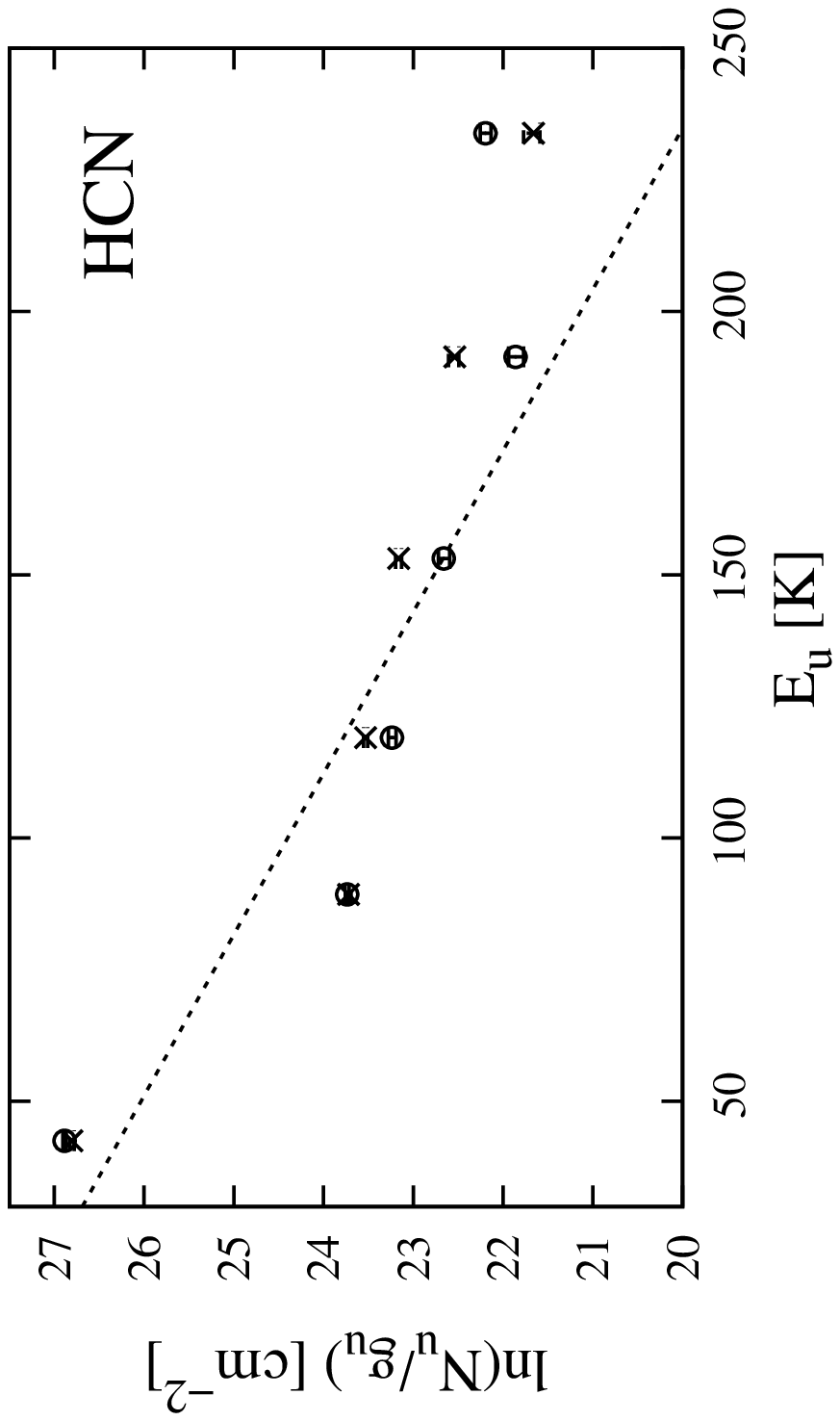}
\includegraphics[width=3.65cm]{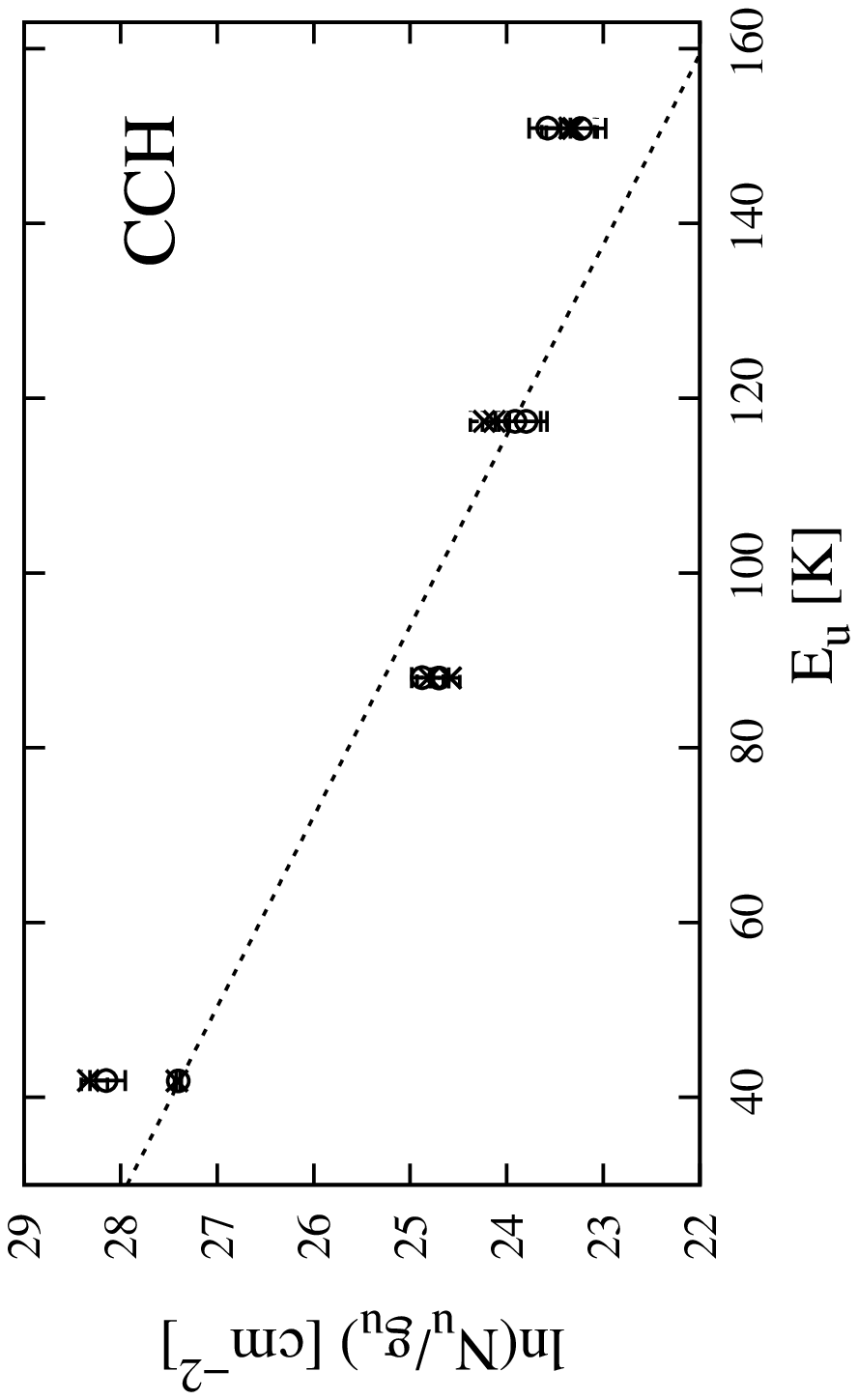}
\includegraphics[width=3.6cm]{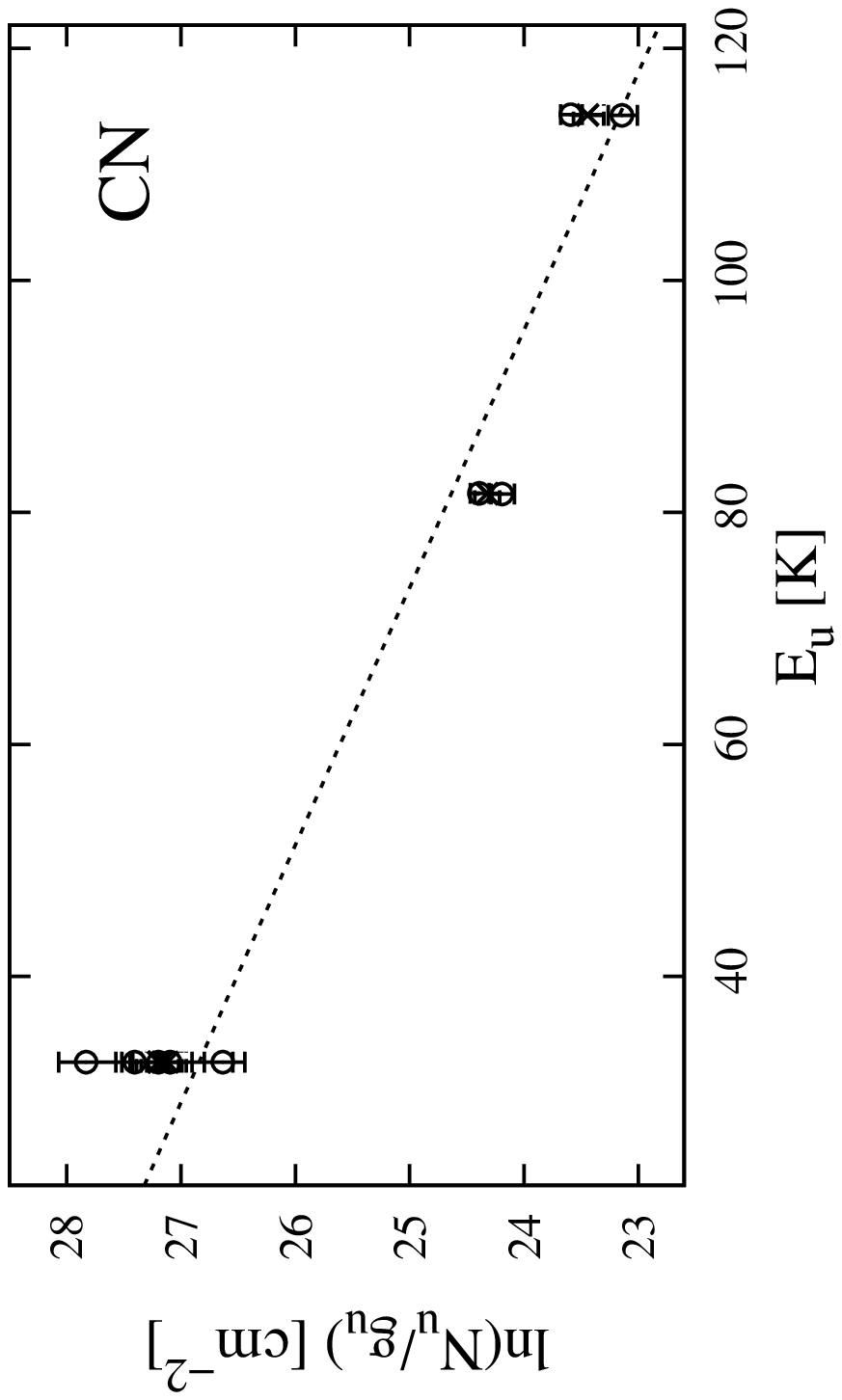}\\
\includegraphics[width=3.6cm]{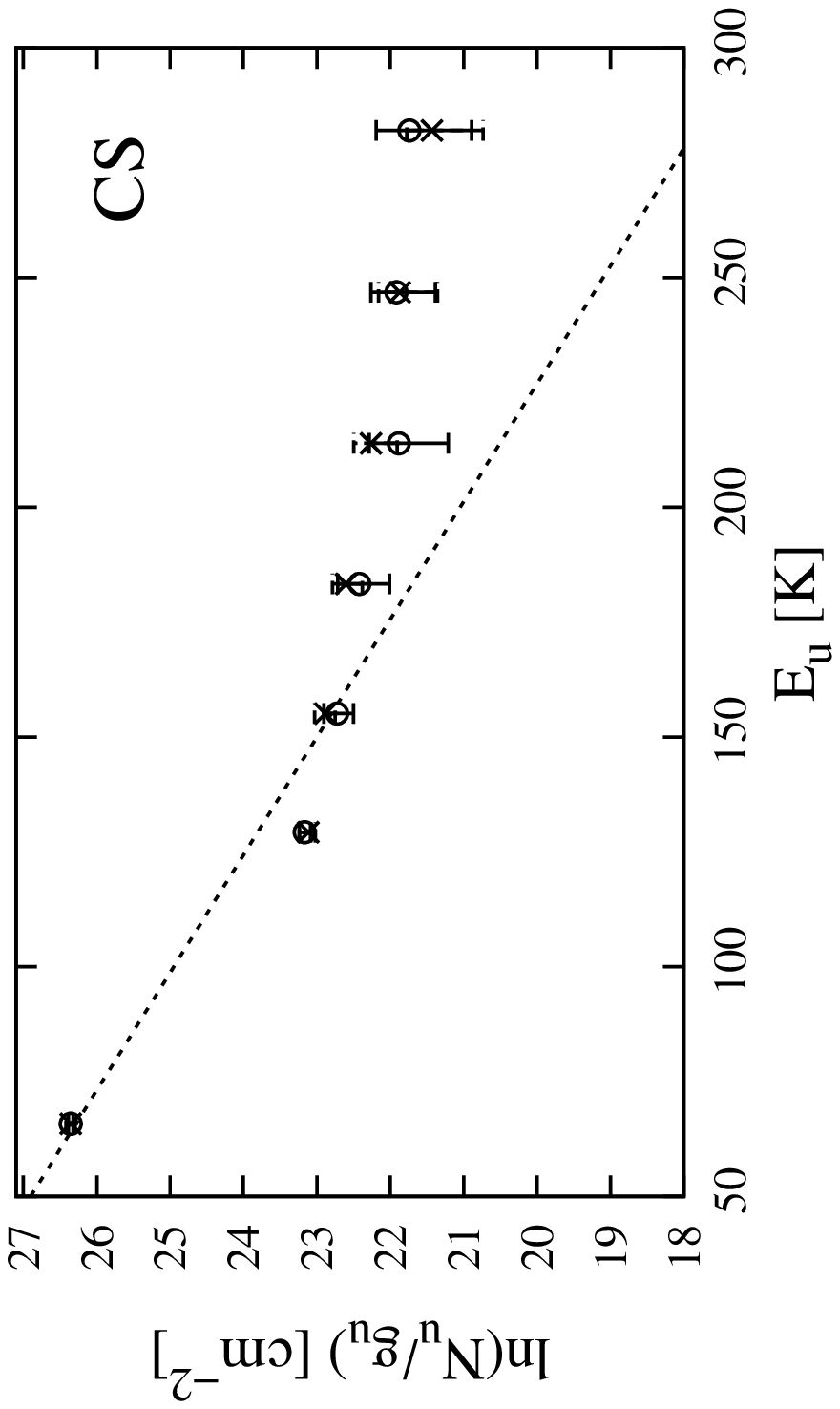}
\includegraphics[width=3.6cm]{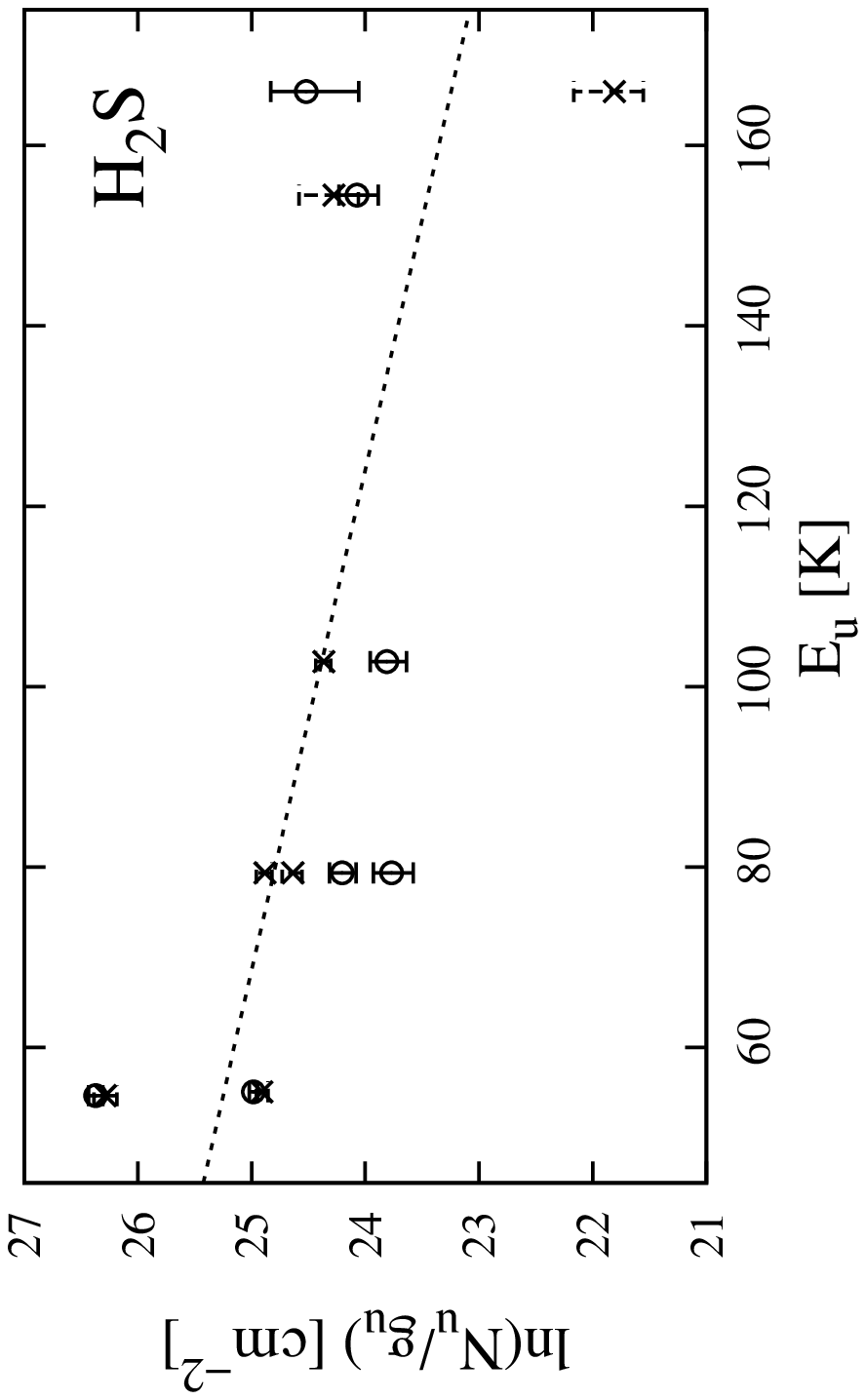}
\includegraphics[width=3.6cm]{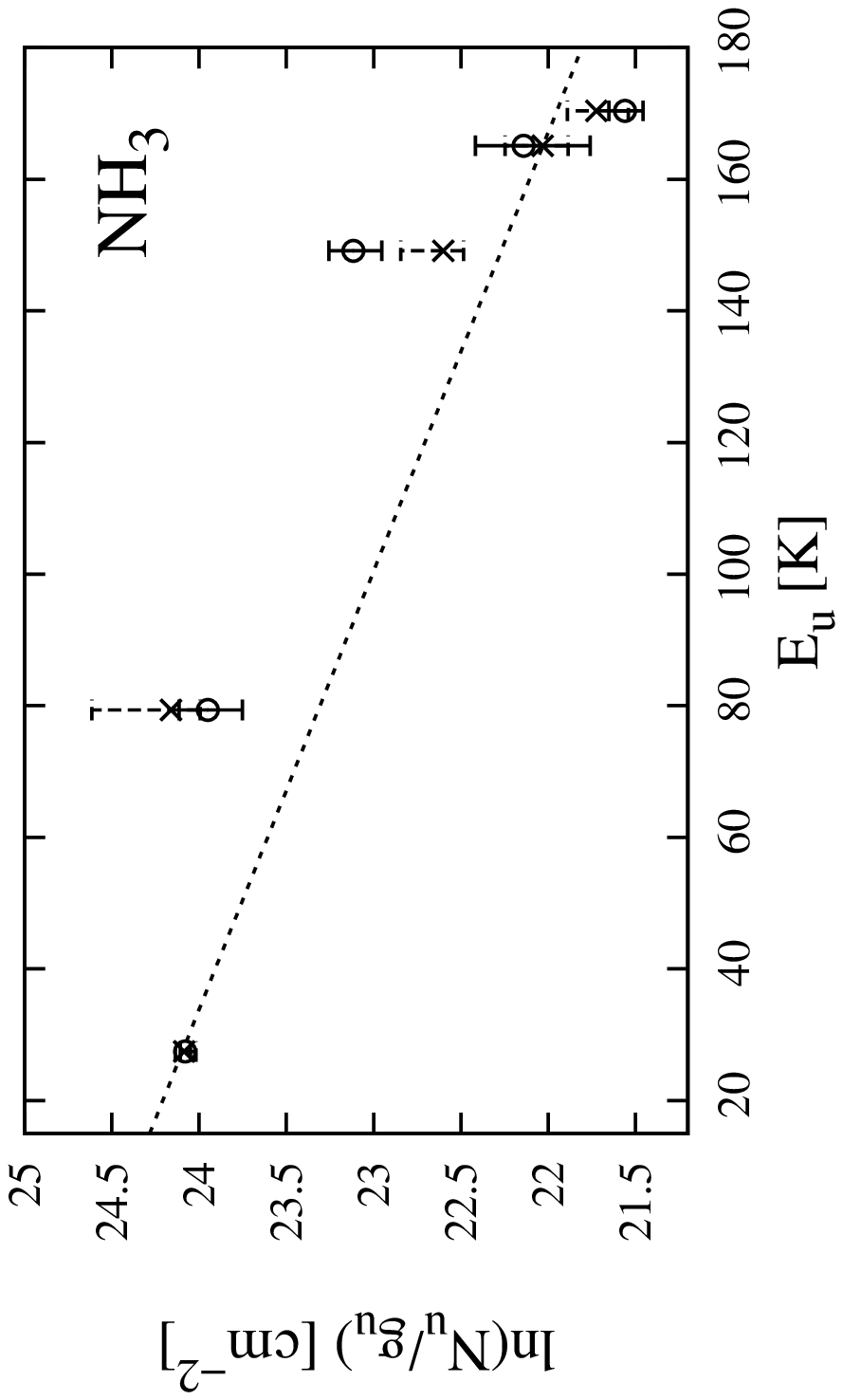}
\includegraphics[width=3.6cm]{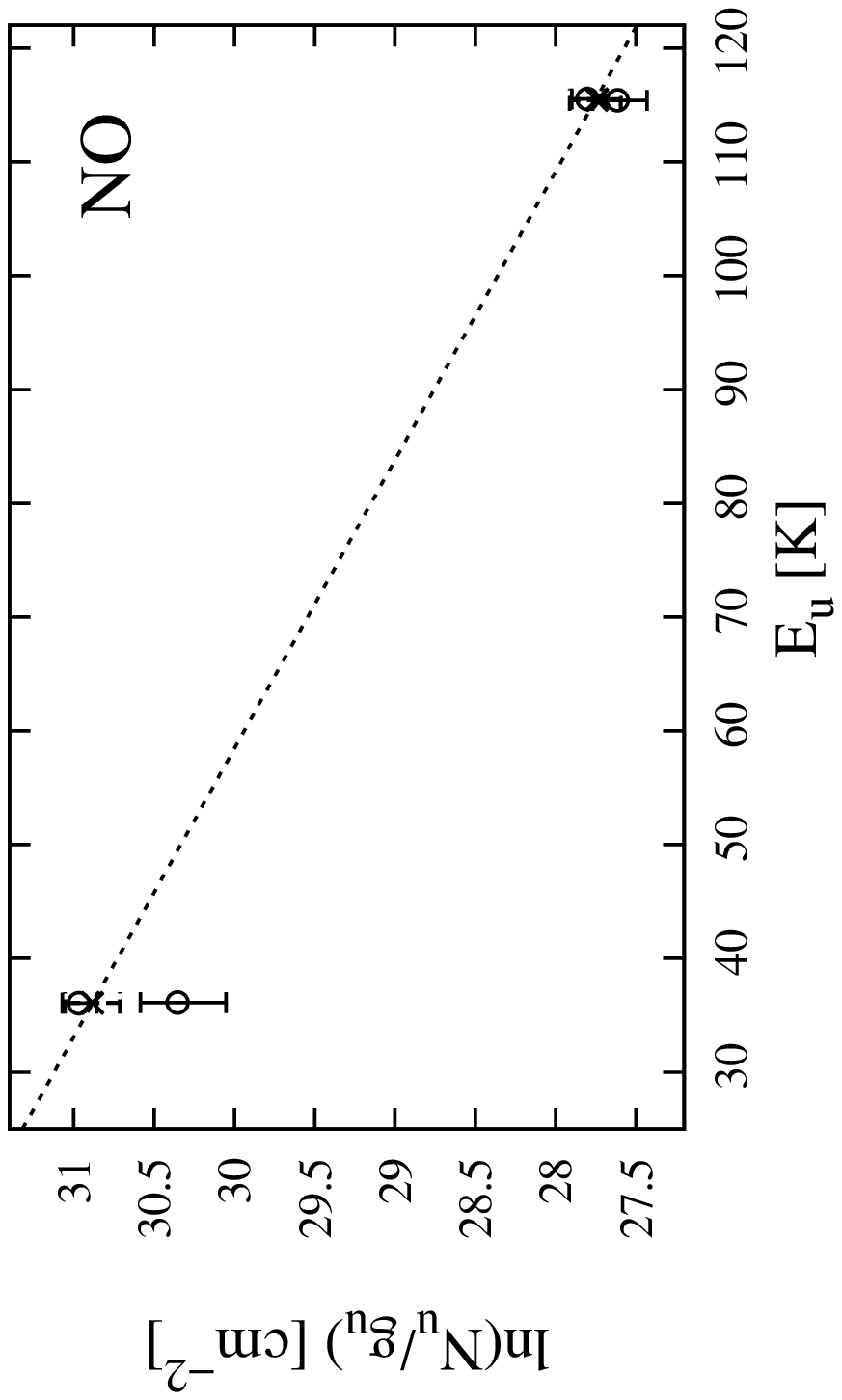}
\includegraphics[width=3.6cm]{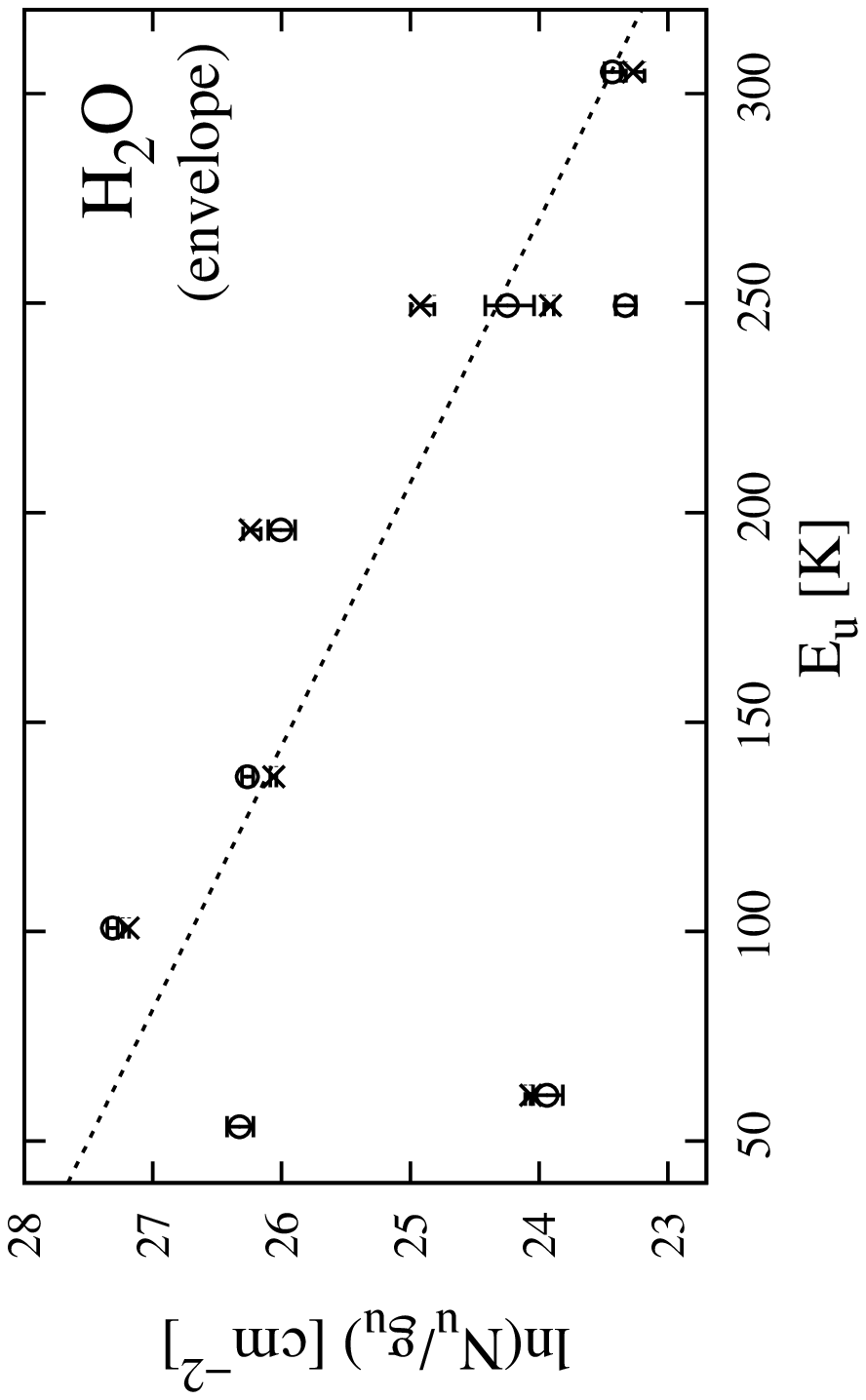}\\
\includegraphics[width=3.6cm]{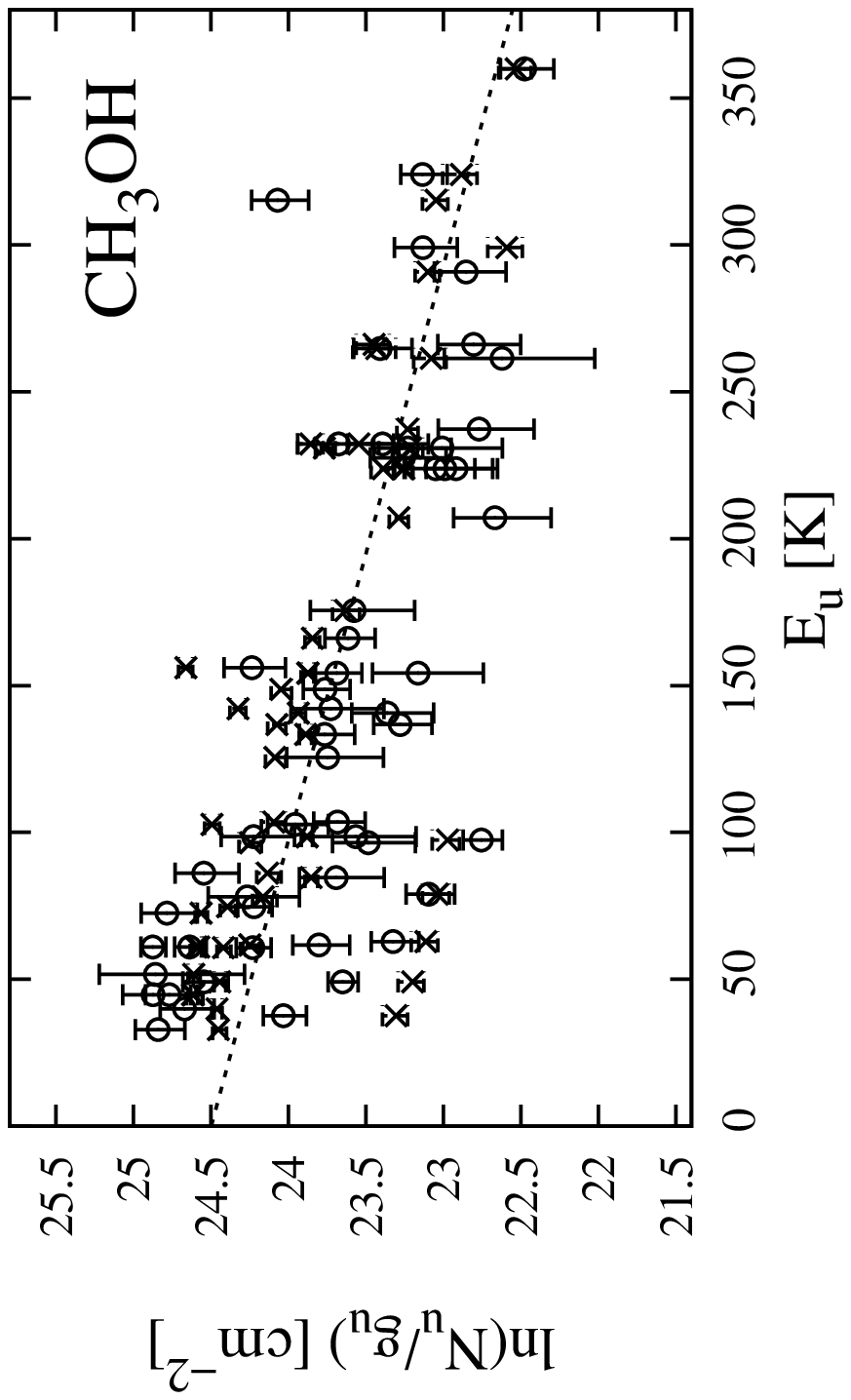}
\includegraphics[width=3.6cm]{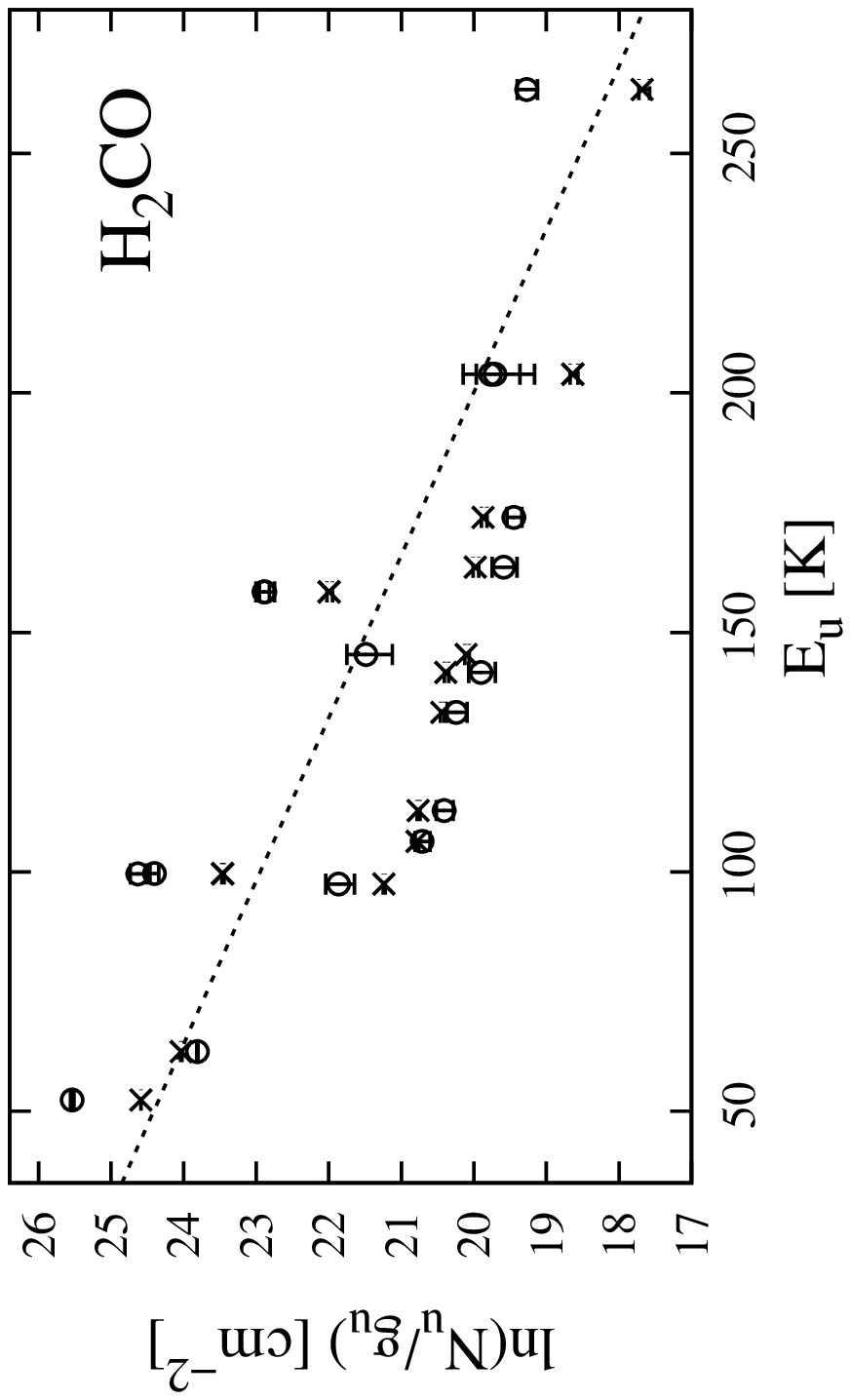}
\includegraphics[width=3.6cm]{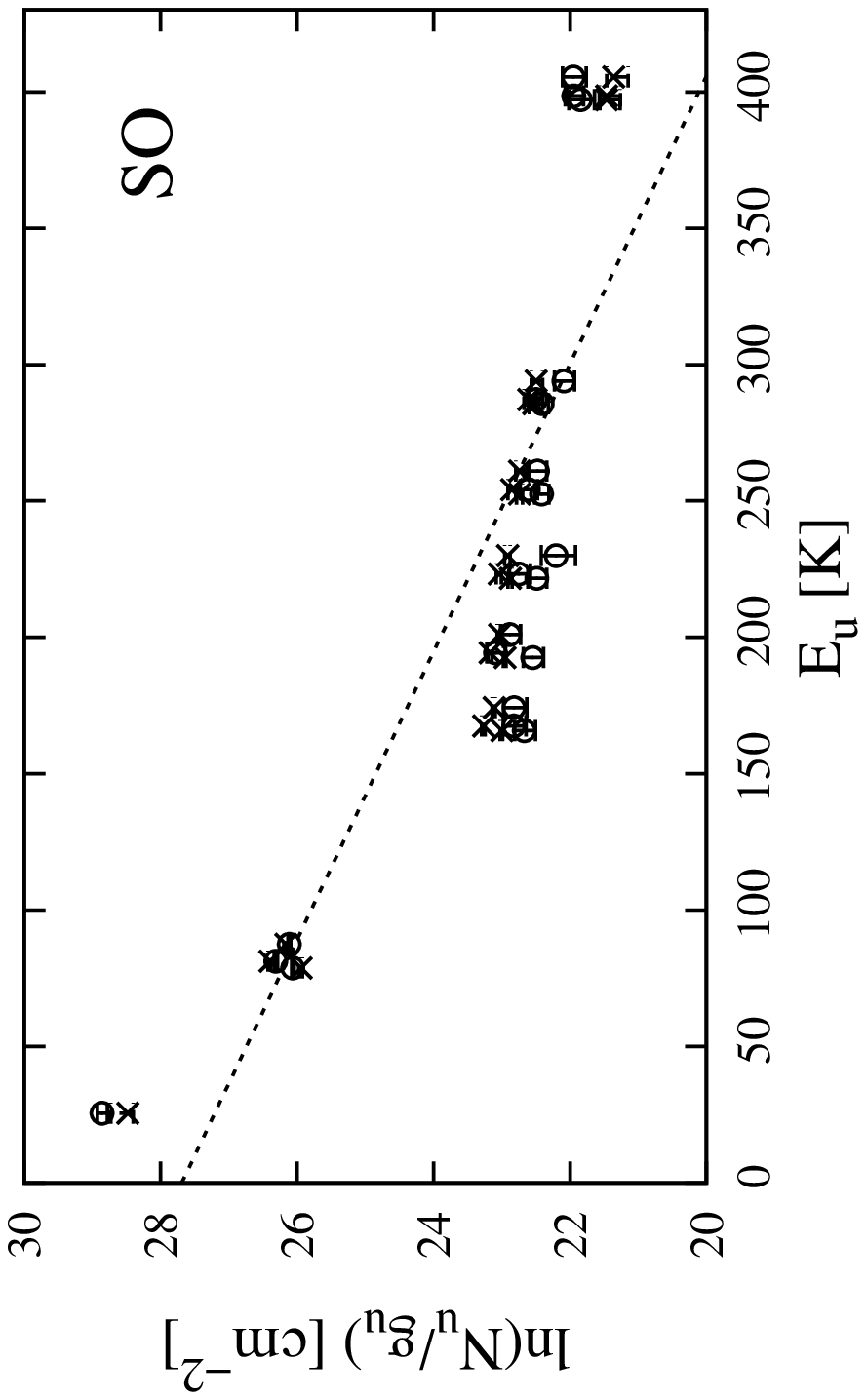}
\includegraphics[width=3.6cm]{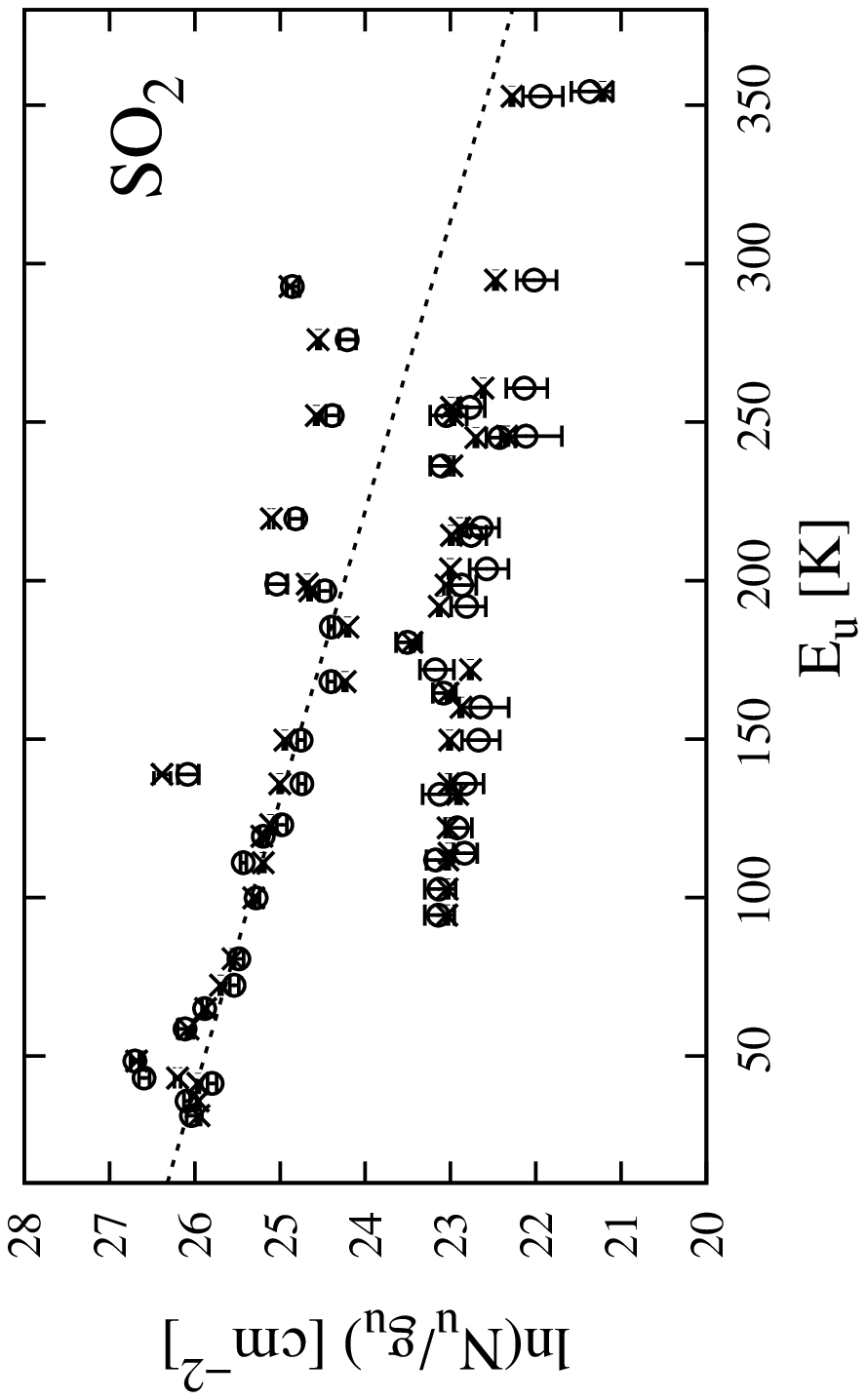}
\caption{Population diagrams. Open circles represent the observational data and crosses are the best-fit model from population diagram analysis. Dotted lines correspond to a linear fit to the rotational diagram.}
\label{figure:pop1}%
\end{figure*}

To estimate column densities and excitation temperatures from the observed emissions we constructed rotational diagrams
which assume that all lines for a given molecule have the same excitation temperature.
The rotational diagram method is a useful tool for the estimation of the column densities and the excitation temperatures when many transitions of particular species are observed. However, in many cases its accuracy is limited since it is based on the assumptions that the emission lines are optically thin and the emissions fill the beam.

\citet{Goldsmith1999} improved this excitation analysis method by introducing correction factors for the effects of the 
beam dilution and optical depth. 
Using this \textit{population diagram method} we estimated the column density, the excitation temperature and the emission extent for each molecule with the observed multiple transitions.
Having three free parameters (column density, excitation temperature and the beam filling factor) we used this method only when at least 4 lines for a given molecule were observed. Otherwise, only the rotational method was applied.
The rotation diagram gives beam-averaged column densities, while the population diagram gives source-averaged values.
Hereafter, all stated column densities (N$_{\mbox{\scriptsize col}}$) or excitation temperatures (T$_{\mbox{\scriptsize ex}}$) were derived from the population diagrams, except those of HNC and N$_2$H$^+$ which were estimated from the rotational diagrams.
At this point, the complementary JCMT data were crucial to increase the number of observed transitions for a given molecule.

The column densities of CO, HCN and HCO$^+$ were obtained from their isotopologues ($^{13}$CO, C$^{18}$O, C$^{17}$O, H$^{13}$CN, HC$^{15}$N, H$^{13}$CO$^+$) using the standard isotopic ratios: \mbox{$^{12}$C/$^{13}$C = 60,} \mbox{$^{16}$O/$^{18}$O = 500,} \mbox{$^{16}$O/$^{17}$O = 2500} and \mbox{$^{14}$N/$^{15}$N = 270} \citep{Wilson1994}.

All column densities and excitation temperatures values based on the rotational and population diagrams methods are given in Table~\ref{table:rot-diag}.
The opacities and emission sizes for each molecule derived from the population diagrams are listed in Table~\ref{table:rot-diag} as well. Table~\ref{table:rot-diag} contains also information about the covered energy \Eup\,range for a given species and the number of lines from different energy levels which were used for the analysis.
The values of the excitation temperatures and column densities are plotted in Fig.~\ref{figure:tkin-survey}, excluding the uncertain measurements (i.e.~HNC, N$_2$H$^+$, CN and NO).

Based on the optical depths values from Table~\ref{table:rot-diag}, lines of CN, CS, NO and H$_2$CO can be characterised as optically thin ($\tau <$~0.6).
However, results of CN and NO are uncertain because of only a few observed lines.
For optically thin lines calculations based on the rotational diagrams 
resulted in good approximations of the column densities and the excitation temperatures. 
The other molecular lines were characterised as optically thick. For those molecular species the population diagram method was more accurate.

The emission extent of analysed molecules associated with AFGL~2591 ranges from around 2$''$ (species like SO, SO$_2$ and CH$_3$OH) up to 23$''$ (CN). 
For most species emission sizes are smaller than 17$''$.

From the comparison of the temperatures derived from the population diagrams (see the bottom panel of Fig.~\ref{figure:tkin-survey}) it is possible to distinguish warm (e.g.~CH$_3$OH, SO$_2$) and cold (e.g.~HCN, H$_2$S, NH$_3$) species.
As cold species we classify these having excitation temperatures up to 70\,K.
Warm molecules have higher temperatures, up to 175~K for~SO$_2$. 
It~is difficult to give an accurate borderline here and classify all species, however, the large range of excitation temperatures seems significant.
Moreover, it was shown before by \citet{Bisschop2007} that some of the complex organic species can be classified as both, warm and cold, which may indicate that they are present in multiple physical components.

The population diagrams are presented in Fig.~\ref{figure:pop1}. They show evidence for excitation gradient of several species (HCO$^+$, HCN, CS, SO), which means that the population diagram method may be not enough to analyse all observed molecules.
This is a motivation to use in the near future more sophisticated method (i.e.,~radiative transfer modeling) to study our spectral survey.

%
\section{Discussion} \label{discussion-survey}
%
%
%
\subsection{CI and CII} \label{discussion-c}
%
%
C and C$^+$ are the only atomic species found in our HIFI spectral survey of AFGL~2591. Both fine-structure transitions of neutral carbon, $^3$P$_1-^3$P$_0$ at 492\,GHz and $^3$P$_2-^3$P$_1$ at 809\,GHz, were observed towards AFGL~2591. These transitions consist of two components originating from the envelope and the outflow, similar to the CO lines (see Fig.~\ref{figure:profiles1}).
CI was observed previously in AFGL~2591 by \citet{vanderTak1999}, but [CII] was observed for the first time with \textit{Herschel}.
\mbox{The [CII] $^2$P$_{3/2}-^2$P$_{1/2}$} line, an important interstellar coolant, shows several velocity components, two of them correspond to the ones in CI and CO. The [CII] line profile is distorted by a contamination from the off-position even after applying corrections within HIPE (Fig.~\ref{figure:profiles1}).

%
%
\subsection{CO and its isotopologues} \label{discussion-co}
%
%
CO is one of the most studied molecules \citep[e.g.][]{Mitchell1989,Black1990,Hasegawa1995}. 
Based on CO observation, \citet{Lada1984} found an extended bipolar outflow associated with AFGL 2591.
Many strong lines of CO and its isotopologues ($^{13}$CO, C$^{18}$O, C$^{17}$O) were also detected in our HIFI spectra showing clearly the envelope and outflow components. C$^{17}$O lines are weaker, and show only the envelope components.
The abundance of \mbox{CO = 3 $\times$ 10$^{-5}$} was calculated in Paper~I.
The CO column density in this work was estimated at \mbox{1.2 $\times$ 10$^{19}$ cm$^{-2}$}, \citet{vanderTak2000} derived a similar value of \mbox{3.4 $\times$ 10$^{19}$ cm$^{-2}$}.

%
%
\subsection{HCO$^+$} \label{discussion-hco+}
%
%
HCO$^+$ was identified by intense lines in the HIFI and JCMT spectra. Moreover, three lines of H$^{13}$CO$^+$ were also positively detected.
The abundance of HCO$^+$ was estimated at \mbox{9 $\times$ 10$^{-9}$} (Paper~I) and column density at~\mbox{1.0 $\times$ 10$^{14}$ cm$^{-2}$}.
\citet{Carr1995} estimated the abundance of \mbox{4 $\times$ 10$^{-10}$} and \citet{vanderTak1999} using a model with lower $H_2$ column density derived \mbox{[HCO$^+$] = 1 $\times$ 10$^{-8}$}.

%
%
\subsection{N-bearing species} \label{discussion-n}
%
%
Six N-bearing species were observed in the HIFI spectra: HCN, HNC, CN, NO, N$_2$H$^+$ and NH$_3$. All of these molecules have been seen before in AFGL~2591 \citep[e.g.][]{Takano1986,Carr1995,Boonman2001}.
Lines of N-bearing species observed with the \textit{Herschel}/HIFI are weak in comparison to CO and 
were sufficiently fitted with a single Gaussian profile revealing these species to be components of the protostellar envelope, centered at $-$5.5\,\kms. 
Only o-NH$_3$ shows a tentative absorption feature from a foreground cloud at V$_{\mbox{\scriptsize lsr}}$~=~0\,\kms. 
Two features observed with the JCMT, HCN \mbox{4-3} and HNC \mbox{4-3} show a contribution from the outflow and double Gaussian profiles were fitted to these lines.
We did not find NH and NH$_2$, which were seen in other HIFI spectral surveys \citep[e.g.][]{Zernickel2012}. Upper limits are 0.8~\Kkms\,for the NH 1-0 line near 946\,\GHz\, and 0.6\,\Kkms\,for the NH$_2$ 1-0 line near 953\,\GHz. Upper limits were measured in the same way as in \mbox{Paper I,} i.e. considering 3 \kms\,a typical line width, hence using 5${\sigma}_{rms}$ $\times$ 3\kms.
Among the observed features, two lines of vibrationally excited HCN \mbox{4-3,} $\upsilon$=1c and $\upsilon$=1d are found (JCMT data). Line $\upsilon$=1c was observed before by \citet{vanderTak1999}. \citet{Boonman2001} analysed excited HCN, the 4-3 and 9-8 transitions. 
The interferometric observations from \citet{Veach2013} showed vibrationally excited $\upsilon$=1 and also $\upsilon$=2 HCN 4-3 lines. These authors suggest that the $\upsilon$=2 HCN lines may be a useful tool to study a protostellar disk.
\citet{Takano1986} observed ammonia transitions (1,1) and (2,2) with the Effelsberg 100~m telescope. They found a compact NH$_3$ cloud of around 0.6\,pc diameter around the central source. These authors estimated a column density of \mbox{8 $\times$ 10$^{13}$cm$^{-2}$}. 
In comparison, calculations of our work gave a column density of \mbox{4.8 $\times$ 10$^{13}$cm$^{-2}$}.

%
%
\subsection{S-bearing species} \label{discussion-s}
%
%
From the S-bearing molecules we detected with HIFI: CS, H$_2$S, H$_2$$^{34}$S, SO and SO$_2$. 
All of these molecules have been seen before in AFGL 2591 \citep[e.g.][]{Yamashita1987,vanderTak2003,Bruderer2009c}.
Additionally, from JCMT dataset we have several lines of the mentioned above molecules and also isotopologues of CS, SO and SO$_2$ ($^{13}$CS and C$^{34}$S, $^{34}$SO, $^{34}$SO$_2$), as well as OCS and o-H$_2$CS. 
SO and SO$_2$ show many weak lines of the envelope component.
SO$_2$ is the example of warm species with the excitation temperature of 175~K,  whereas H$_2$S is classified as colder species with the excitation temperature of 26~K. CS and SO have similar excitation temperatures, 61~K and 64~K, respectively.
\citet{vanderTak2003} studied the sulphur chemistry in the envelopes of massive star-forming regions and found the excitation temperatures of 185~K for SO$_2$, which is a similar results to the one calculated in this work. However, the column density of SO$_2$ varies a lot, \mbox{5.2 $\times$ 10$^{14}$cm$^{-2}$} and \mbox{5.4 $\times$ 10$^{17}$cm$^{-2}$}, \citet{vanderTak2003} and our work, respectively.
Results of column density of CS also differ in one order of magnitude, \mbox{3 $\times$ 10$^{13}$cm$^{-2}$} and \mbox{4.9 $\times$ 10$^{14}$cm$^{-2}$}, \citep{vanderTak2003} and our work, respectively.
The population diagram method is a good first step for the spectral surveys analysis, but in some cases more advanced method is needed. Especially, when there are not enough observed transitions from the lower energy levels for a given molecule, e.g. SO or CS and the excitation gradient is visible (see Fig.~\ref{figure:pop1}).
We are planning for a near future to use radiative transfer modeling and estimate molecular abundances. 

%
%
\subsection{CCH, CH, CH$^+$, OH and OH$^+$} \label{discussion-ch}
%
%
Our spectra also revealed lines from the protostellar envelope and foreground clouds belonging to CCH, CH, CH$^+$, OH and OH$^+$.
CCH and CH show three absorption lines \mbox{at $\sim$0\,\kms} while OH$^+$ three absorptions at $\sim$3.6\,\kms.
Using HIFI, \citet{Bruderer2010b,Bruderer2010} found lines of CH, CH$^+$, NH, OH$^+$ and H$_2$O$^+$, while lines of NH$^+$ and SH$^+$ have not been detected.
\citet{Bruderer2010b} concluded that absorption lines of NH, OH$^+$ and H$_2$O$^+$ originate from a foreground cloud and an outflow lobe, while emission lines of CH and CH$^+$ are connected with the protostellar envelope (compare Sect.~\ref{subsection-absorptions}). 

%
%
\subsection{Water} \label{discussion-h2o}
%
%
Water lines have also been detected in our spectra. We found 4 transitions of o-H$_2$O ($1_{10}-1_{01}$ at 557\,GHz, $3_{12}-3_{03}$ at 1097\,GHz, $3_{12}-2_{21}$ at 1153\,GHz and $3_{21}-3_{12}$ at 1163\,GHz) and 4 transitions of p-H$_2$O ($2_{11}-2_{02}$ at 752\,GHz, $2_{02}-1_{11}$ at 988\,GHz, $1_{11}-0_{00}$ at 1113\,GHz and $2_{20}-2_{11}$ at 1229\,GHz). 
They show different profiles, mostly the envelope and outflow components, but also some absorptions (see Fig.~\ref{figure:profiles1}).
For the envelope component we estimated a column density of 2.4 $\times$ 10$^{15}$cm$^{-2}$, an excitation temperature of 38\,K and an emission extent of 9.1$''$.
The full analysis of water lines in AFGL~2591 as part of the WISH Project (Water In Star-forming regions with Herschel) will be presented in a forthcoming paper of \citet{Choi2014}.

%
%
\subsection{HF} \label{discussion-hf}
%
%
HF is the only detected fluorine-bearing species in AFGL~2591. Its \mbox{1-0} transition at 1233\,GHz was observed and analysed by \citet{Emprechtinger2012}. They calculated HF column density of\, \mbox{2 $\times$ 10$^{14}$cm$^{-2}$} and\, \mbox{4 $\times$ 10$^{13}$cm$^{-2}$}, for emission and absorption respectively.

%
%
\subsection{HCl} \label{discussion-hcl}
%
%
Thanks to HIFI many chlorine-bearing molecules (e.g. HCl, H$^{37}$Cl, H$_2$Cl$^+$, H$_2$$^{37}$Cl$^+$) were observed in different environments, e.g. toward protostellar shocks \citep{Codella2012}, diffuse clouds \citep{Monje2013} and star-forming regions \citep{Neufeld2012}.
HCl and H$^{37}$Cl are the only observed chlorine-bearing species in our HIFI spectra of AFGL~2591. Three hyperfine components of HCl from the energy level of \mbox{\Eup $=$ 30\,K} and two from the higher state \mbox{\Eup $=$ 90.1\,K} were detected. 
In agreement with \citep{Neufeld2012} 
neither lines of H$_2$Cl$^+$ nor lines of H$_2$$^{37}$Cl$^+$ toward AFGL~2591 were found.

%
%
\subsection{Complex species} \label{discussion-complex}
%
%
From the HIFI spectral survey we found only two molecules (i.e. methanol and formaldehyde) which belong to complex organics. \citet{Bisschop2007} showed before that AFGL~2591 is a line-poor source. These authors analysed complex organic molecules in massive young stellar objects and found only a few of them in AFGL~2591; all of the intensities of the observed lines were very low.
Many weak CH$_3$OH and H$_2$CO lines were detected in our HIFI spectra.
Their column densities and excitation temperatures are:
\mbox{1.5 $\times$ 10$^{17}$cm$^{-2}$} and 108\,K for CH$_3$OH, and
\mbox{9.9 $\times$ 10$^{13}$cm$^{-2}$} and 41\,K for H$_2$CO.
\citet{vanderTak2000a} estimated:
\mbox{1.2 $\times$ 10$^{15}$cm$^{-2}$} and 163\,K for CH$_3$OH, and
\mbox{8.0 $\times$ 10$^{13}$cm$^{-2}$} and 89\,K for H$_2$CO.
From the rotational diagrams \citet{Bisschop2007} derived \mbox{4.7 $\times$ 10$^{16}$cm$^{-2}$} and 147\,K for methanol. 
All of these results slightly vary, but also suggest that methanol represents warm species.

%
%
\section{Conclusions} \label{Conclusions}
%
%

The main conclusions concerning AFGL~2591 spectral survey are as follows:
\begin{enumerate}
\item In the \textit{Herschel}/HIFI spectral survey of AFGL~2591 we observed 268 lines (excluding blends) of a total 32 species. 
\mbox{No unidentified} features were found in the spectra.
JCMT data supplemented the excitation analysis of several species seen in emissions.
\item Among the observed 268 lines, 16 absorptions were detected. 
Mostly they  belong to the known foreground cloud at V$_{\mbox{\scriptsize lsr}}$~$\sim$~0\,\kms.
Three broad absorptions are associated with the outflow lobe.
The estimated column densities are in good agreement with previous work.
\item Based on the population diagrams method, the column densities and excitation temperatures were estimated. Molecular column densities range from \mbox{6 $\times\, 10^{11}$} to \mbox{1 $\times\, 10^{19}$~cm$^{-2}$} and excitation temperatures range from 19 to 175~K.
We can distinguish between species of higher (e.g. CH$_3$OH, SO$_2$) and lower (e.g. HCN, H$_2$S, NH$_3$) excitation temperature.  
\item The population diagram method is a very useful tool for spectral surveys analysis, however, it is far from being perfect.
Several species (HCO$^+$, HCN, CS, SO) show evidence for excitation gradient, which is a motivation to use in the near future more sophisticated method (i.e.,~radiative transfer modeling) to study molecules observed in the protostellar envelope of AFGL~2591.
\end{enumerate}

%
%
\begin{acknowledgements}
%
%
We thank Matthijs van der Wiel for providing JCMT data and useful discussions.

HIFI has been designed and built by a consortium of institutes and university departments from across Europe, Canada and the United States under the leadership of SRON Netherlands Institute for Space Research, Groningen, The Netherlands and with major contributions from Germany, France and the US. Consortium members are: Canada: CSA, U.Waterloo; France: CESR, LAB, LERMA, IRAM; Germany: KOSMA, MPIfR, MPS; Ireland, NUI Maynooth; Italy: ASI, IFSI-INAF, Osservatorio Astrofisico di Arcetri-INAF; Netherlands: SRON, TUD; Poland: CAMK, CBK; Spain: Observatorio Astron{\'o}mico Nacional (IGN), Centro de Astrobiología (CSIC-INTA). Sweden: Chalmers University of Technology - MC2, RSS \& GARD; Onsala Space Observatory; Swedish National Space Board, Stockholm University - Stockholm Observatory; Switzerland: ETH Zurich, FHNW; USA: Caltech, JPL, NHSC.

\end{acknowledgements}

%
%
\bibliographystyle{aa}
\bibliography{afgl-2591-survey}

\Online

\begin{appendix}
\section{HIFI/CHESS spectral survey} \label{spectral_survey}

\begin{figure*}
\subfloat{\includegraphics[width=4.43cm]{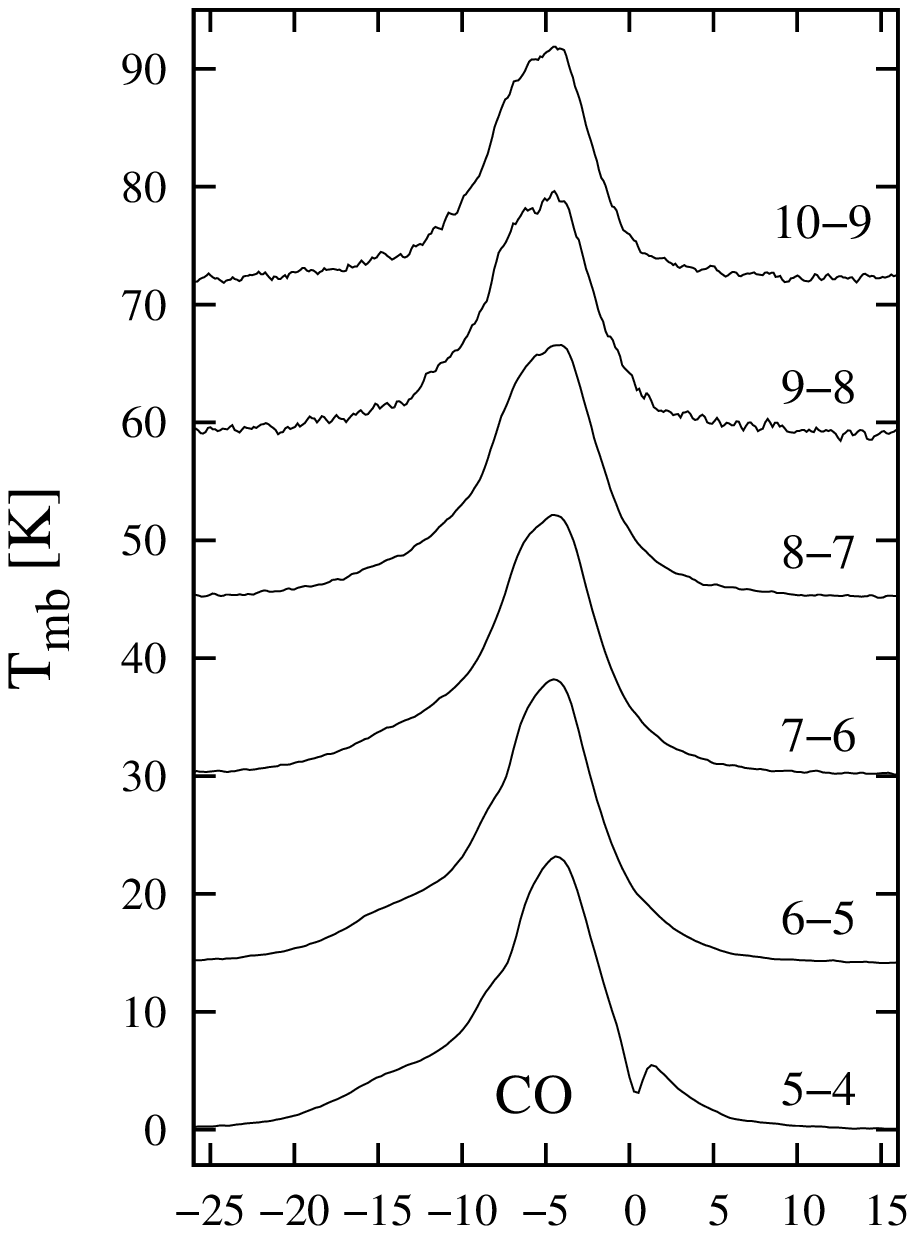}}
\subfloat{\includegraphics[width=4.43cm]{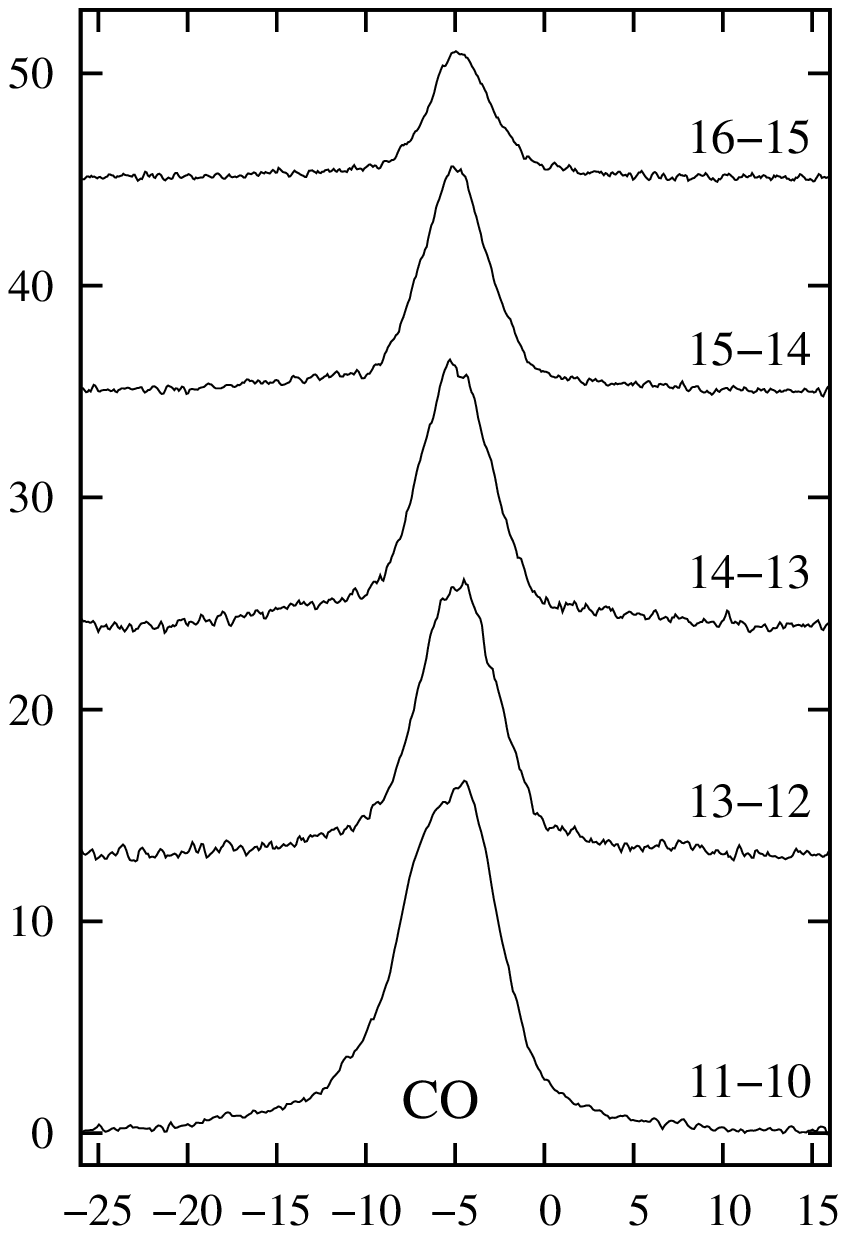}}
\subfloat{\includegraphics[width=4.43cm]{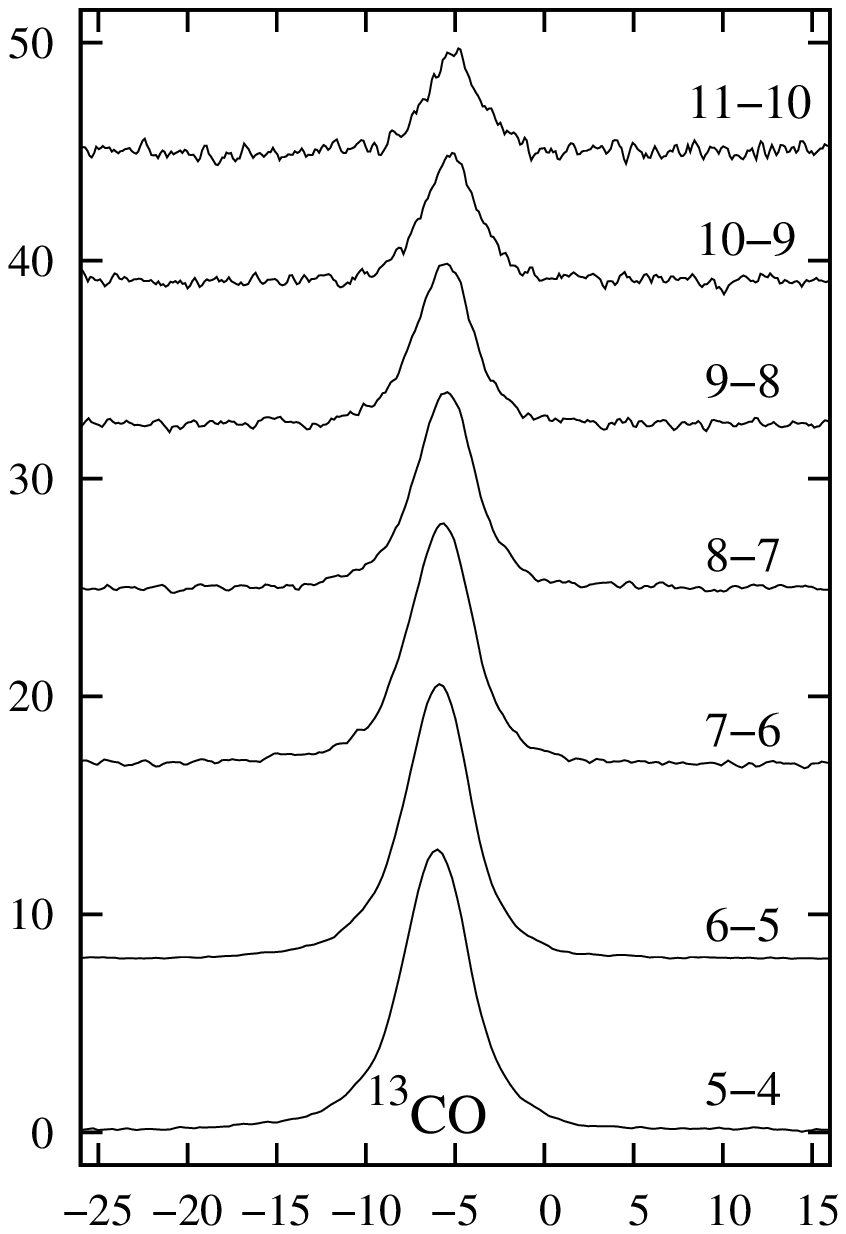}}
\subfloat{\includegraphics[width=4.43cm]{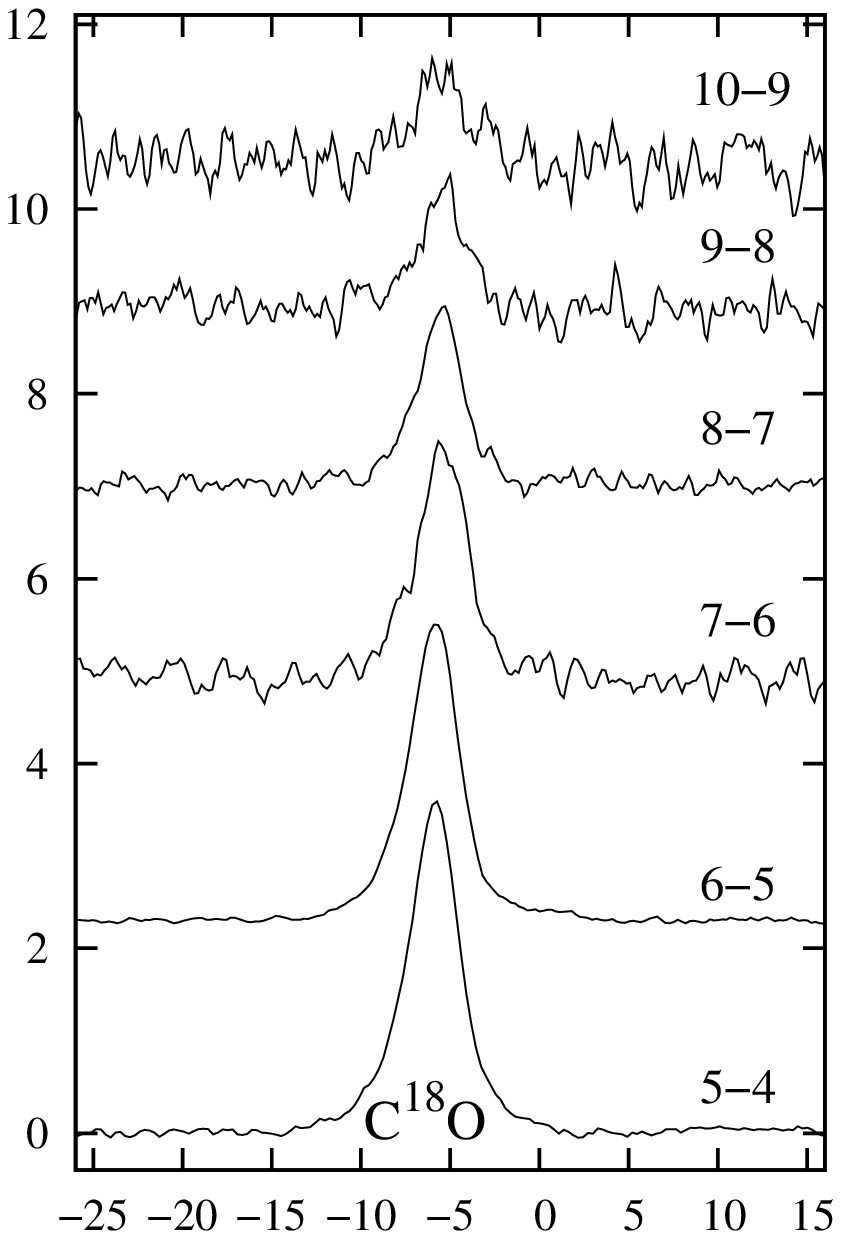}}\\
\subfloat{\includegraphics[width=4.43cm]{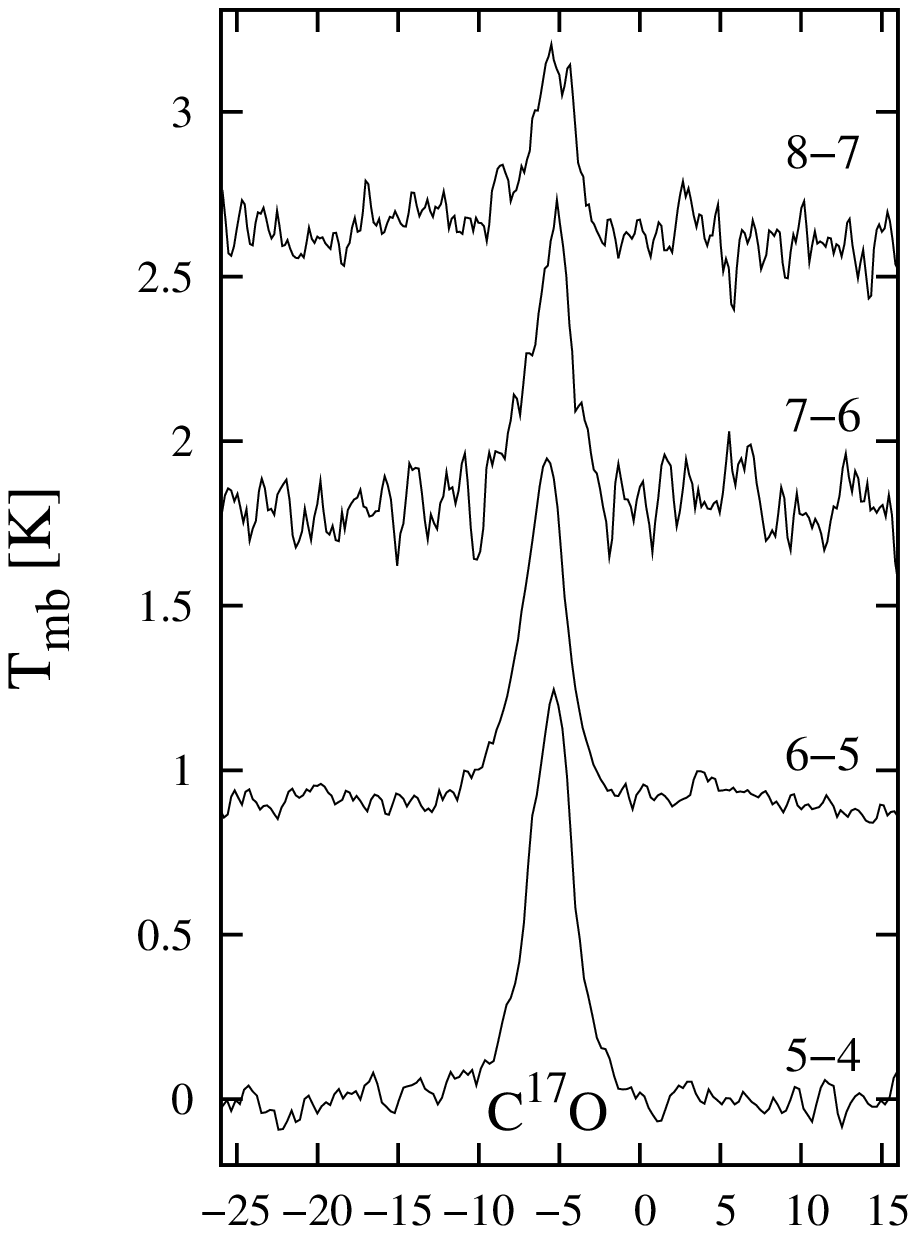}}
\subfloat{\includegraphics[width=4.43cm]{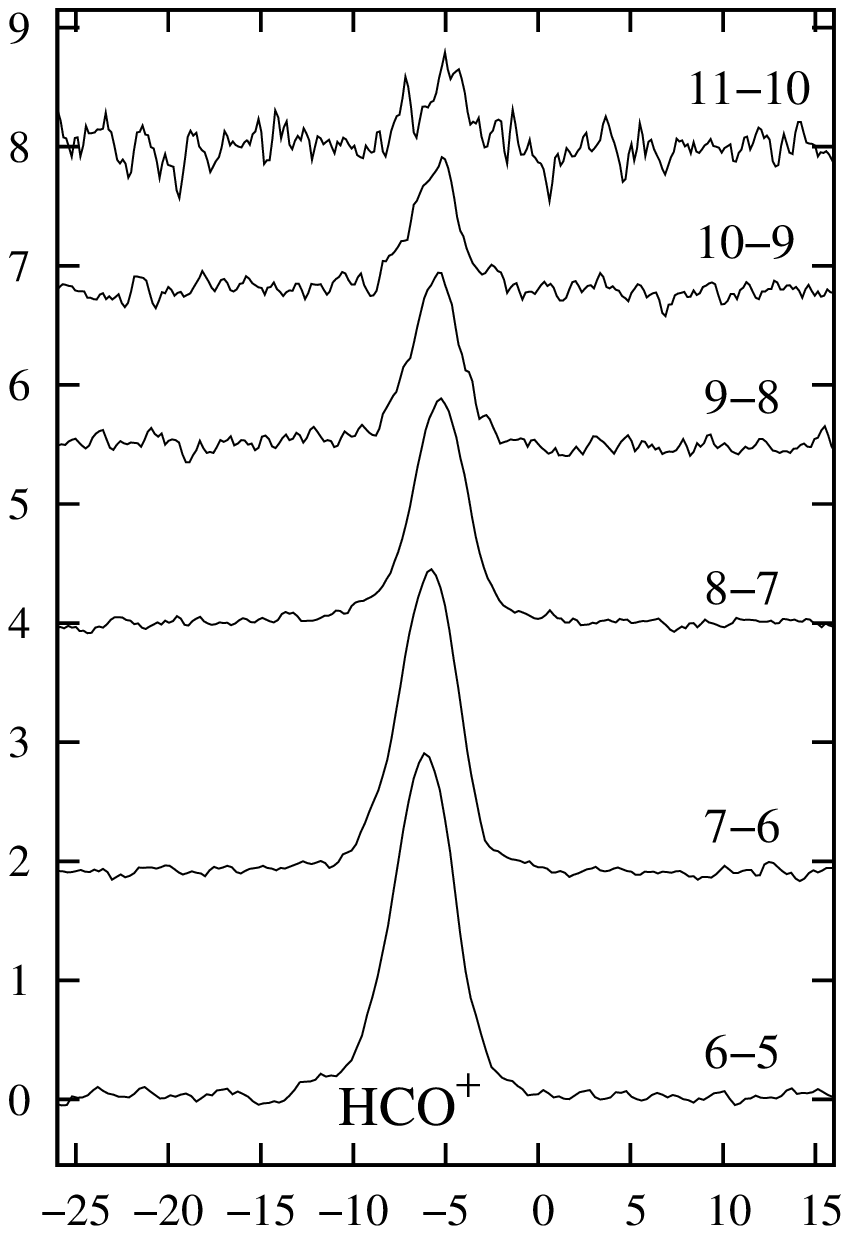}}
\subfloat{\includegraphics[width=4.43cm]{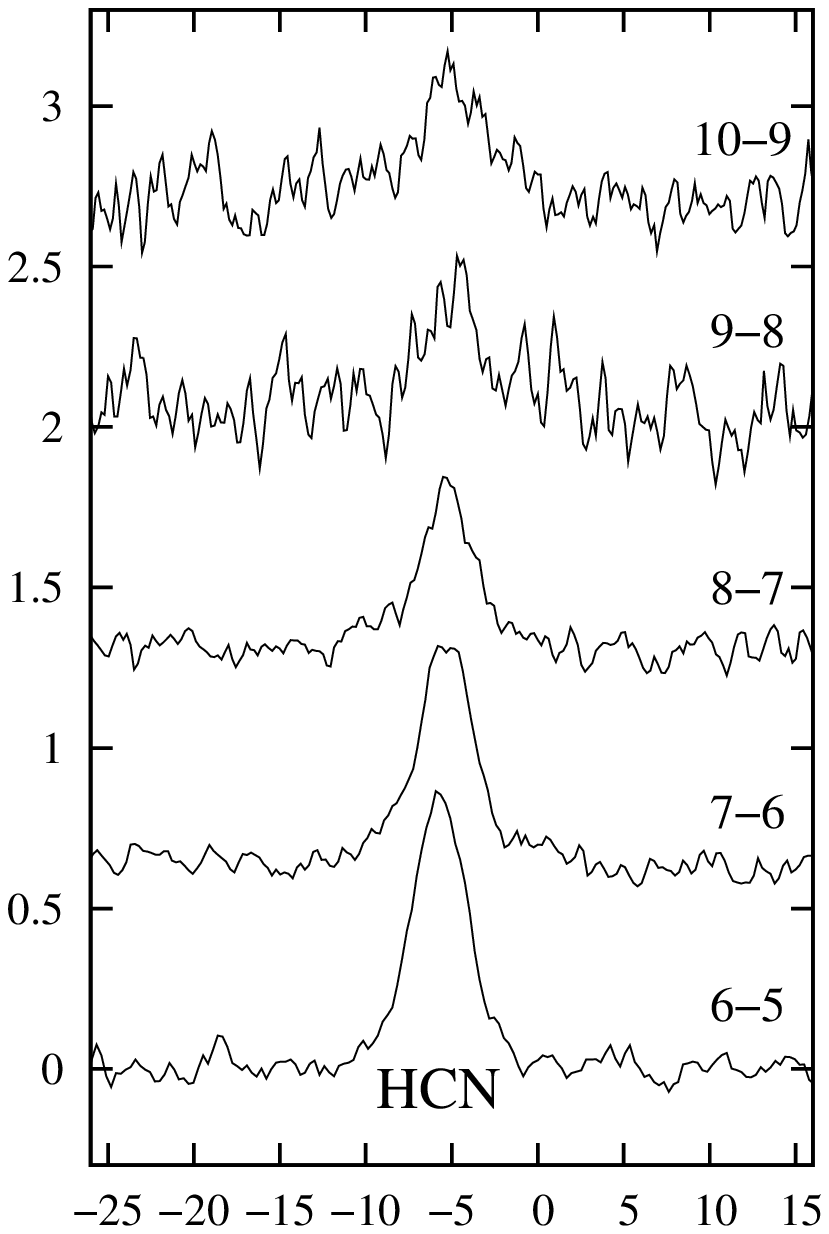}}
\subfloat{\includegraphics[width=4.43cm]{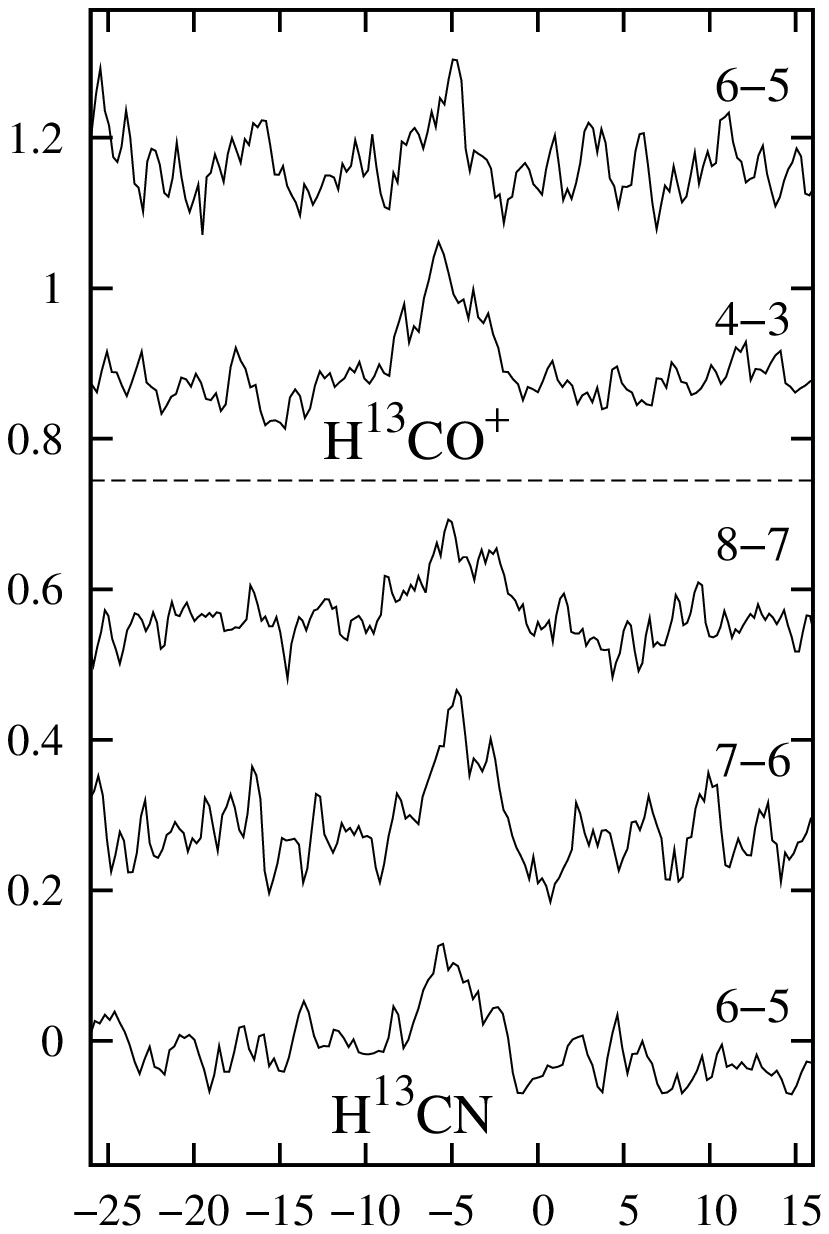}}\\
\subfloat{\includegraphics[width=4.43cm]{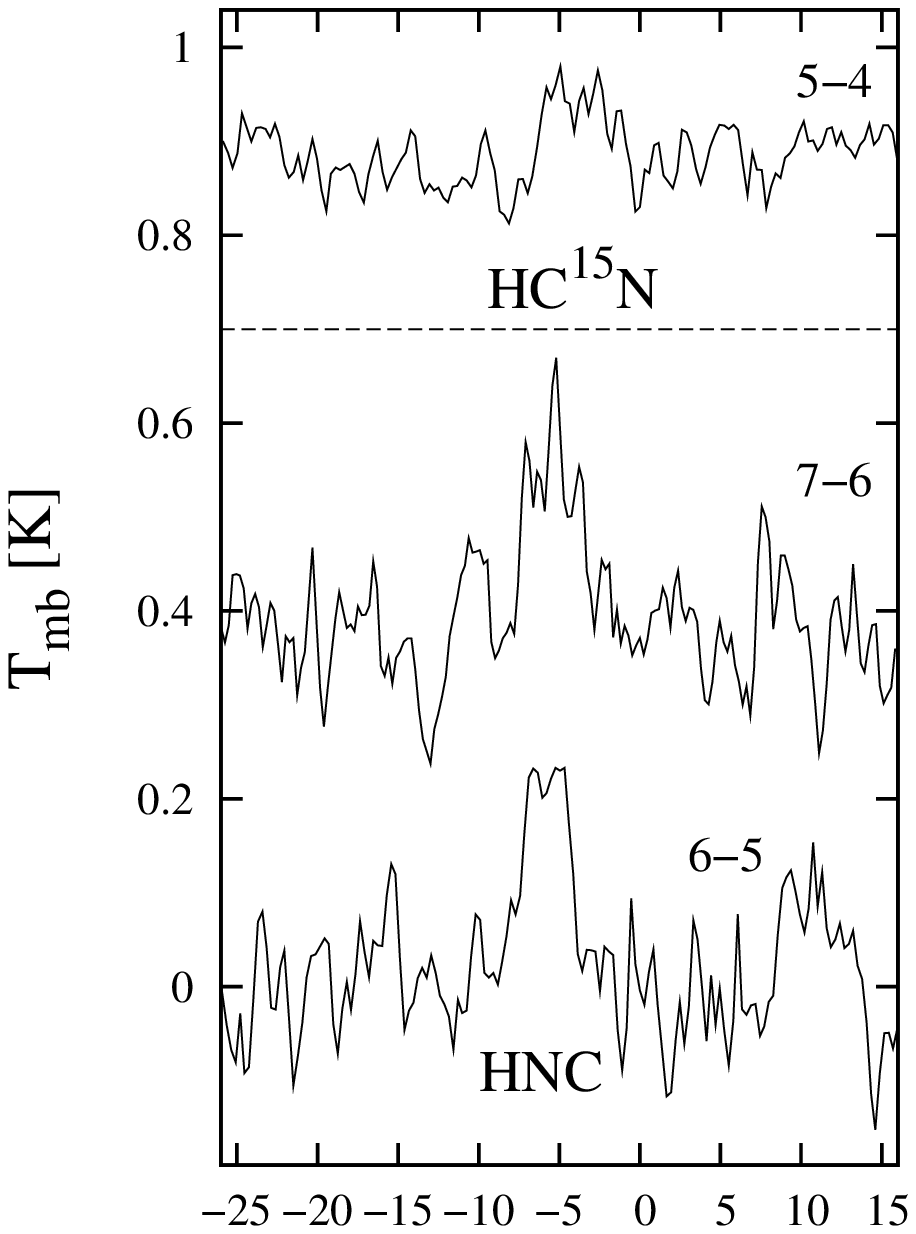}}
\subfloat{\includegraphics[width=4.43cm]{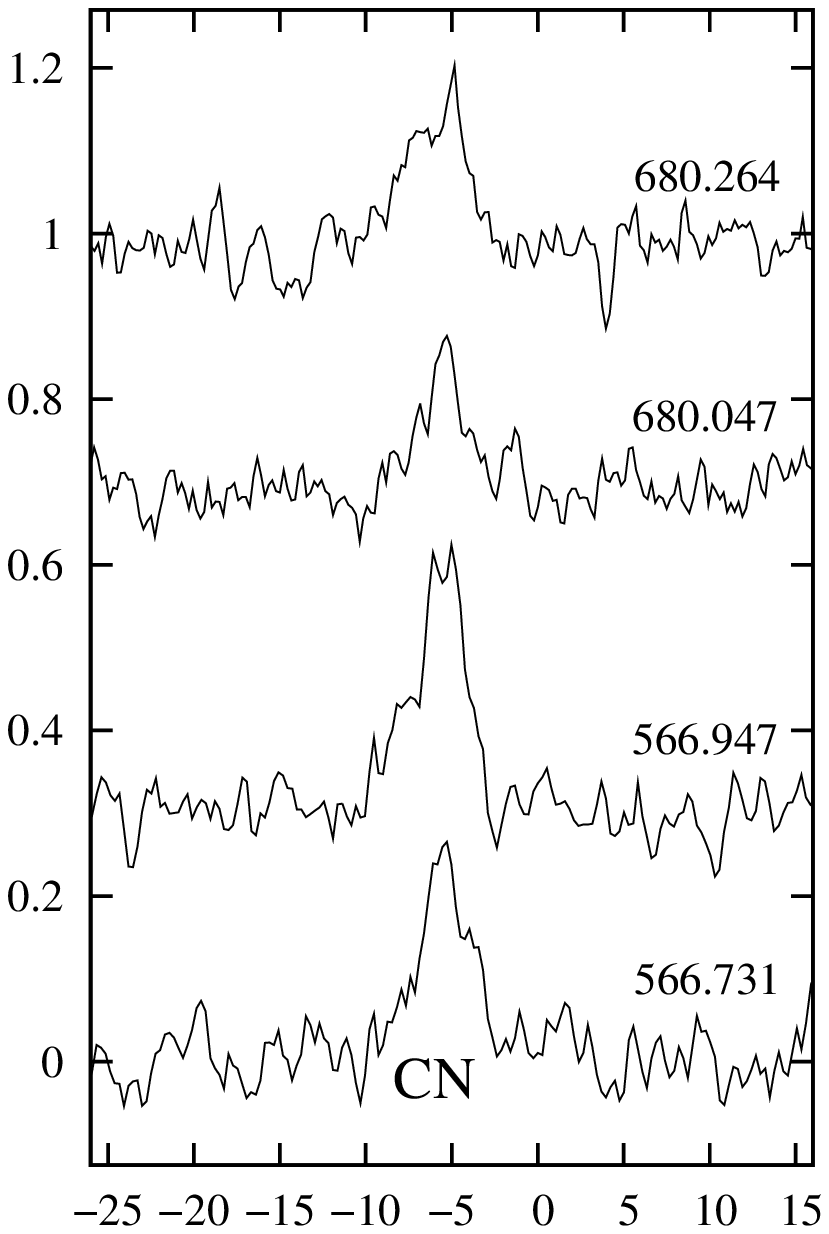}}
\subfloat{\includegraphics[width=4.43cm]{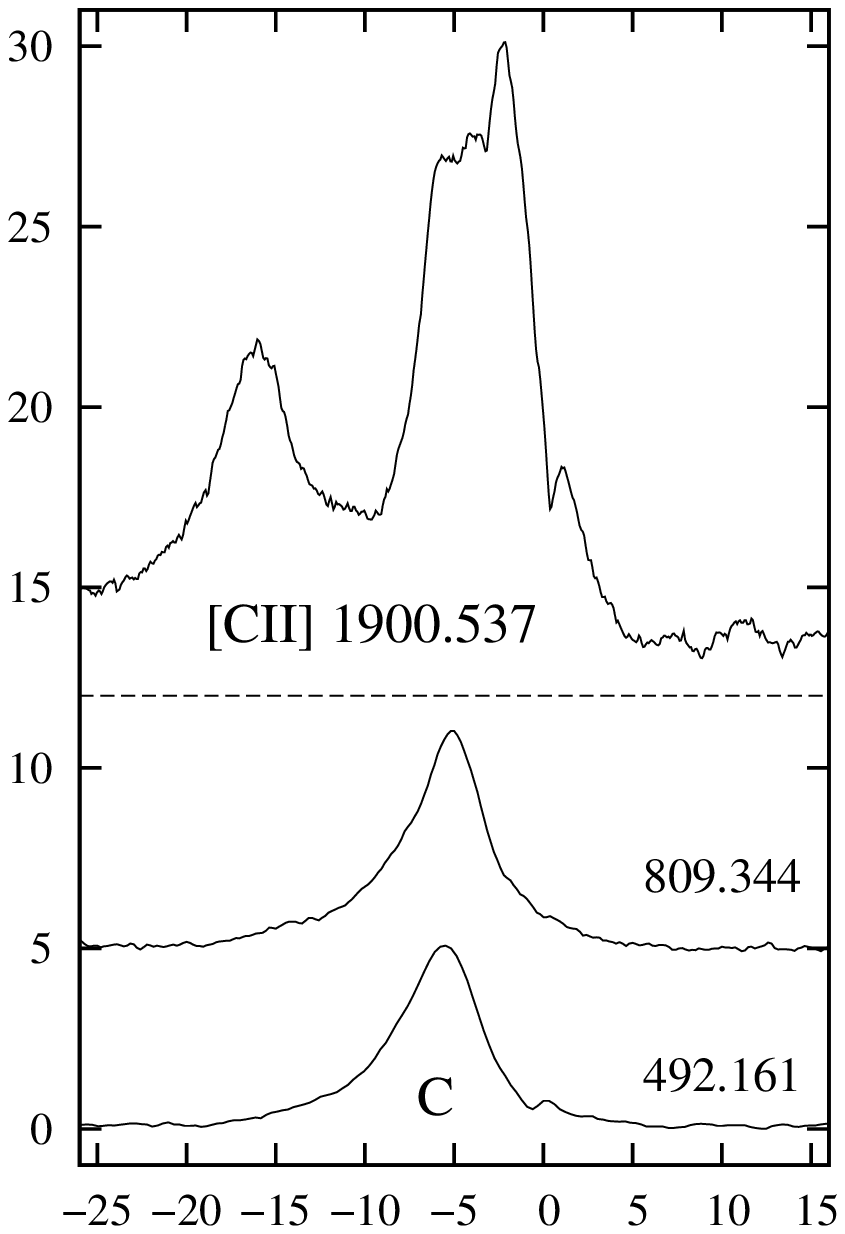}}
\subfloat{\includegraphics[width=4.43cm]{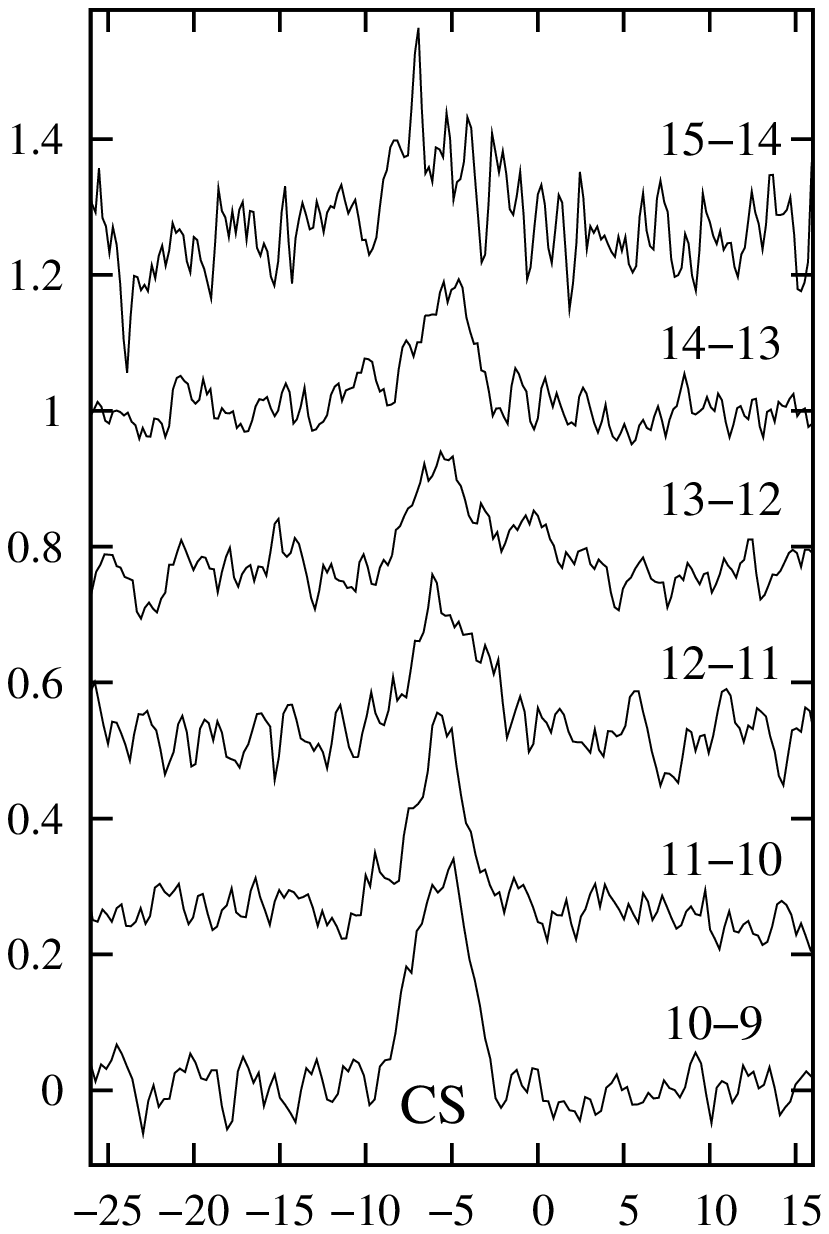}}\\
\subfloat{\includegraphics[width=4.43cm]{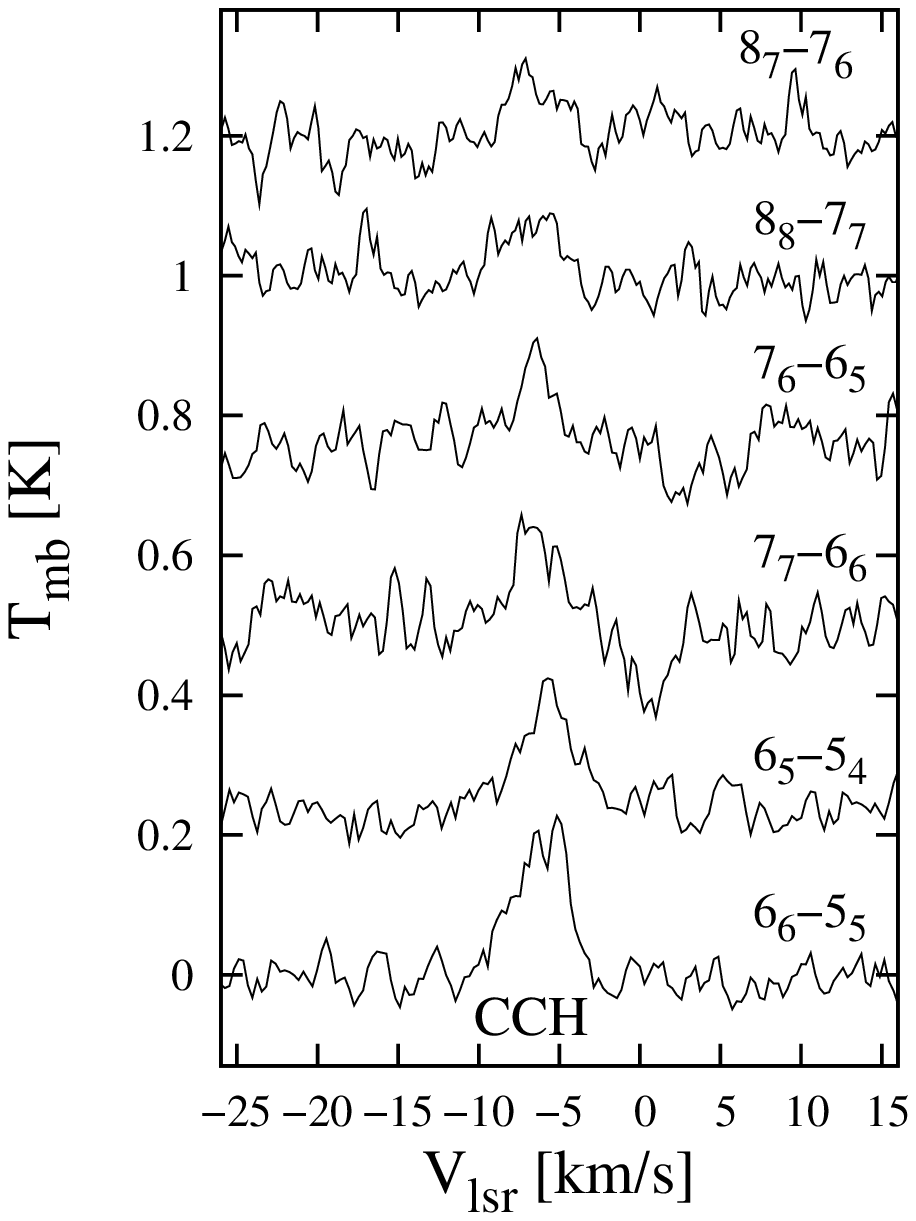}}
\subfloat{\includegraphics[width=4.43cm]{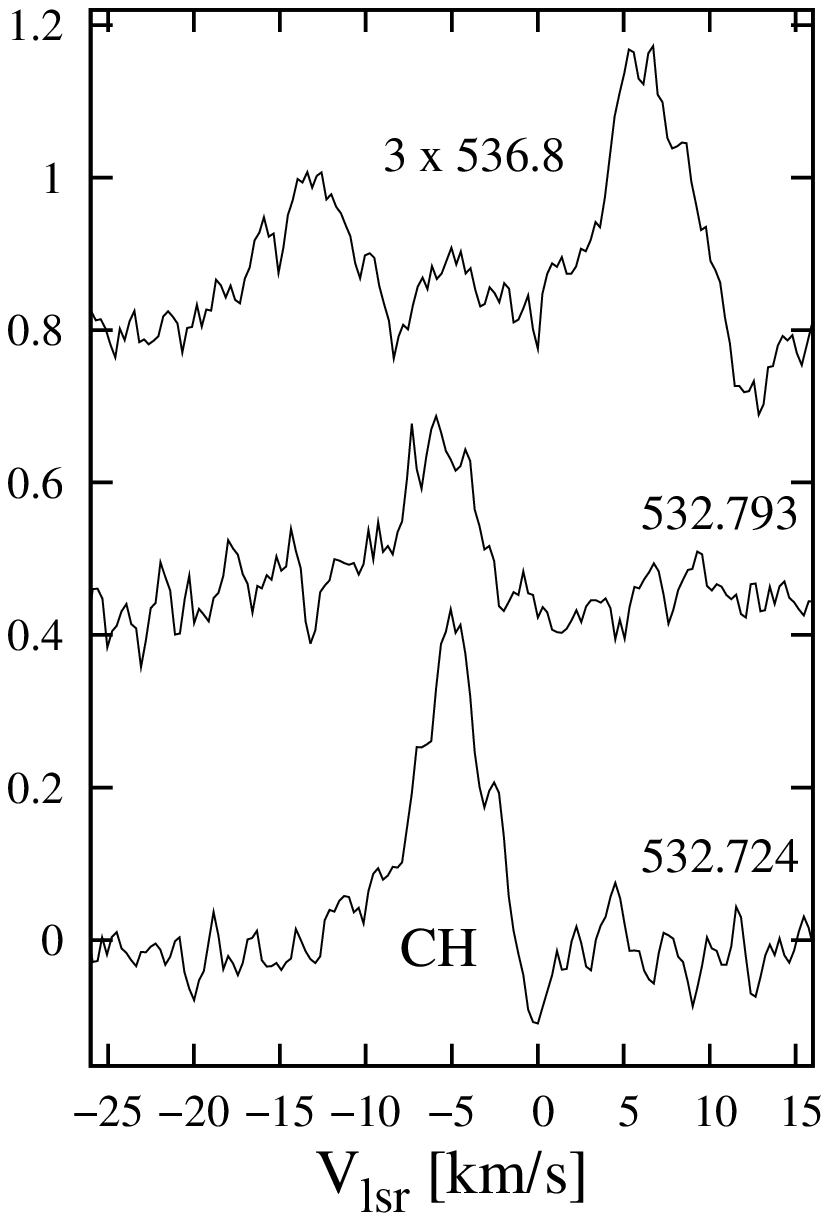}}
\subfloat{\includegraphics[width=4.43cm]{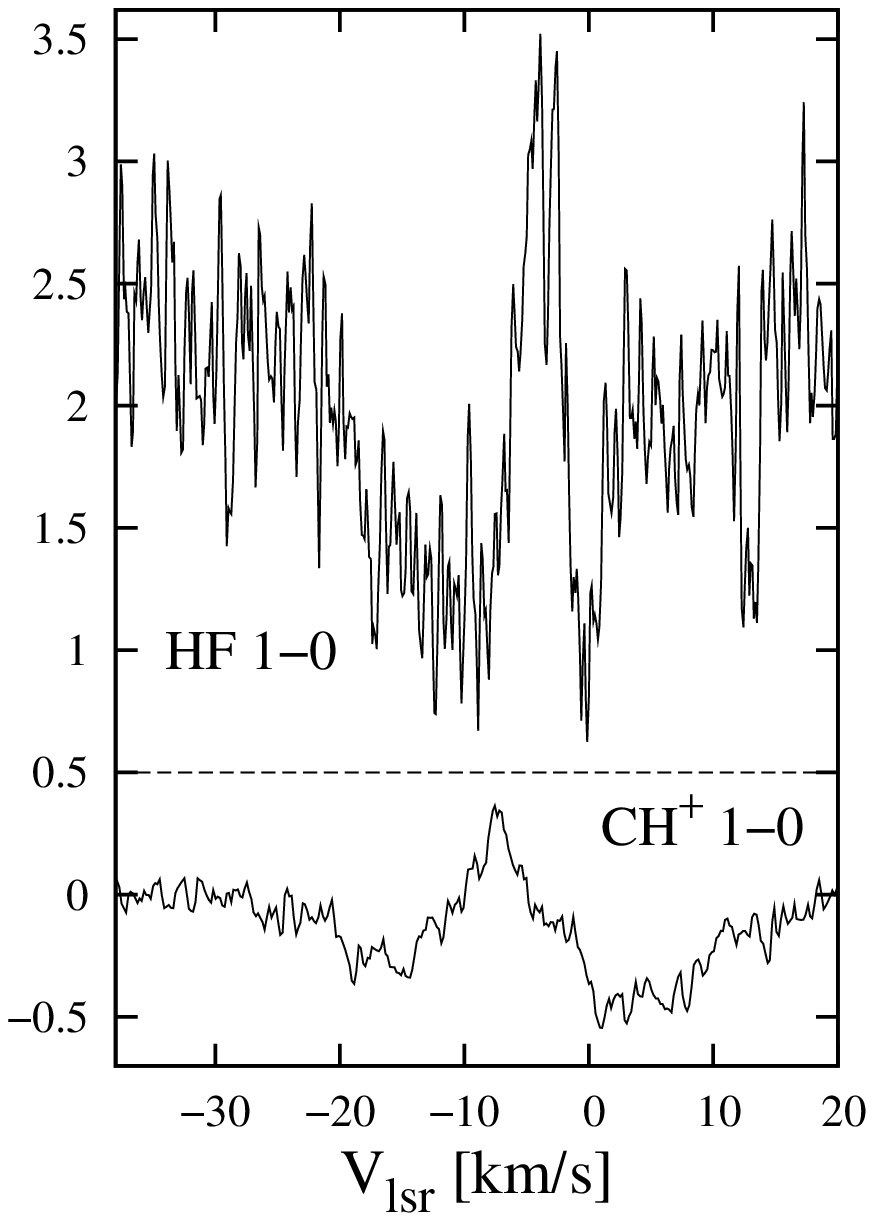}}
\subfloat{\includegraphics[width=4.43cm]{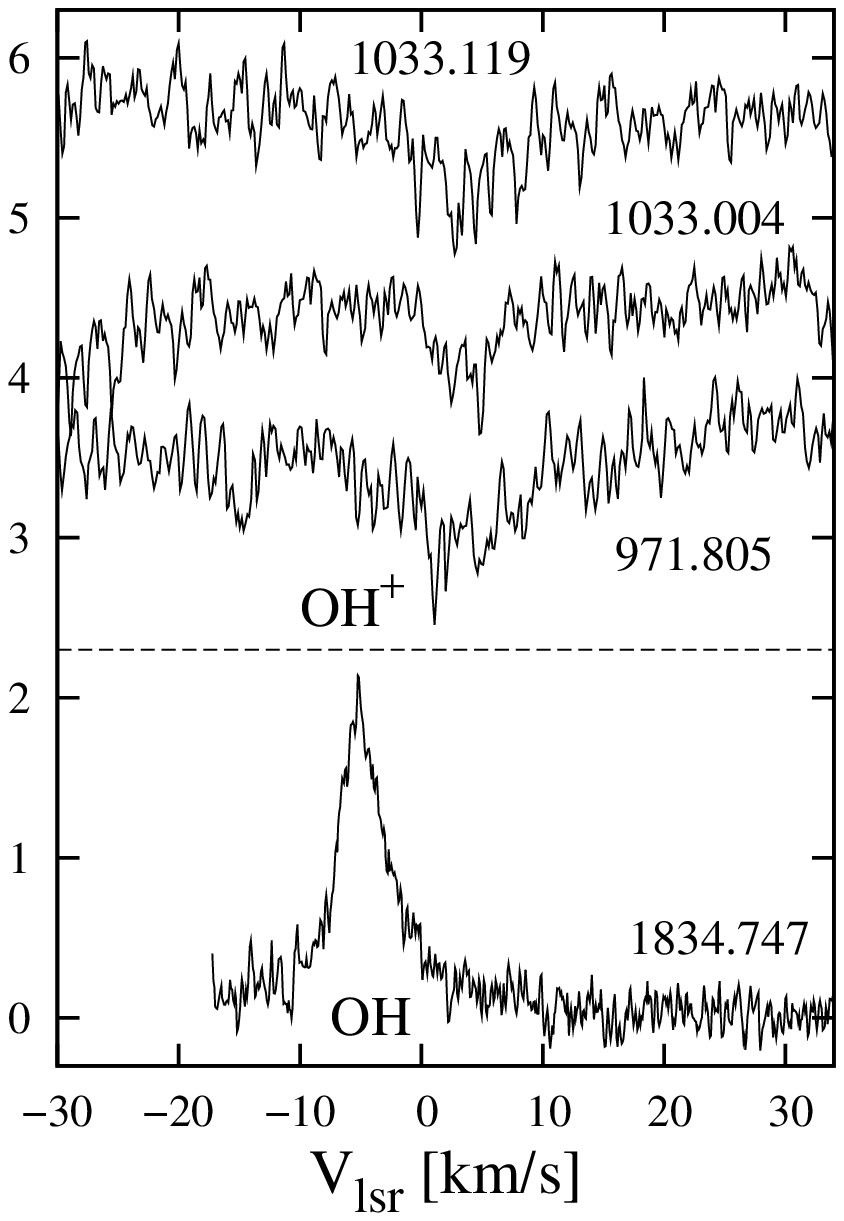}}
\caption[]{Line profiles of all identified species in HIFI spectral survey of AFGL~2591.
Molecules (e.g. HCl, CH) for which hyperfine components were detected are plotted all together and centered at the velocity of the middle line. Lines which show multiply profiles (e.g. CO -- envelope and outflow components; HF -- emission and absorptions) are also presented in one figure and are centered at the velocity of the envelope component; please compare with Table A.1.  
}
\label{figure:profiles1}
\end{figure*}

\begin{figure*}
\ContinuedFloat
\subfloat{\includegraphics[width=4.43cm]{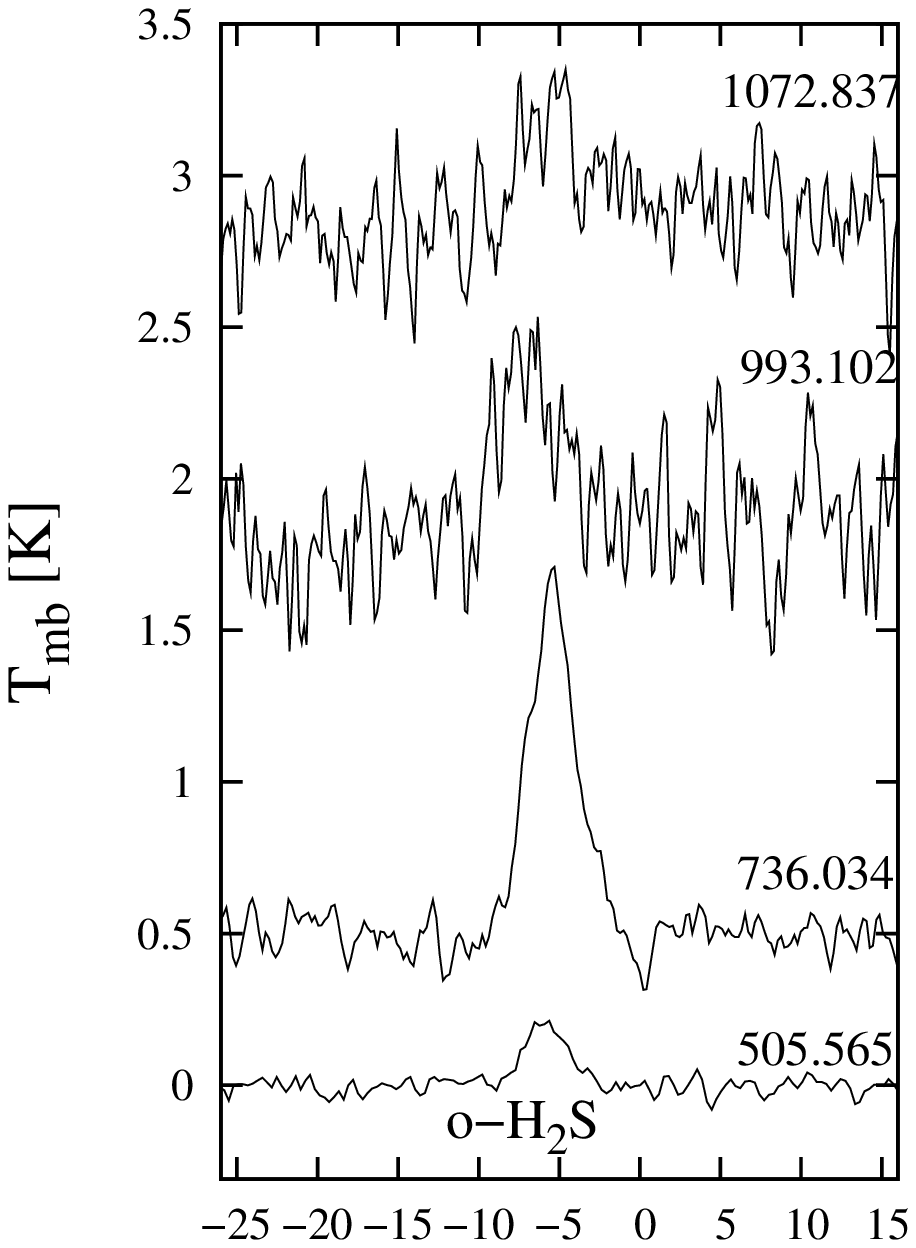}}
\subfloat{\includegraphics[width=4.43cm]{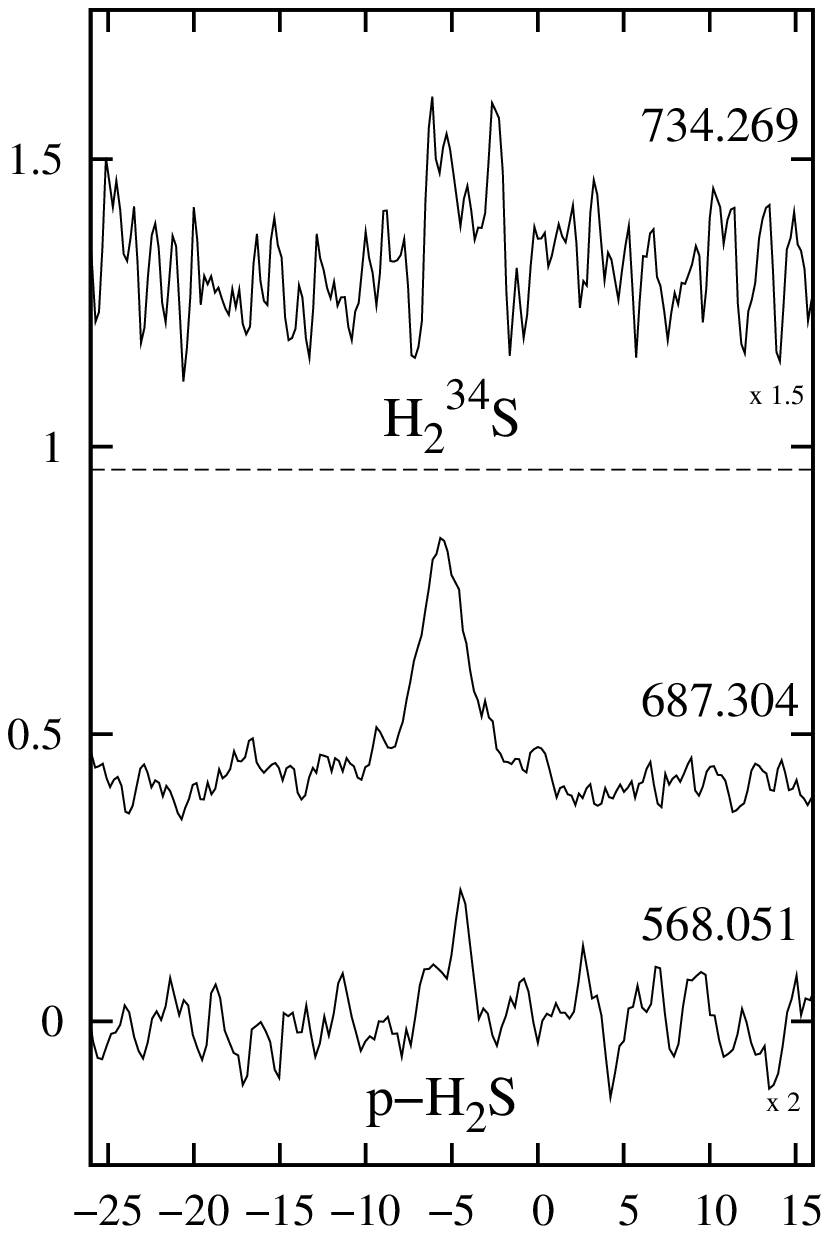}}
\subfloat{\includegraphics[width=4.43cm]{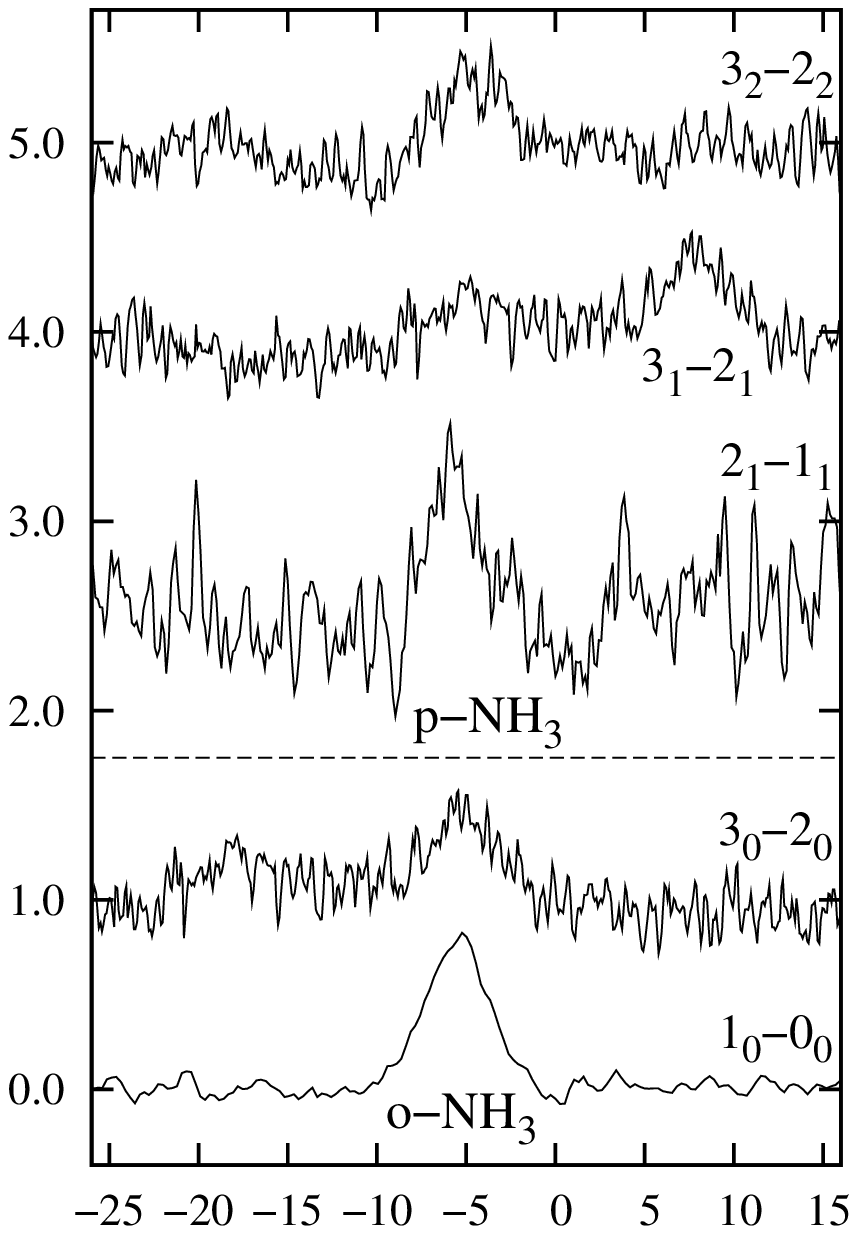}}
\subfloat{\includegraphics[width=4.43cm]{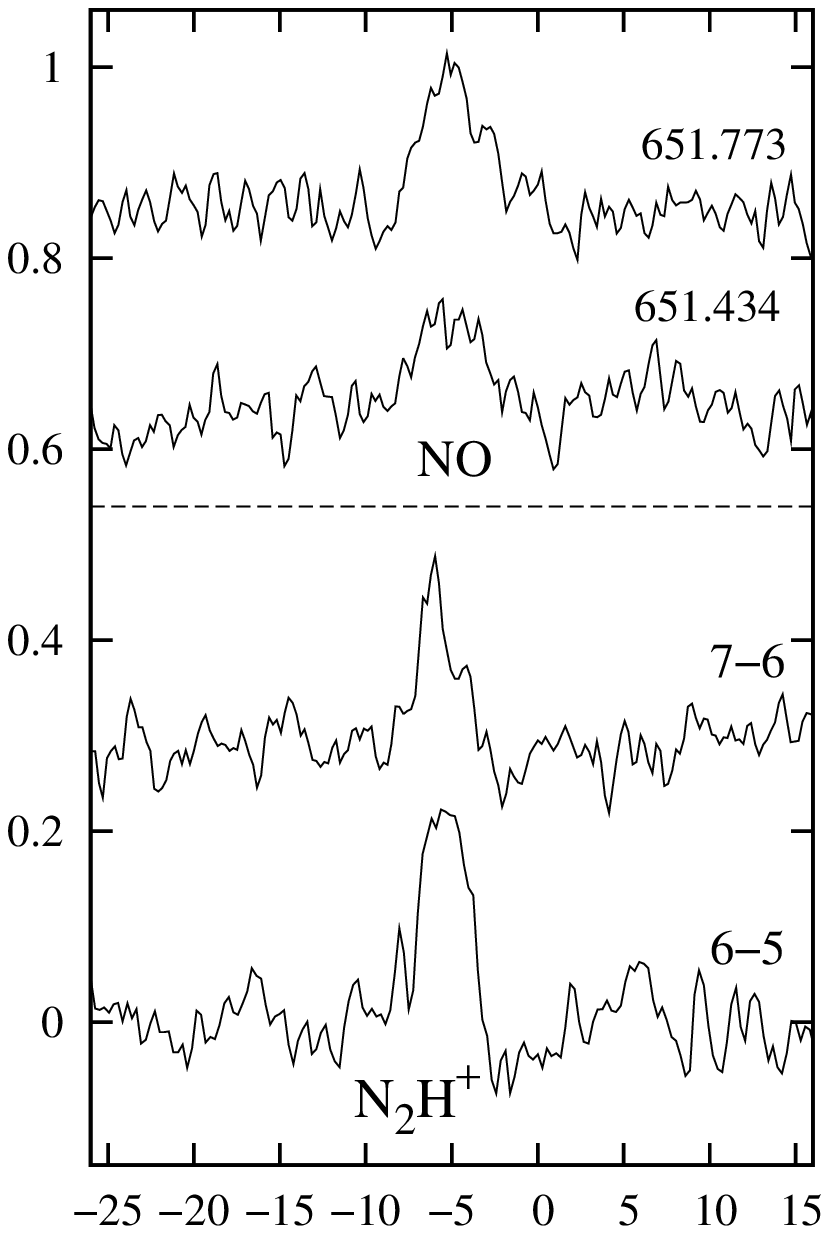}}\\
\subfloat{\includegraphics[width=4.43cm]{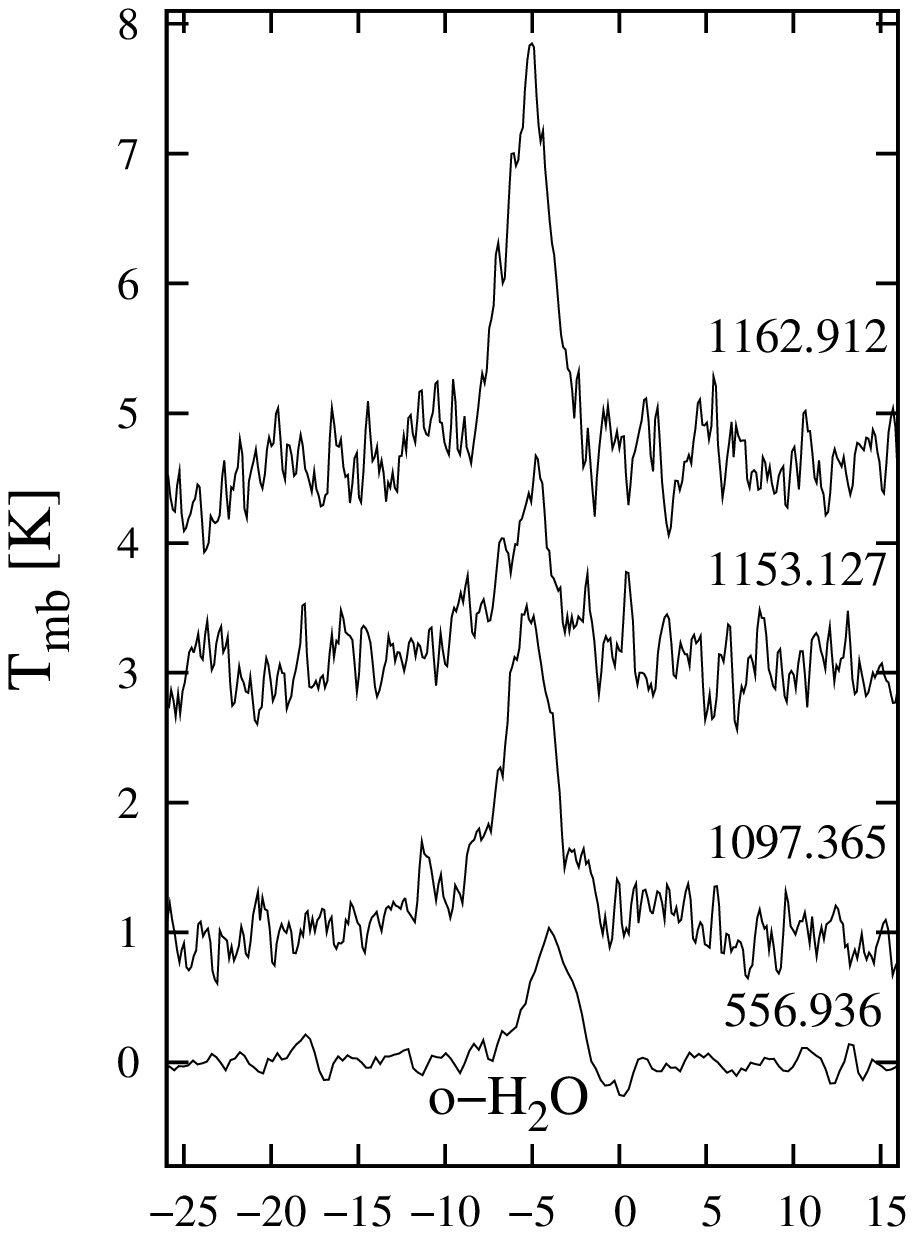}}
\subfloat{\includegraphics[width=4.43cm]{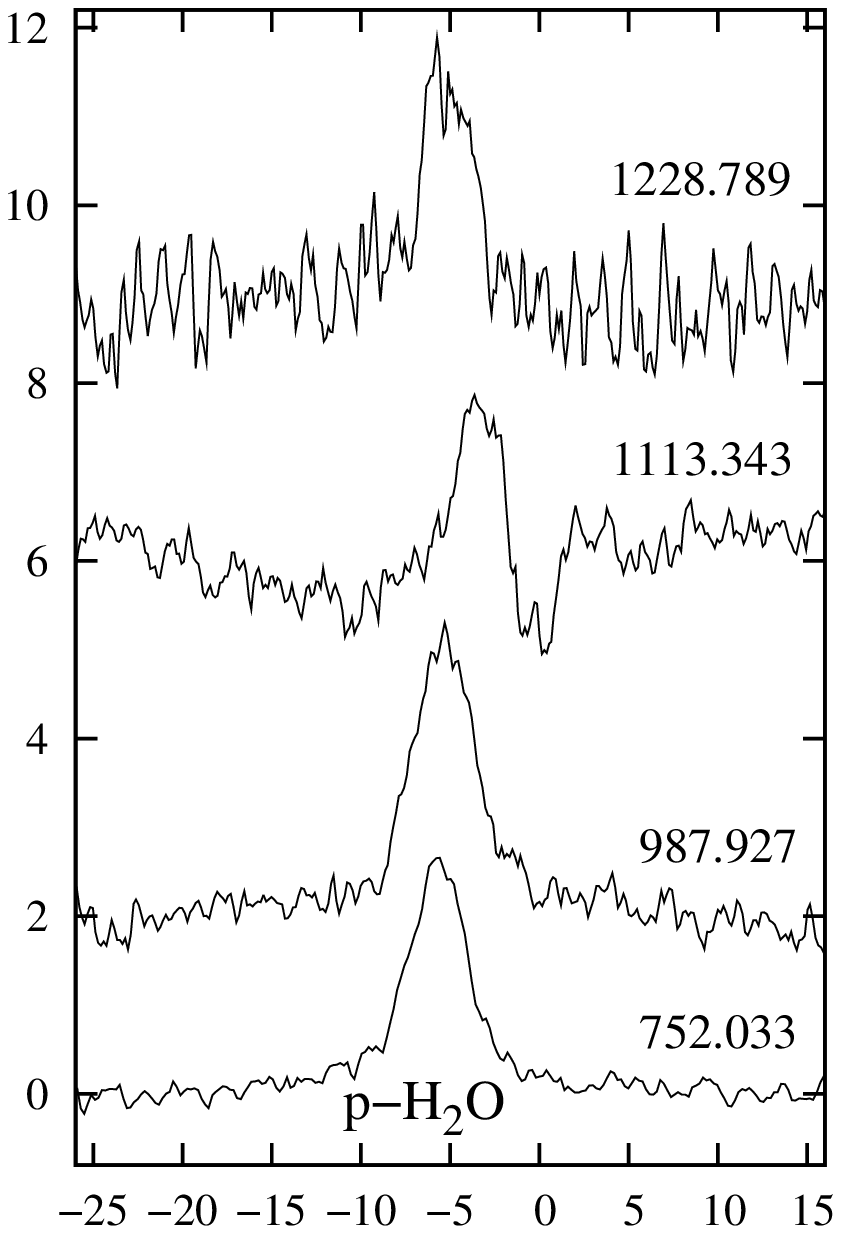}}
\subfloat{\includegraphics[width=4.43cm]{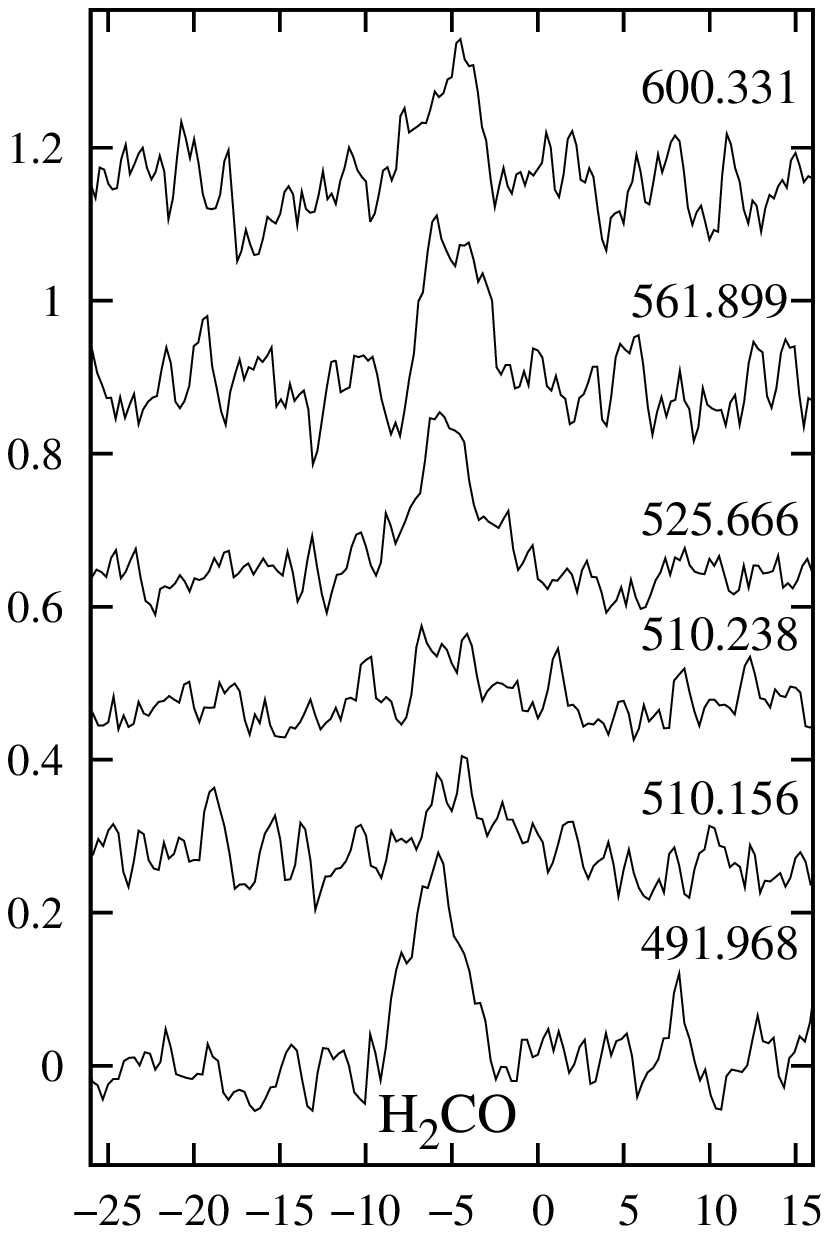}}
\subfloat{\includegraphics[width=4.43cm]{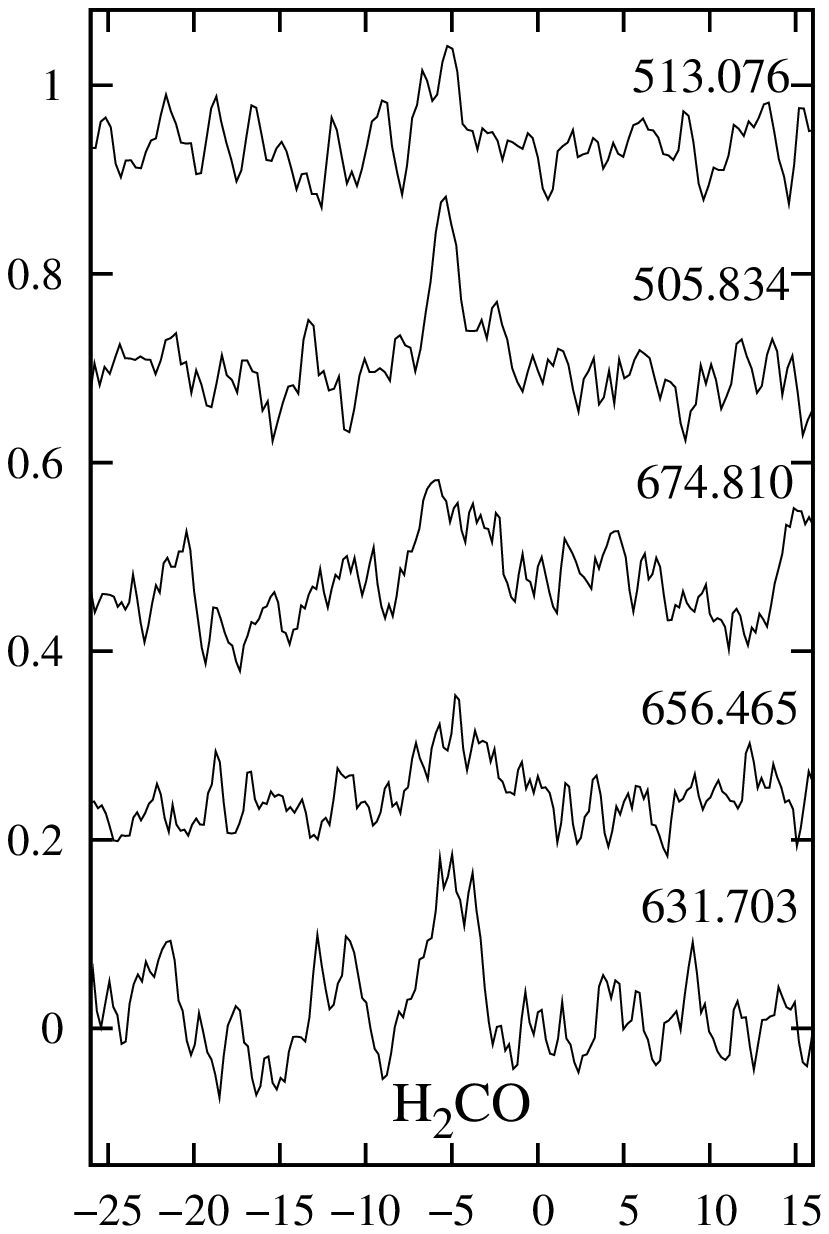}}\\
\subfloat{\includegraphics[width=4.43cm]{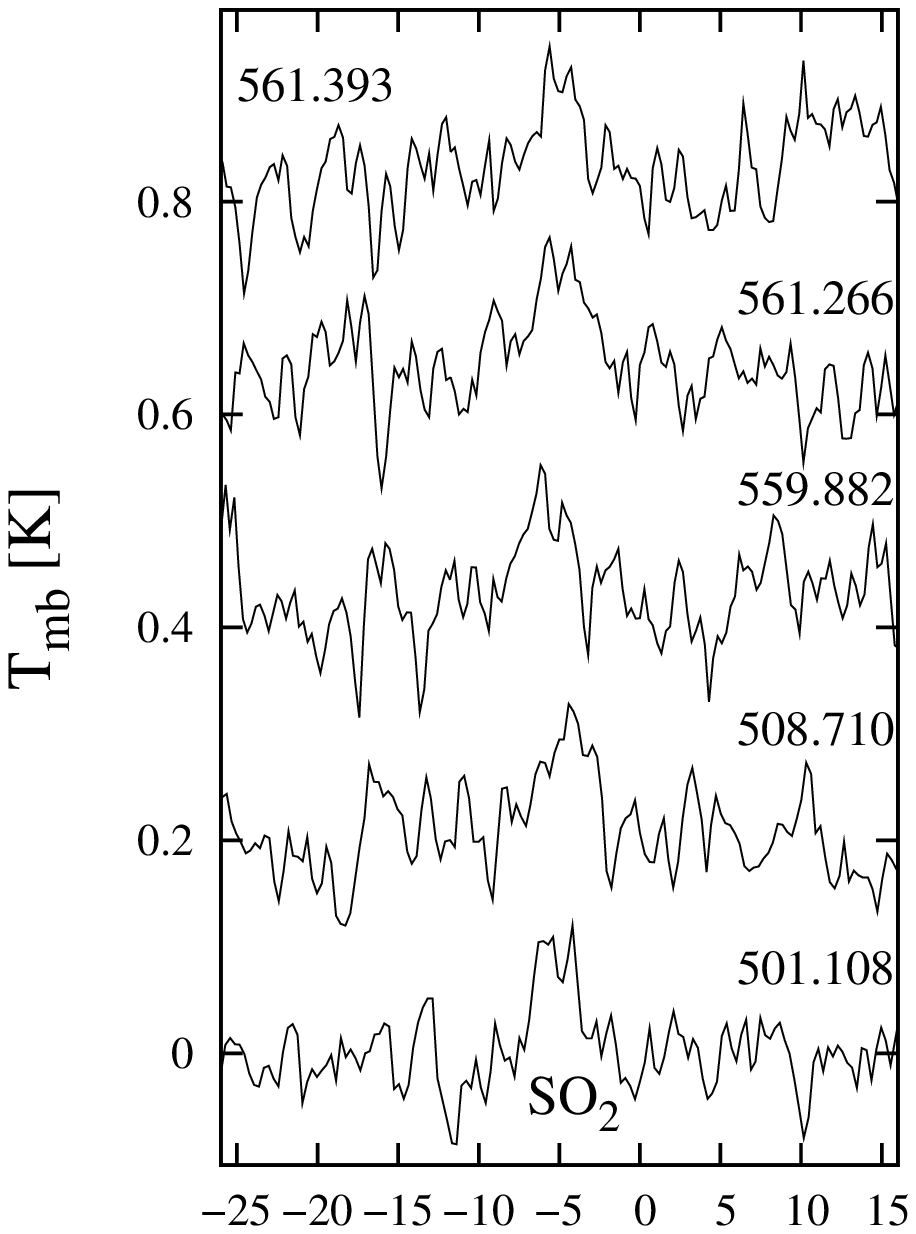}}
\subfloat{\includegraphics[width=4.43cm]{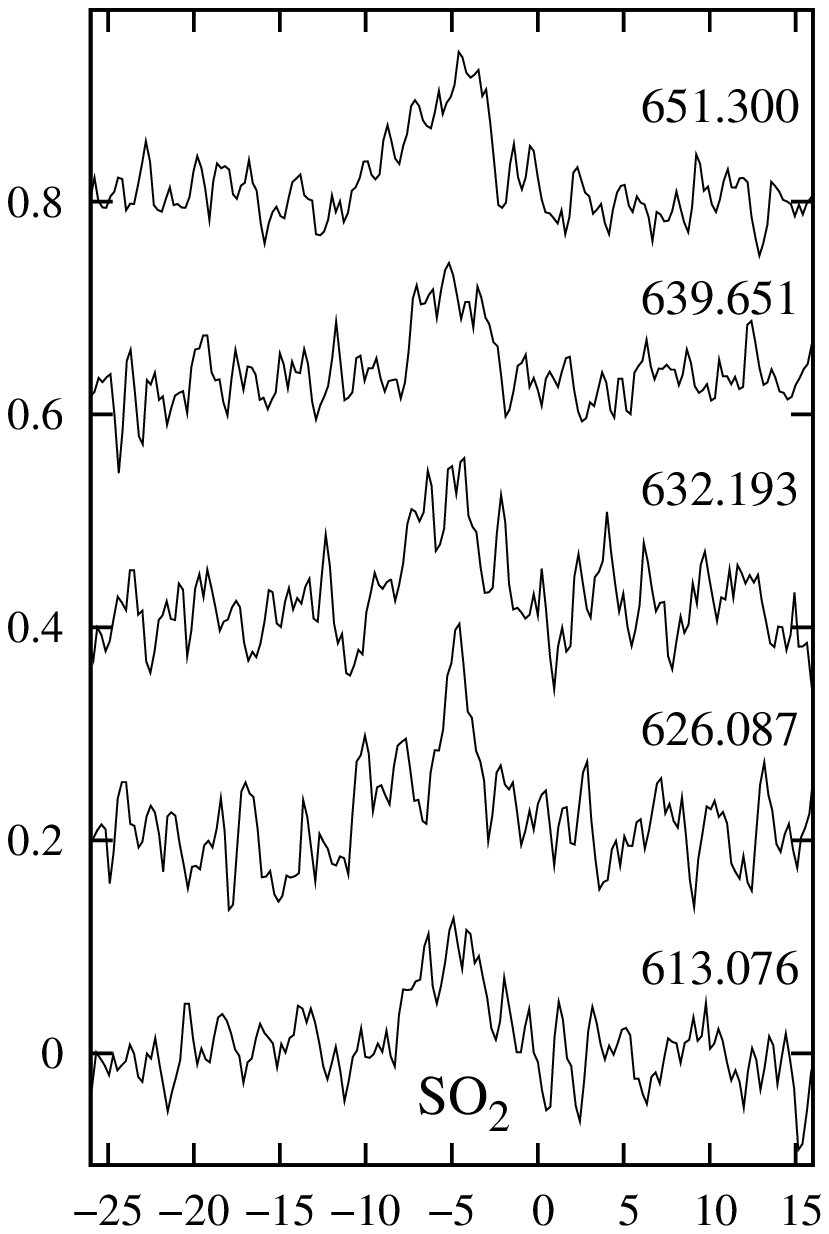}}
\subfloat{\includegraphics[width=4.43cm]{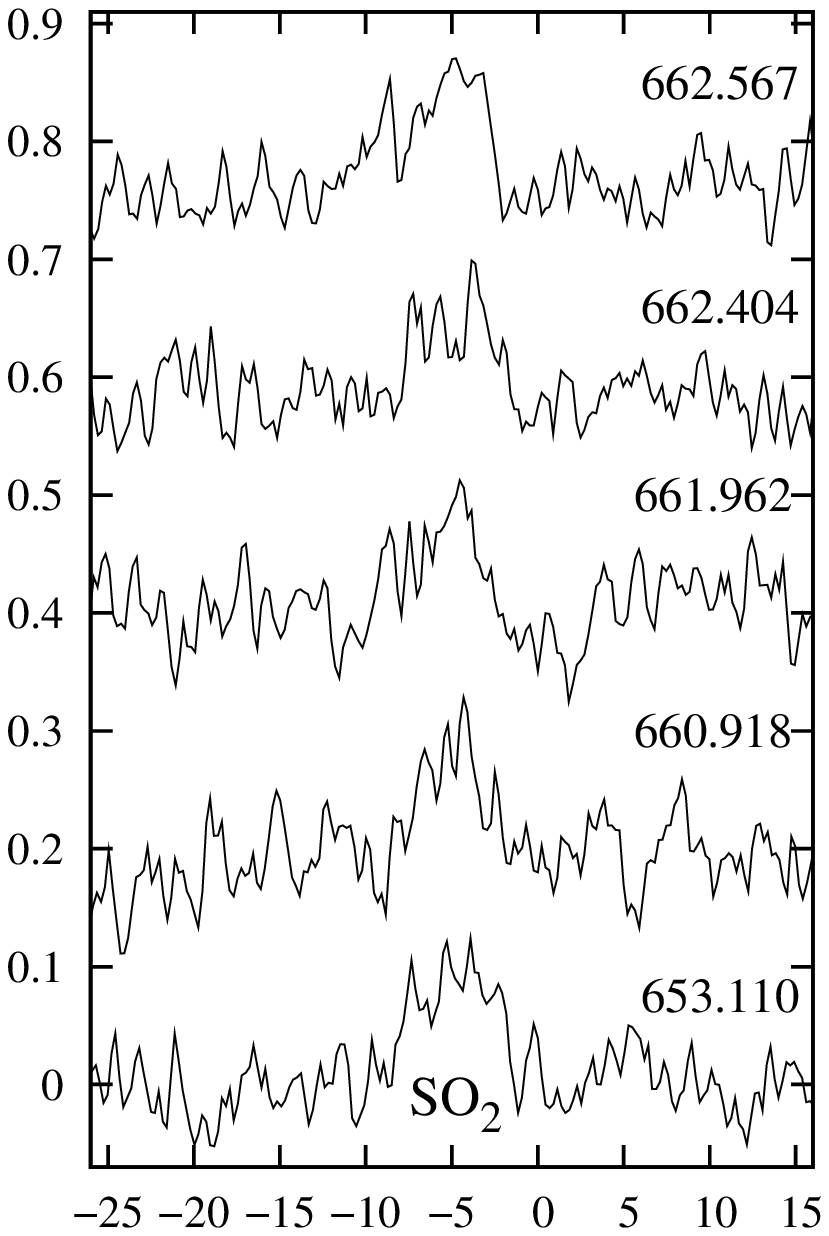}}
\subfloat{\includegraphics[width=4.43cm]{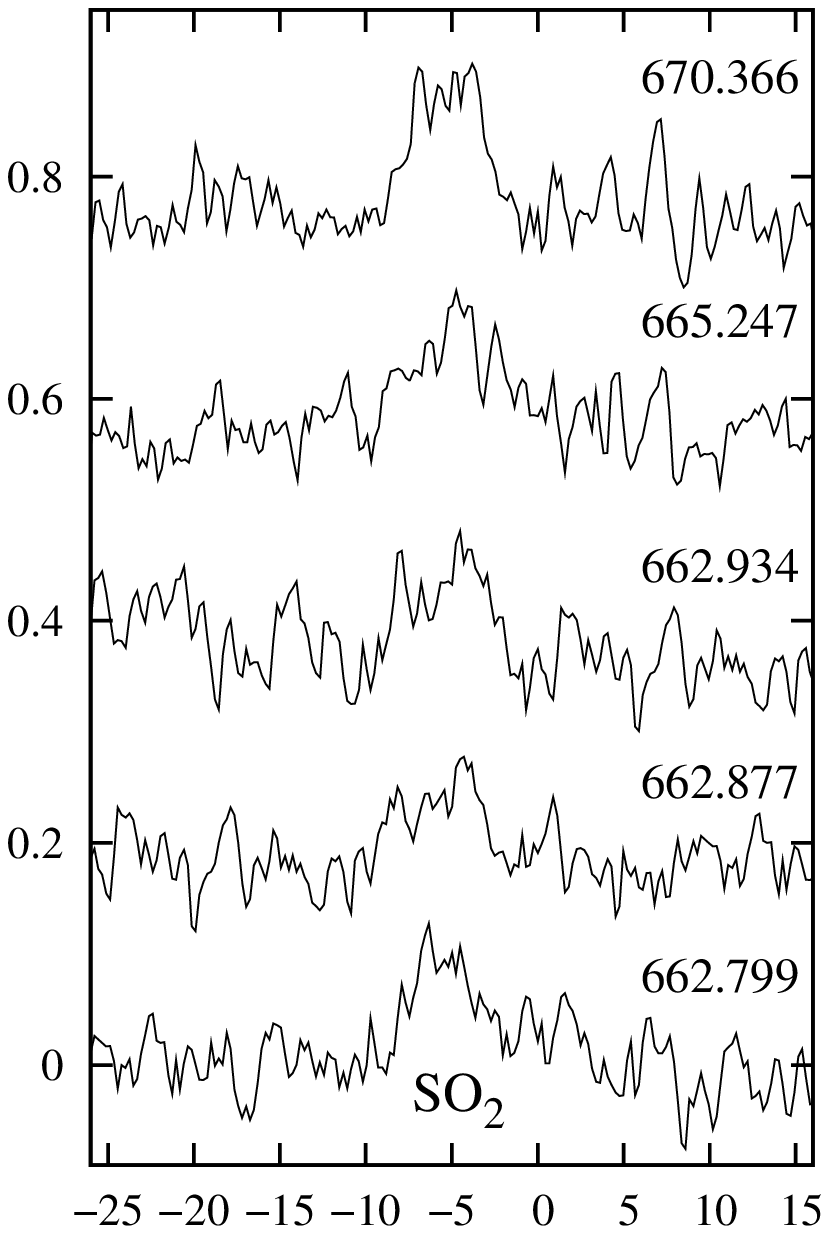}}\\
\subfloat{\includegraphics[width=4.43cm]{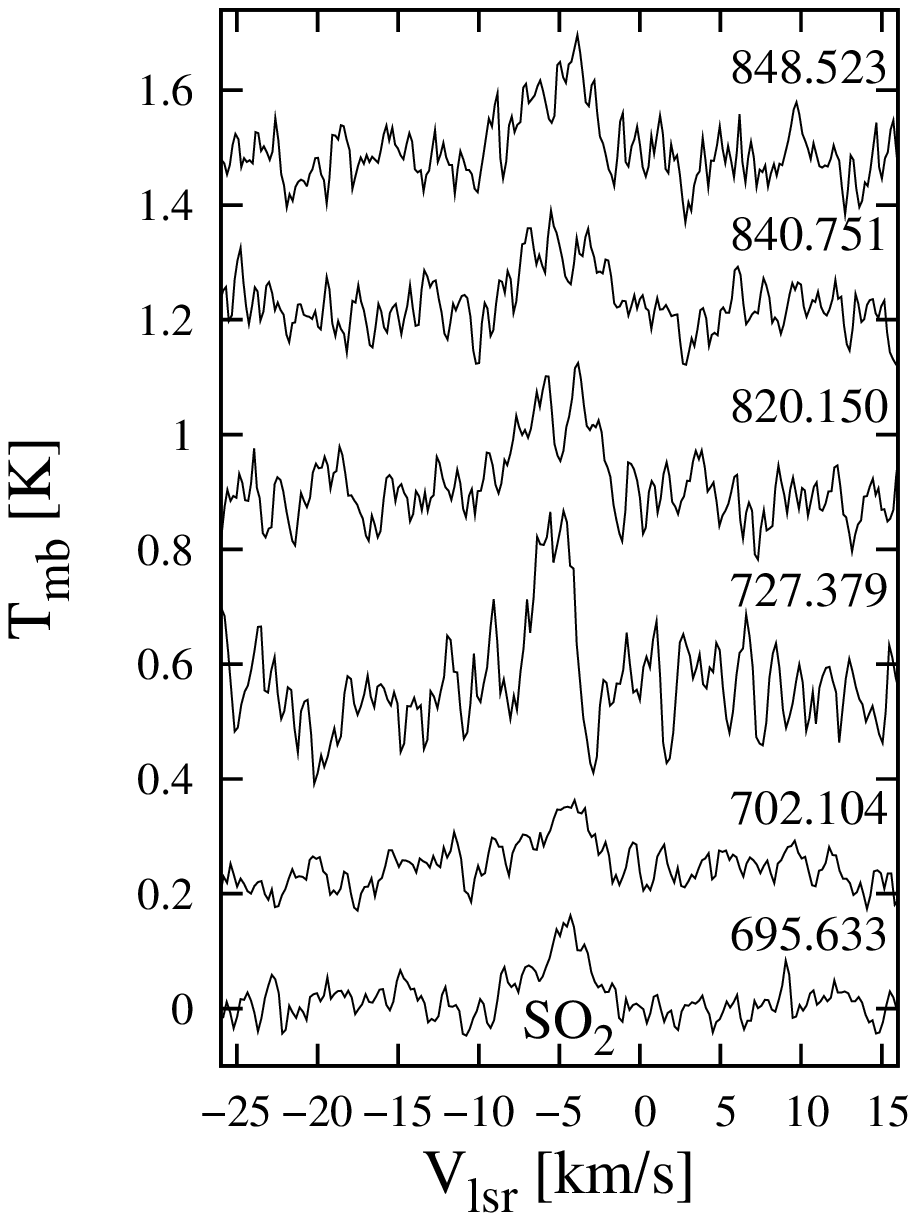}}
\subfloat{\includegraphics[width=4.43cm]{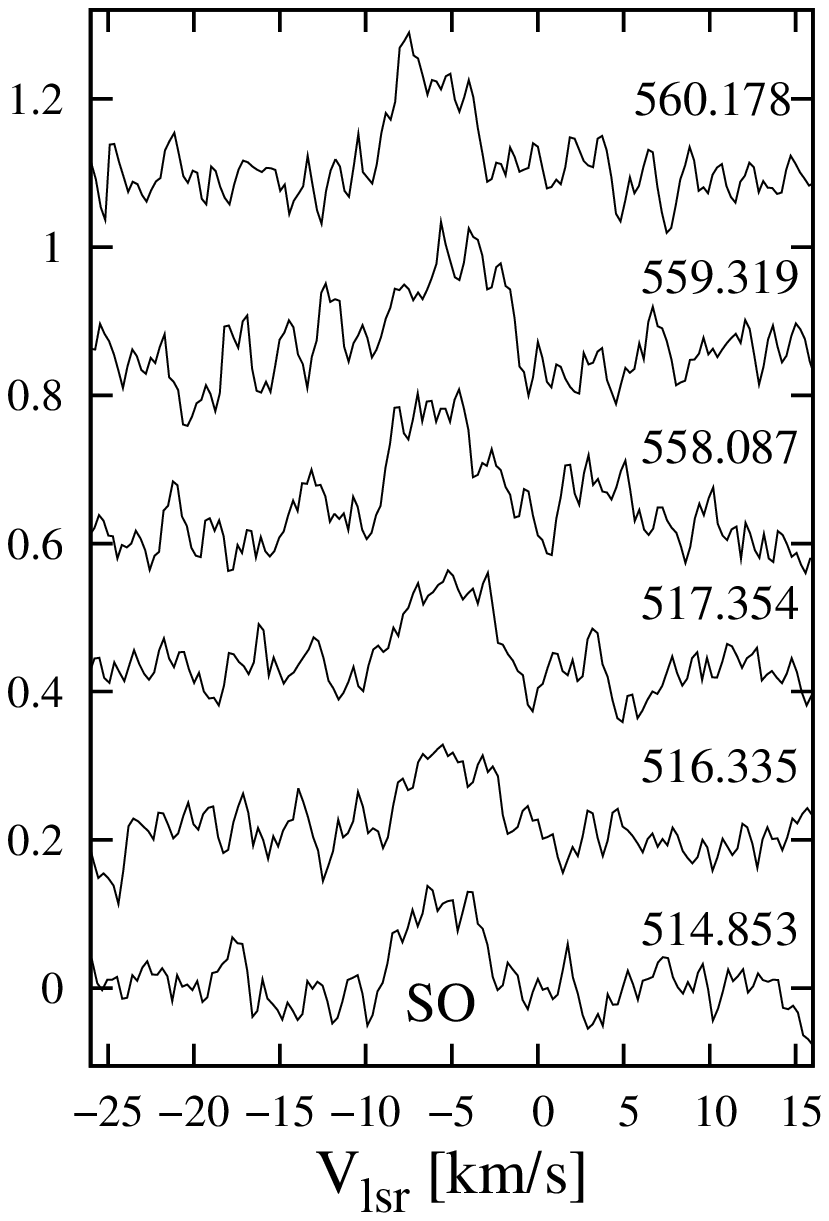}}
\subfloat{\includegraphics[width=4.43cm]{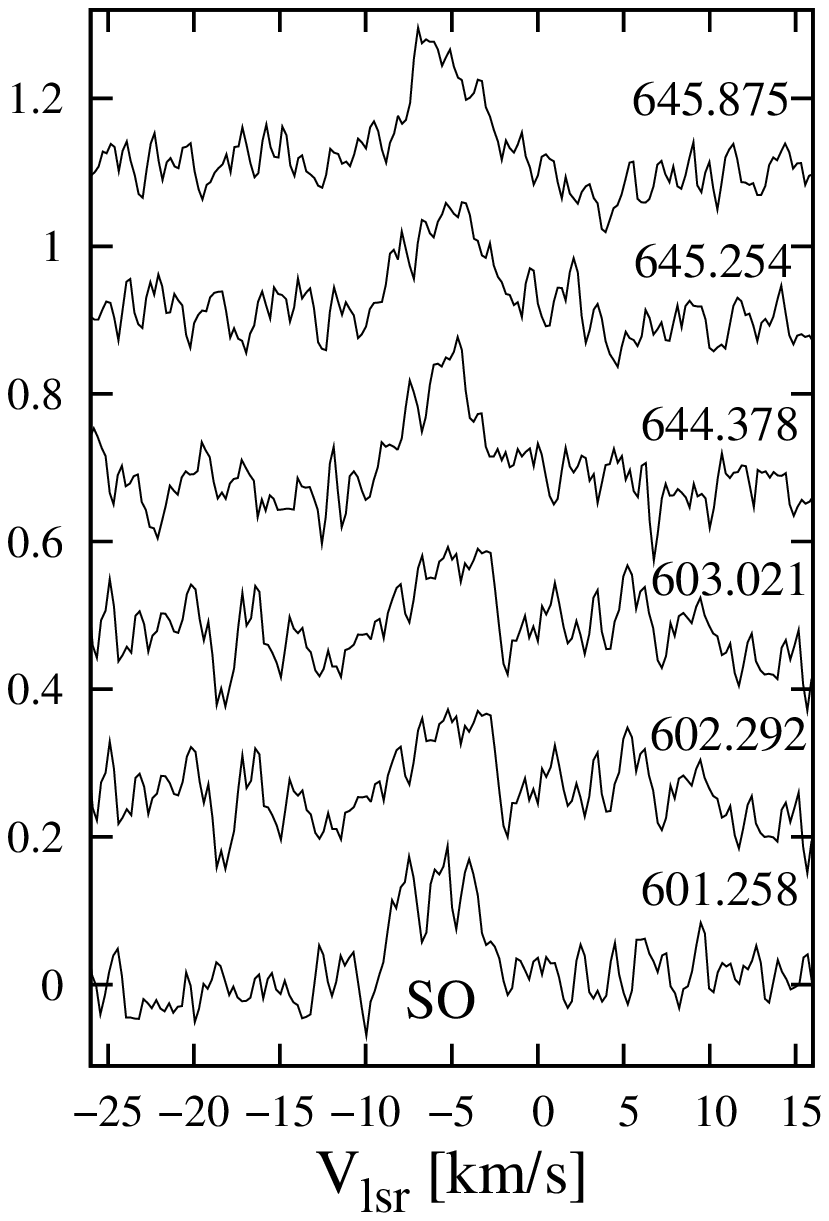}}
\subfloat{\includegraphics[width=4.43cm]{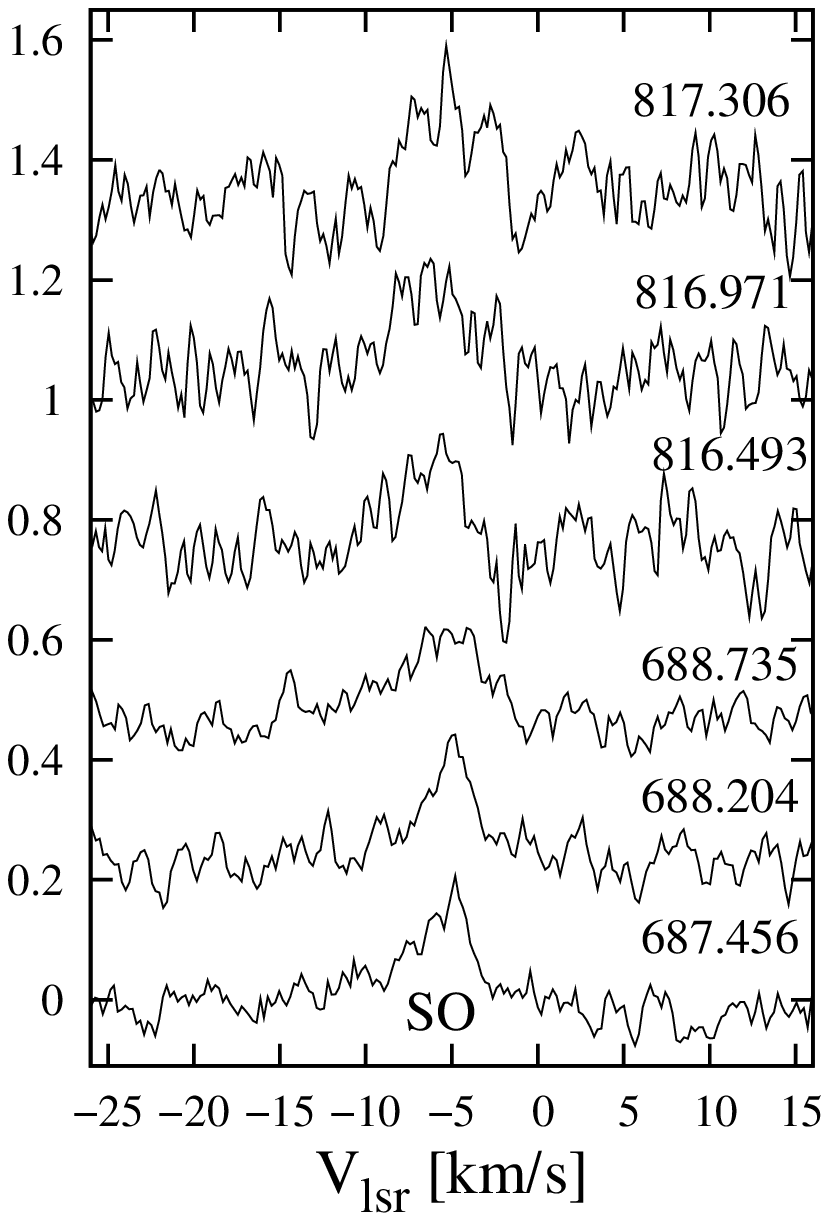}}
\caption[]{Continued.}
\end{figure*}

\begin{figure*}
\ContinuedFloat
\subfloat{\includegraphics[width=4.43cm]{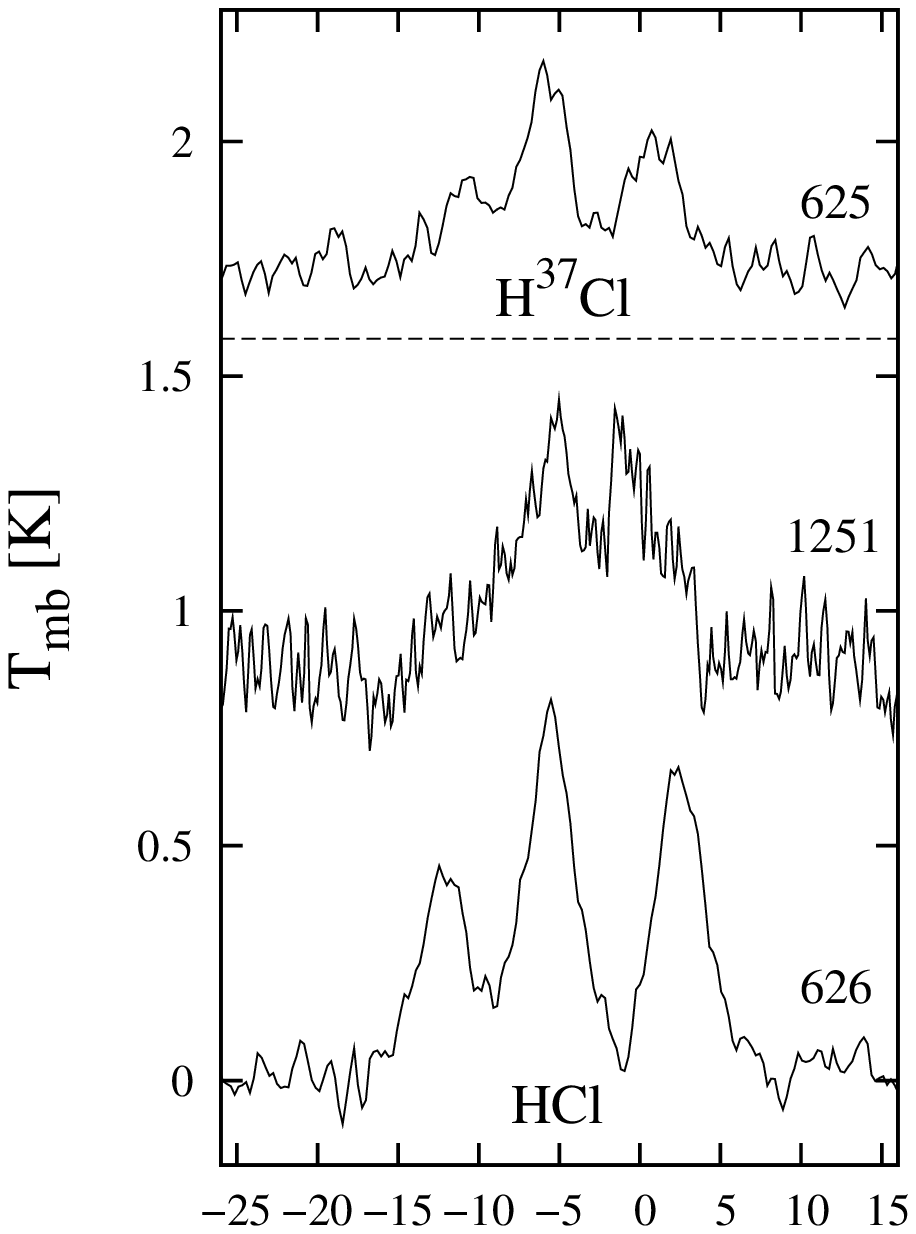}}
\subfloat{\includegraphics[width=4.43cm]{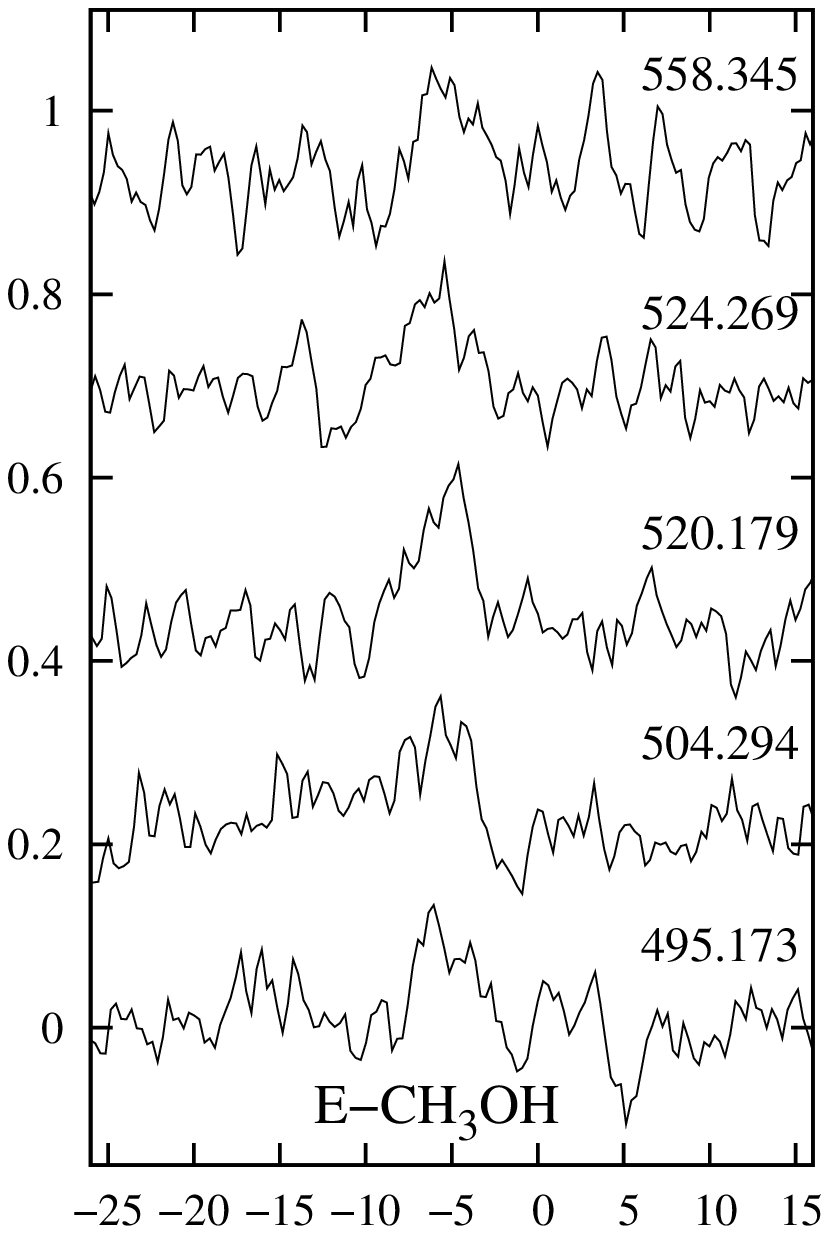}}
\subfloat{\includegraphics[width=4.43cm]{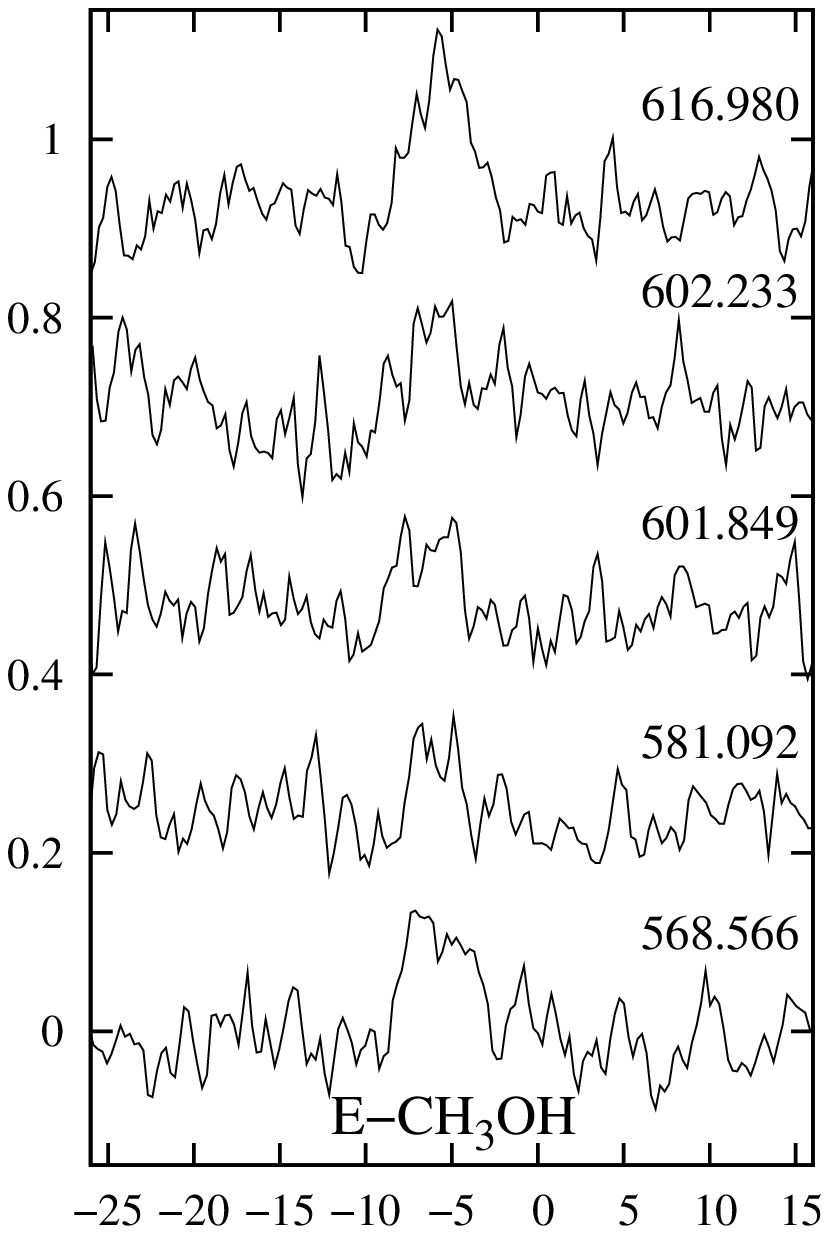}}
\subfloat{\includegraphics[width=4.43cm]{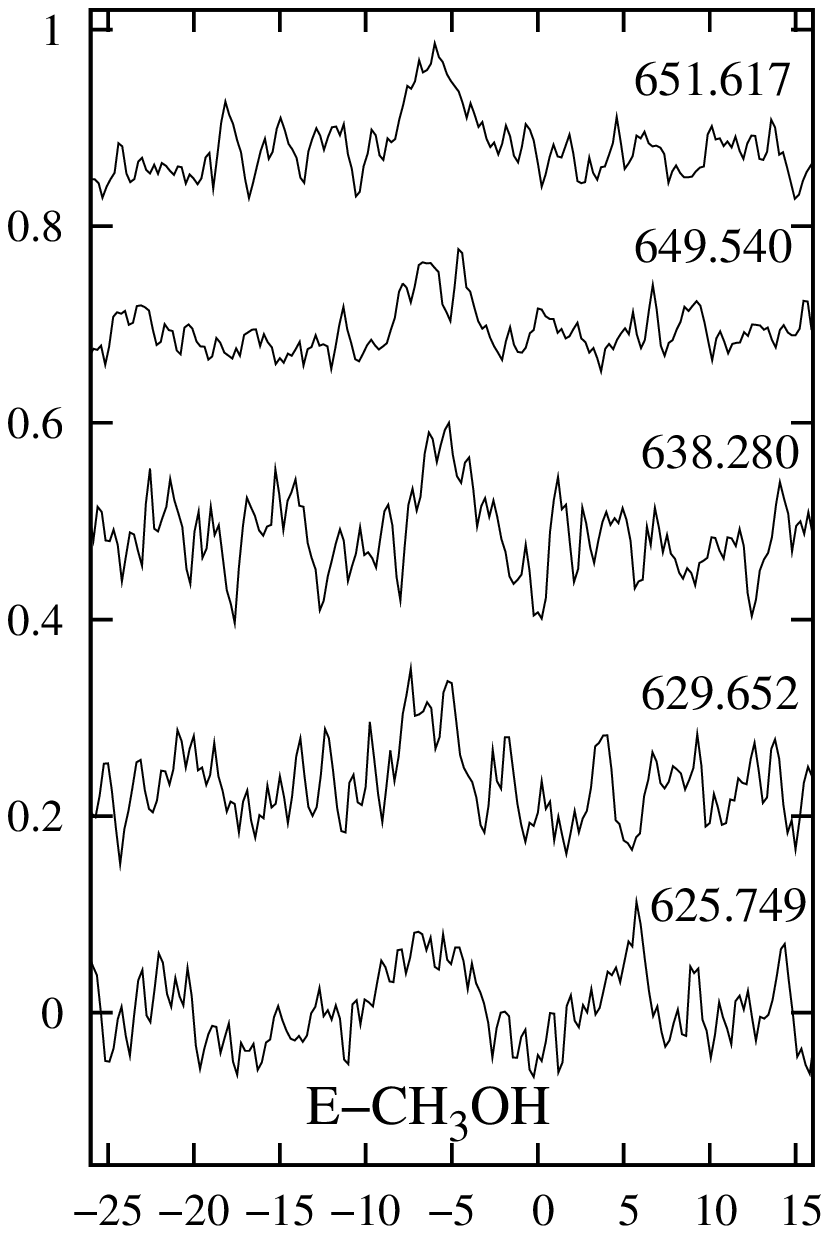}}\\
\subfloat{\includegraphics[width=4.43cm]{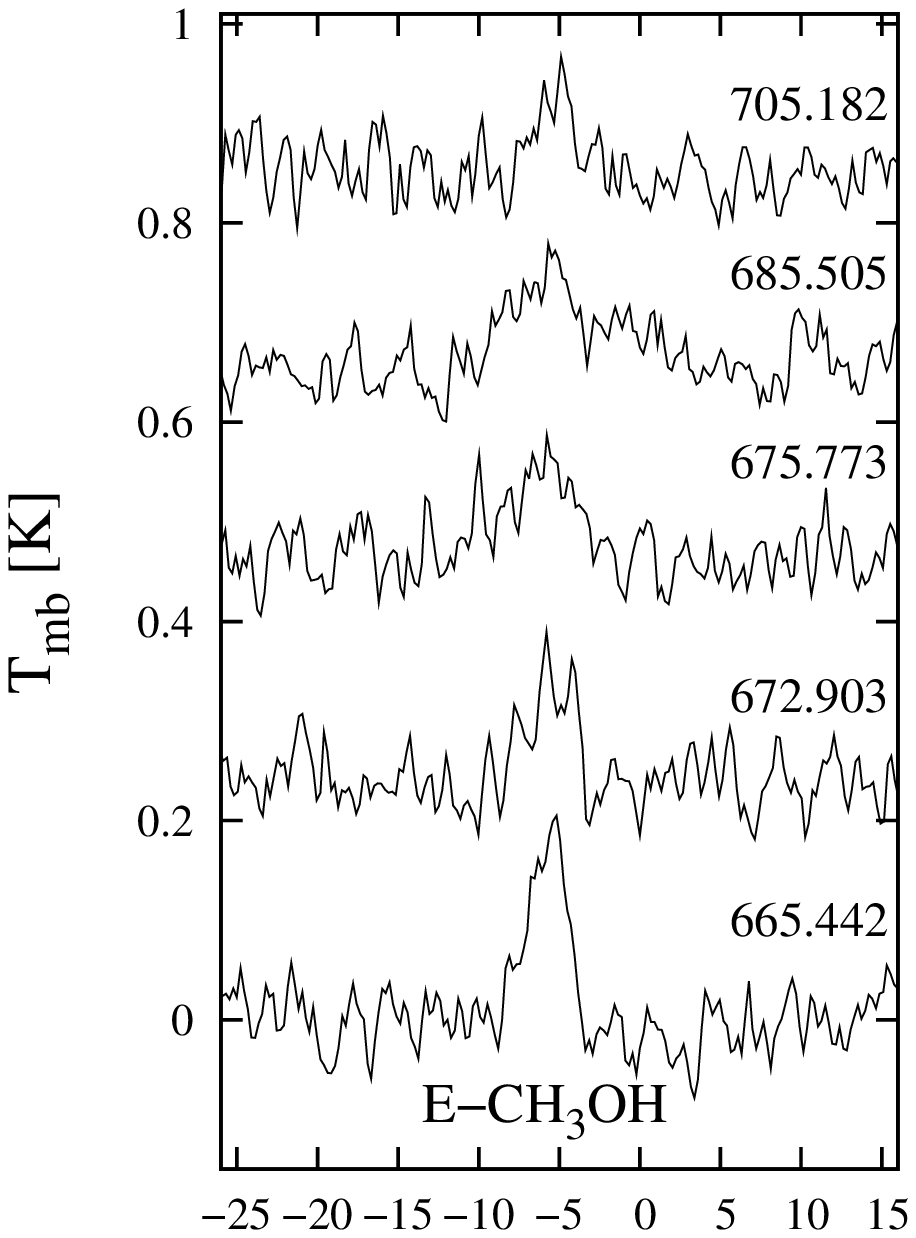}}
\subfloat{\includegraphics[width=4.43cm]{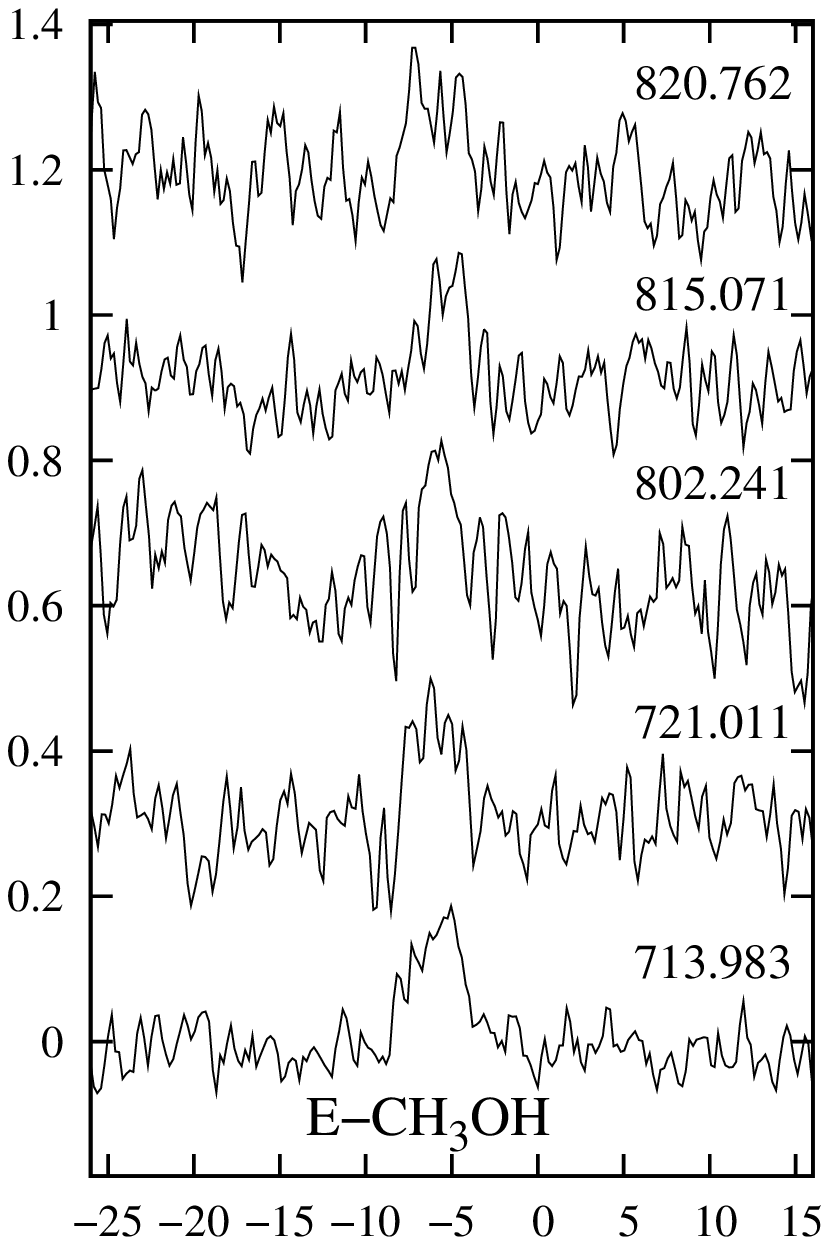}}
\subfloat{\includegraphics[width=4.43cm]{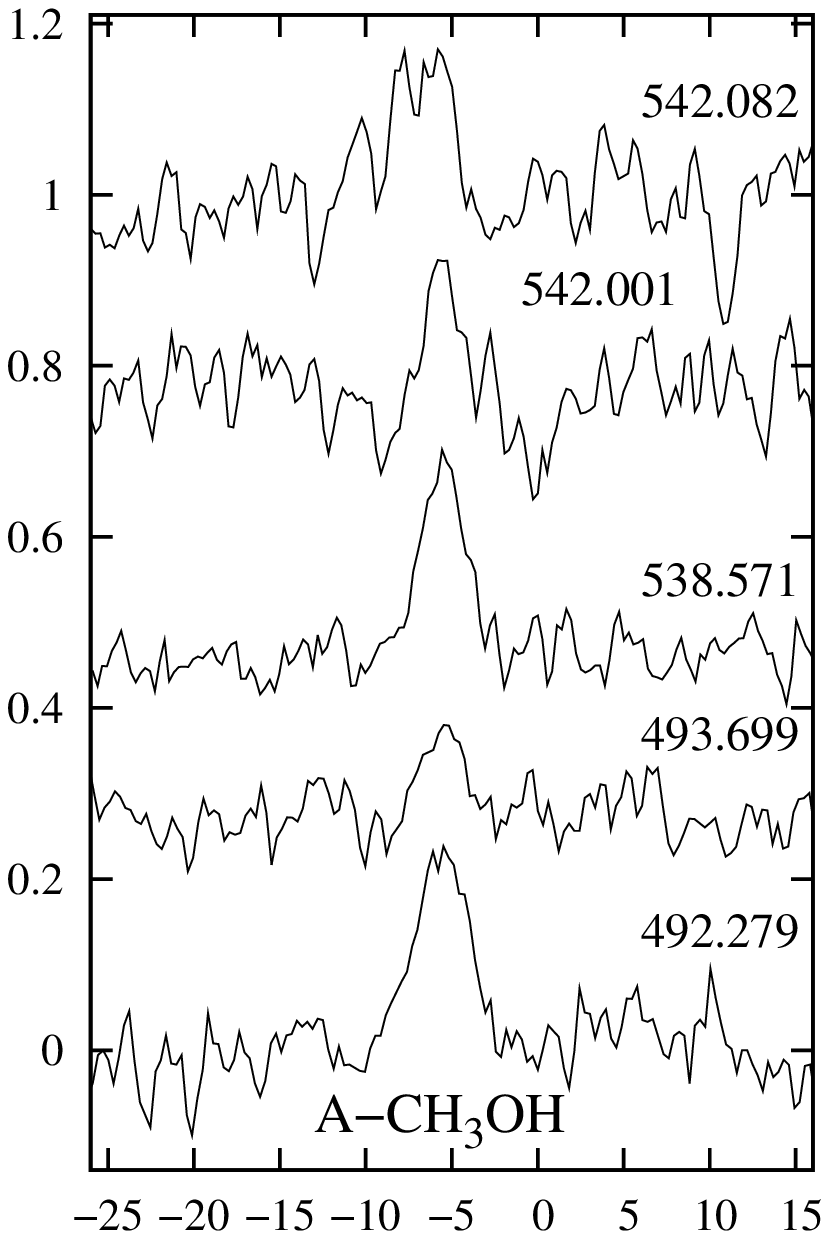}}
\subfloat{\includegraphics[width=4.43cm]{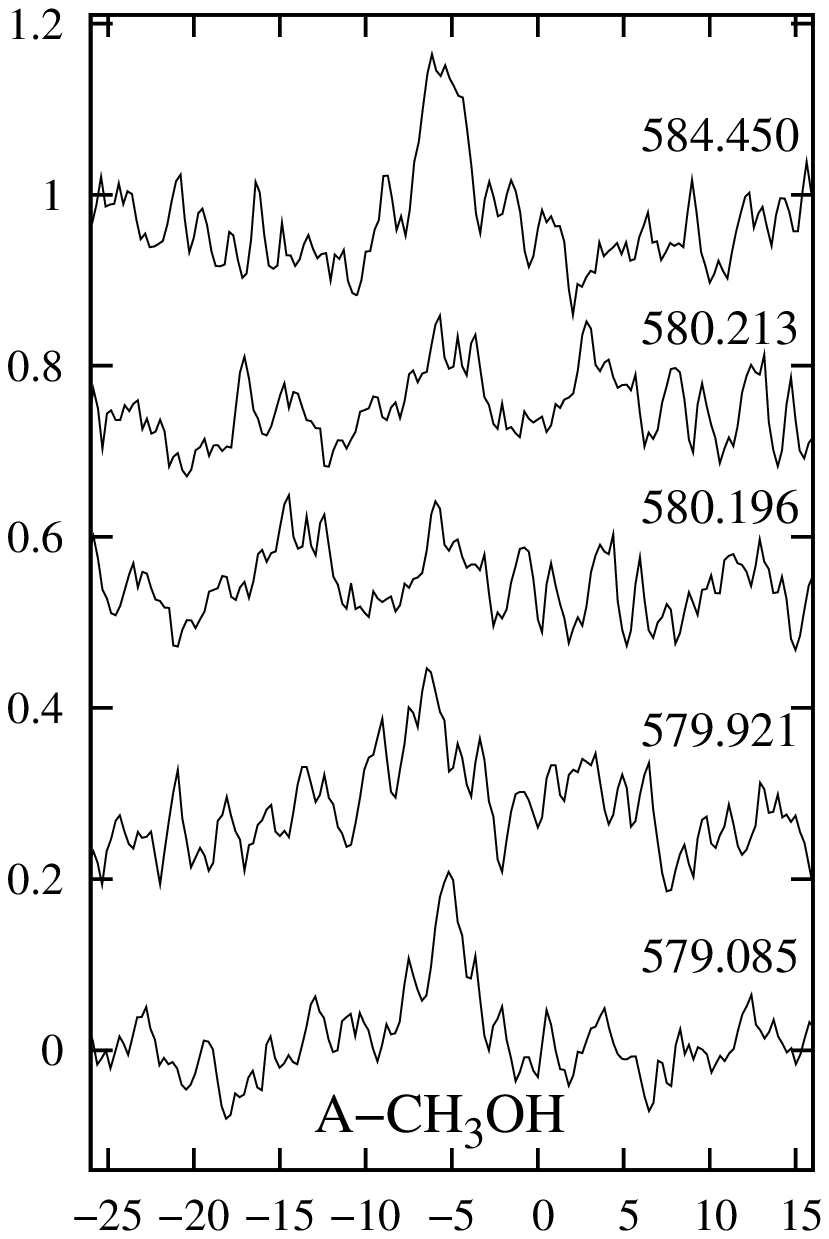}}\\
\subfloat{\includegraphics[width=4.43cm]{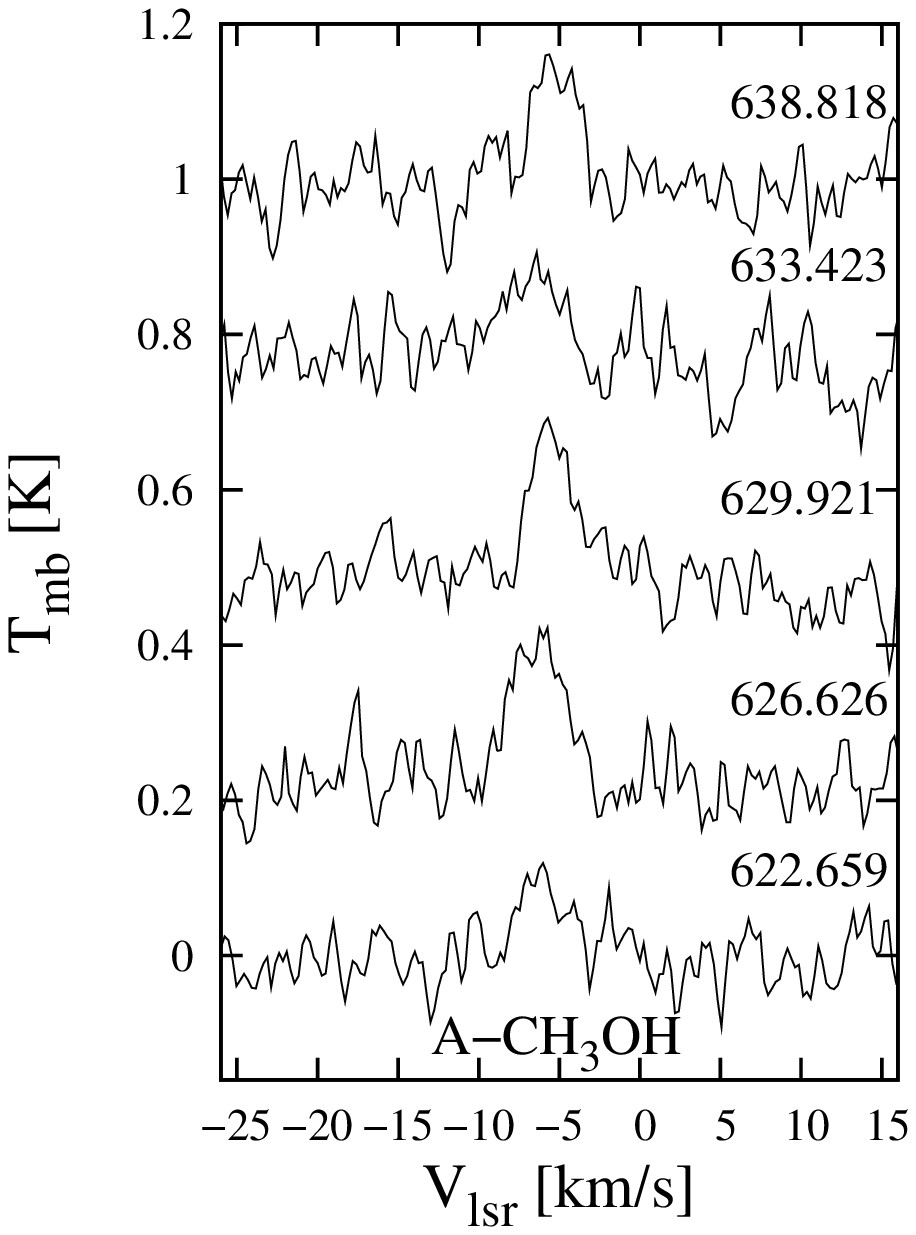}}
\subfloat{\includegraphics[width=4.43cm]{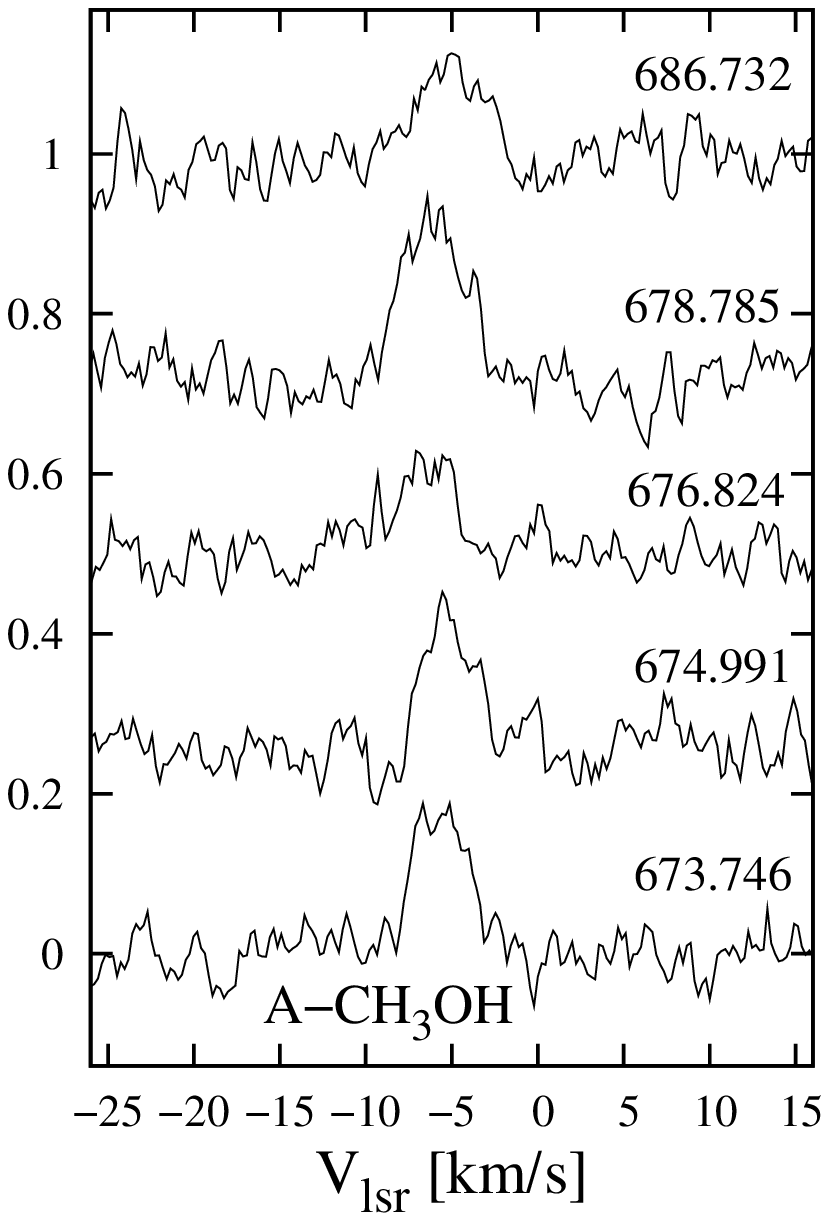}}
\subfloat{\includegraphics[width=4.43cm]{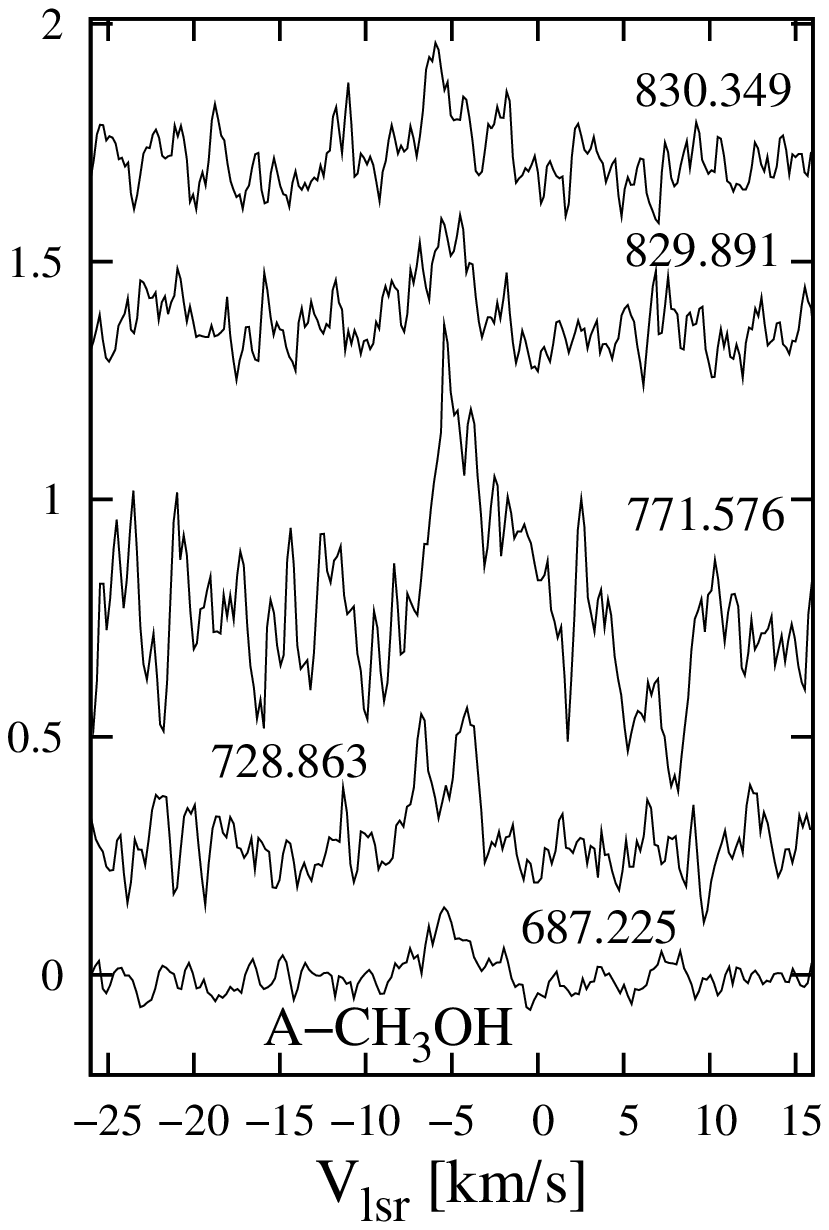}}
\subfloat{\includegraphics[width=4.43cm]{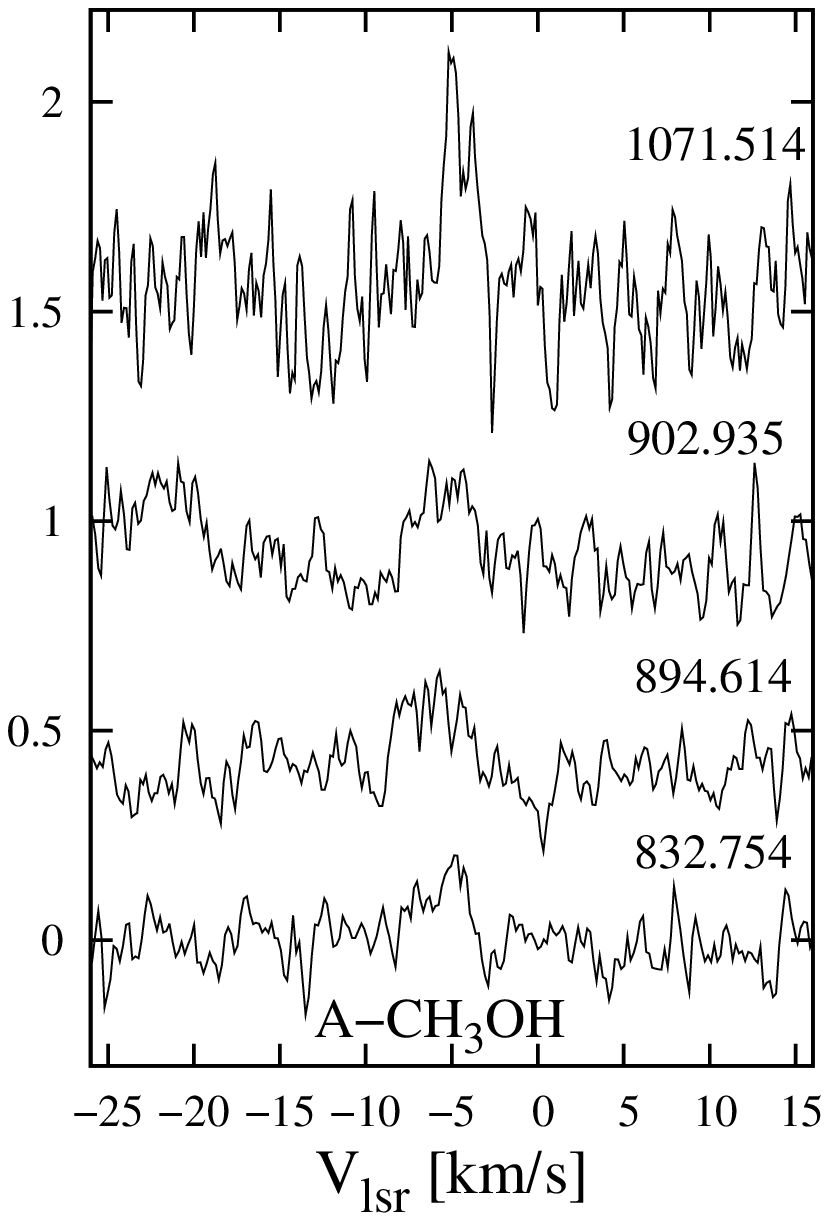}}
\caption{Continued.}
\end{figure*}

\longtab{1}{
\begin{longtable}{ccrcccc}
\caption{\label{table:measurement} Results of line measurements of all identified species from HIFI spectral survey of AFGL~2591 (~$^*$ indicates JCMT data).}\\
\hline\hline
Transition & Frequency & \Eupk & \Vlsr& $\Delta$V& \flux & \Tp\\ 
 & [\GHz] & [\K] &[\kms] & [\kms] &[\Kkms]& [\K]\\
\hline
\endfirsthead
\caption{continued.}\\
\hline\hline
Transition & Frequency & \Eupk & \Vlsr& $\Delta$V& \flux & \Tp\\ 
 & [\GHz] & [\K] &[\kms] & [\kms] &[\Kkms]& [\K]\\
\hline
\endhead
\hline
\endfoot
\multicolumn{7}{c}{CO}\\
\hline
3-2$^*$  & 345.796& 33.2& -4.16$\pm$0.01& 4.13$\pm$0.01&195.3$\pm$0.2& 44.40$\pm$0.02\\
     &        &     &-14.03$\pm$0.01&11.40$\pm$0.02&170.8$\pm$0.4& 14.08$\pm$0.01\\
5-4  & 576.268& 83.0& -4.42$\pm$0.02& 4.98$\pm$0.05& 76.6$\pm$0.1& 14.45$\pm$0.14\\
     &        &     & -6.70$\pm$0.06&15.85$\pm$0.12&155.7$\pm$0.2&  9.23$\pm$0.13\\
6-5  & 691.473&116.2& -4.61$\pm$0.01& 5.39$\pm$0.04& 92.1$\pm$0.1& 16.07$\pm$0.12\\
     &        &     & -7.23$\pm$0.05&15.70$\pm$0.10&145.5$\pm$0.1&  8.71$\pm$0.11\\
7-6  & 806.652&154.9& -4.78$\pm$0.01& 5.52$\pm$0.03& 90.8$\pm$0.1& 15.45$\pm$0.09\\
     &        &     & -7.36$\pm$0.04&15.09$\pm$0.09&119.5$\pm$0.1&  7.44$\pm$0.09\\
8-7  & 921.800&199.1& -4.90$\pm$0.02& 5.96$\pm$0.06&100.1$\pm$0.2& 15.77$\pm$0.17\\
     &        &     & -7.21$\pm$0.09&15.17$\pm$0.18&105.9$\pm$0.2&  6.56$\pm$0.17\\
9-8  &1036.912&248.9& -5.02$\pm$0.03& 6.43$\pm$0.09&109.9$\pm$0.3& 16.06$\pm$0.27\\
     &        &     & -7.22$\pm$0.18&15.96$\pm$0.42& 77.8$\pm$0.5&  4.58$\pm$0.27\\
10-9 &1151.985&304.2& -4.97$\pm$0.02& 6.01$\pm$0.07&102.9$\pm$0.2& 16.08$\pm$0.23\\
     &        &     & -7.29$\pm$0.17&14.64$\pm$0.35& 66.1$\pm$0.4&  4.24$\pm$0.23\\
11-10&1267.014&365.0& -5.05$\pm$0.02& 5.71$\pm$0.06&100.8$\pm$0.2& 10.10$\pm$0.12\\ 
     &        &     & -7.21$\pm$0.17&14.87$\pm$0.40& 58.1$\pm$0.4&  2.23$\pm$0.12\\
13-12&1496.923&503.1& -4.94$\pm$0.01& 4.82$\pm$0.03& 69.5$\pm$0.1&  9.34$\pm$0.05\\
     &        &     & -4.95$\pm$0.30&18.24$\pm$0.84& 34.7$\pm$0.8&  1.23$\pm$0.06\\ 
14-13&1611.794&580.5& -5.00$\pm$0.01& 4.54$\pm$0.02& 62.2$\pm$0.1&  8.80$\pm$0.03\\
     &        &     & -5.98$\pm$0.11&17.47$\pm$0.36& 24.7$\pm$0.4&  0.91$\pm$0.03\\
15-14&1726.603&663.4& -5.08$\pm$0.01& 4.22$\pm$0.02& 46.6$\pm$0.1&  7.01$\pm$0.03\\ 
     &        &     & -6.07$\pm$0.13&16.61$\pm$0.41& 18.0$\pm$0.4&  0.69$\pm$0.03\\
16-15&1841.346&751.7& -4.82$\pm$0.01& 3.80$\pm$0.03& 23.1$\pm$0.1&  3.83$\pm$0.03\\
     &        &     & -4.96$\pm$0.11&13.34$\pm$0.40& 12.1$\pm$0.4&  0.57$\pm$0.03\\
\hline
\multicolumn{7}{c}{$^{13}$CO}\\
\hline
3-2$^*$& 330.588& 31.3& -5.58$\pm$0.01& 3.68$\pm$0.02& 92.3$\pm$0.6&23.56$\pm$0.05\\
     &        &     & -7.33$\pm$0.01& 9.13$\pm$0.04& 97.3$\pm$1.1&10.01$\pm$0.02\\ 
5-4  & 550.926& 79.3& -5.85$\pm$0.01& 3.78$\pm$0.03& 38.7$\pm$0.1& 9.61$\pm$0.12\\
     &        &     & -6.34$\pm$0.04& 9.06$\pm$0.14& 31.6$\pm$0.2& 3.27$\pm$0.12\\
6-5  & 661.067&111.1& -5.76$\pm$0.01& 3.58$\pm$0.03& 34.6$\pm$0.1& 9.07$\pm$0.11\\
     &        &     & -6.41$\pm$0.03& 8.21$\pm$0.11& 31.0$\pm$0.2& 3.54$\pm$0.11\\    
7-6  & 771.184&148.1& -5.66$\pm$0.01& 3.63$\pm$0.04& 32.1$\pm$0.2& 8.31$\pm$0.15\\
     &        &     & -6.34$\pm$0.06& 8.06$\pm$0.19& 22.4$\pm$0.2& 2.61$\pm$0.16\\
8-7  & 881.273&190.4& -5.55$\pm$0.01& 3.18$\pm$0.05& 21.0$\pm$0.2& 6.21$\pm$0.16\\
     &        &     & -5.92$\pm$0.04& 7.03$\pm$0.17& 20.8$\pm$0.2& 2.78$\pm$0.17\\
9-8  & 991.329&237.9& -5.43$\pm$0.02& 2.94$\pm$0.10& 14.1$\pm$0.3& 4.51$\pm$0.28\\
     &        &     & -5.77$\pm$0.05& 6.03$\pm$0.21& 18.5$\pm$0.4& 2.88$\pm$0.29\\
10-9 &1101.350&290.8& -5.33$\pm$0.04& 2.32$\pm$0.18&  6.2$\pm$0.4& 2.53$\pm$0.32\\
     &        &     & -5.41$\pm$0.04& 5.19$\pm$0.21& 18.1$\pm$0.4& 3.28$\pm$0.33\\
11-10&1211.330&348.9& -5.22$\pm$0.03& 3.81$\pm$0.07& 17.4$\pm$0.1& 4.29$\pm$0.07\\
\hline
\multicolumn{7}{c}{C$^{18}$O}\\
\hline
5-4  & 548.831& 79.0& -5.79$\pm$0.01& 2.55$\pm$0.05& 6.2$\pm$0.1& 2.30$\pm$0.07\\
     &        &     & -6.30$\pm$0.04& 5.72$\pm$0.12& 7.8$\pm$0.1& 1.29$\pm$0.07\\
6-5  & 658.553&110.6& -5.69$\pm$0.01& 2.63$\pm$0.06& 6.3$\pm$0.1& 2.26$\pm$0.08\\
     &        &     & -6.23$\pm$0.06& 5.84$\pm$0.19& 6.2$\pm$0.2& 0.99$\pm$0.08\\
7-6  & 768.252&147.5& -5.47$\pm$0.05& 2.58$\pm$0.19& 4.6$\pm$0.3& 1.69$\pm$0.19\\
     &        &     & -6.01$\pm$0.17& 5.71$\pm$0.51& 5.3$\pm$0.5& 0.87$\pm$0.20\\ 
8-7  & 877.922&189.6& -5.41$\pm$0.04& 1.79$\pm$0.15& 1.7$\pm$0.2& 0.88$\pm$0.10\\
     &        &     & -5.82$\pm$0.06& 4.32$\pm$0.18& 4.9$\pm$0.2& 1.07$\pm$0.10\\
9-8  & 987.560&237.0& -5.30$\pm$0.03& 3.24$\pm$0.06& 4.1$\pm$0.1& 1.19$\pm$0.07\\ 
10-9 &1097.163&289.7& -5.51$\pm$0.08& 4.00$\pm$0.19& 3.8$\pm$0.2& 0.90$\pm$0.05\\
\hline
\multicolumn{7}{c}{C$^{17}$O}\\
\hline
3-2$^*$& 337.061& 32.4& -5.64$\pm$0.04& 2.80$\pm$0.10& 9.8$\pm$0.6& 3.29$\pm$0.08\\
   &        &     & -6.90$\pm$0.20& 5.70$\pm$0.30& 7.3$\pm$1.0& 1.20$\pm$0.04\\
5-4& 561.713& 80.9& -5.63$\pm$0.03& 3.44$\pm$0.07& 4.3$\pm$0.1& 1.17$\pm$0.02\\
6-5& 674.009&113.2& -5.81$\pm$0.03& 3.51$\pm$0.08& 3.6$\pm$0.1& 0.97$\pm$0.02\\
7-6& 786.281&150.9& -5.60$\pm$0.07& 3.37$\pm$0.16& 2.8$\pm$0.2& 0.79$\pm$0.03\\
8-7& 898.523&194.1& -5.44$\pm$0.06& 3.04$\pm$0.15& 1.7$\pm$0.1& 0.51$\pm$0.02\\ 
\hline
\multicolumn{7}{c}{C}\\
\hline
$^3$P$_1-^3$P$_0$& 492.161& 23.6& -5.72$\pm$0.02& 3.97$\pm$0.06& 13.66$\pm$0.39& 3.23$\pm$0.05\\
                 &        &     & -7.17$\pm$0.06&10.19$\pm$0.15& 19.87$\pm$0.79& 1.83$\pm$0.05\\
$^3$P$_2-^3$P$_1$& 809.344& 62.5& -5.30$\pm$0.02& 3.48$\pm$0.05& 13.33$\pm$0.30& 3.59$\pm$0.04\\
                 &        &     & -6.55$\pm$0.05&10.96$\pm$0.12& 27.63$\pm$0.71& 2.37$\pm$0.04\\ 
\hline
\multicolumn{7}{c}{C$^{+}$}\\
\hline
$^2$P$_{3/2}-^2$P$_{1/2}$&1900.537&91.2&-16.04$\pm$0.05& 4.95$\pm$0.24& 30.59$\pm$4.62& 5.81$\pm$0.65\\
                         &        &    & -5.42$\pm$0.05& 3.11$\pm$0.12& 29.51$\pm$2.56& 8.92$\pm$0.51\\
                         &        &    & -2.05$\pm$0.03& 3.21$\pm$0.09& 44.82$\pm$2.94&13.10$\pm$0.58\\
                         &        &    &-19.48$\pm$2.60&12.79$\pm$2.79& 17.55$\pm$7.91& 1.29$\pm$0.38\\
                         &        &    & -6.64$\pm$0.71& 9.05$\pm$0.75& 33.74$\pm$6.24& 3.51$\pm$0.33\\
                         &        &    &  1.59$\pm$0.04& 1.75$\pm$0.13&  4.86$\pm$0.65& 2.61$\pm$0.20\\
\hline
\multicolumn{7}{c}{HCO$^+$}\\
\hline
4-3$^*$& 356.734& 42.8& -5.87$\pm$0.01& 4.18$\pm$0.02&61.7$\pm$0.5&13.87$\pm$0.04\\
     &        &     & -7.61$\pm$0.04& 9.46$\pm$0.08&39.8$\pm$0.9& 3.96$\pm$0.02\\
6-5  & 535.061& 89.9& -5.73$\pm$0.03& 3.48$\pm$0.10& 8.5$\pm$0.2& 2.30$\pm$0.14\\
     &        &     & -6.65$\pm$0.21& 6.12$\pm$0.37& 4.0$\pm$0.4& 0.62$\pm$0.15\\
7-6  & 624.208&119.8& -5.55$\pm$0.04& 3.15$\pm$0.14& 6.0$\pm$0.2& 1.80$\pm$0.20\\
     &        &     & -6.42$\pm$0.20& 5.15$\pm$0.27& 4.5$\pm$0.3& 0.82$\pm$0.20\\
8-7  & 713.342&154.1& -5.61$\pm$0.01& 3.10$\pm$0.04& 5.4$\pm$0.1& 1.65$\pm$0.02\\
     &        &     & -6.62$\pm$0.14& 8.29$\pm$0.38& 2.2$\pm$0.4& 0.25$\pm$0.02\\ 
9-8  & 802.458&192.6& -5.47$\pm$0.03& 3.27$\pm$0.06& 4.8$\pm$0.1& 1.37$\pm$0.02\\
10-9 & 891.557&235.3& -5.47$\pm$0.04& 3.11$\pm$0.09& 3.4$\pm$0.1& 1.02$\pm$0.03\\ 
11-10& 980.637&282.4& -5.21$\pm$0.10& 3.52$\pm$0.25& 2.1$\pm$0.3& 0.56$\pm$0.04\\ 
\hline
\multicolumn{7}{c}{H$^{13}$CO$^+$}\\
\hline
4-3$^*$&346.998&41.6&-5.60$\pm$0.10&2.90$\pm$0.10& 5.74$\pm$0.36 & 1.86$\pm$0.05\\
6-5& 520.460& 87.4& -5.49$\pm$0.13& 4.25$\pm$0.31& 0.66$\pm$0.08& 0.15$\pm$0.01\\
7-6& 607.175&116.6& -5.34$\pm$0.15& 1.96$\pm$0.35& 0.25$\pm$0.08& 0.12$\pm$0.02\\ 
\hline
\multicolumn{7}{c}{HCN}\\
\hline
4-3$^*$& 354.506& 42.5& -5.51$\pm$0.02& 4.33$\pm$0.04&36.20$\pm$0.70& 7.86$\pm$0.28\\
    &        &     & -6.52$\pm$0.06& 8.40$\pm$0.20&22.40$\pm$1.30& 2.51$\pm$0.30\\
6-5 & 531.716& 89.3& -5.52$\pm$0.02& 3.91$\pm$0.05& 3.49$\pm$0.08& 0.84$\pm$0.01\\
7-6 & 620.304&119.1& -5.35$\pm$0.04& 4.02$\pm$0.09& 2.88$\pm$0.12& 0.67$\pm$0.02\\
8-7 & 708.877&153.1& -5.30$\pm$0.06& 4.16$\pm$0.14& 2.11$\pm$0.13& 0.48$\pm$0.02\\
9-8 & 797.434&191.4& -5.14$\pm$0.07& 3.46$\pm$0.17& 1.20$\pm$0.10& 0.33$\pm$0.02\\
10-9& 885.971&233.9& -4.98$\pm$0.09& 5.47$\pm$0.20& 2.07$\pm$0.13& 0.36$\pm$0.01\\ 
\hline
HCN $\upsilon$=1c,4-3&354.46043&1066.9&-5.22$\pm$0.18&5.99$\pm$0.45&2.29$\pm$0.14&0.36$\pm$0.10\\
HCN $\upsilon$=1d,4-3&356.25556&1067.1&-5.08$\pm$0.25&5.27$\pm$0.68&1.54$\pm$0.15&0.28$\pm$0.07\\
\hline
\multicolumn{7}{c}{H$^{13}$CN}\\
\hline
4-3$^*$& 345.340& 41.4& -5.10$\pm$0.10& 5.40$\pm$0.20& 7.93$\pm$0.44& 1.38$\pm$0.03\\
6-5& 517.970& 87.0& -5.44$\pm$0.15& 3.54$\pm$0.36& 0.46$\pm$0.08& 0.12$\pm$0.02\\
7-6& 604.268&116.0& -4.82$\pm$0.14& 3.38$\pm$0.33& 0.56$\pm$0.09& 0.16$\pm$0.01\\
8-7& 690.552&149.2& -4.92$\pm$0.18& 4.43$\pm$0.44& 0.64$\pm$0.08& 0.11$\pm$0.01\\
\hline
\multicolumn{7}{c}{HC$^{15}$N}\\
\hline
4-3$^*$& 344.200& 41.3& -5.50$\pm$0.20& 4.30$\pm$0.40&  2.52$\pm$0.30& 0.55$\pm$0.04\\
6-5& 516.262& 86.7& -4.17$\pm$0.23& 3.66$\pm$0.55&  0.32$\pm$0.08& 0.08$\pm$0.01\\
\hline
\multicolumn{7}{c}{HNC}\\
\hline
4-3$^*$& 362.630& 43.5& -5.51$\pm$0.03& 2.50$\pm$0.10& 8.20$\pm$0.70& 3.08$\pm$0.13\\
   &        &     & -6.01$\pm$0.07& 4.80$\pm$0.20&10.00$\pm$2.00& 1.96$\pm$0.12\\
6-5& 543.897& 91.4& -5.37$\pm$0.06& 3.36$\pm$0.14& 0.86$\pm$0.06& 0.24$\pm$0.01\\
7-6& 634.511&121.8& -5.32$\pm$0.08& 3.55$\pm$0.19& 0.79$\pm$0.07& 0.21$\pm$0.01\\
\hline
\multicolumn{7}{c}{CCH}\\
\hline
4$_{5}$-3$_{4}$$^*$&349.338& 41.9& -5.89$\pm$0.04& 3.80$\pm$0.11& 8.50$\pm$0.19&2.12$\pm$0.06\\
4$_{5}$-3$_{2}$$^*$&349.401& 41.9& -5.46$\pm$0.32& 4.01$\pm$0.99& 6.68$\pm$1.21&1.56$\pm$0.31\\    
6$_{6}$-5$_{5}$& 523.971& 88.0& -6.41$\pm$0.09& 3.79$\pm$0.21& 0.85$\pm$0.08& 0.21$\pm$0.01\\
6$_{5}$-5$_{4}$& 524.033& 88.0& -6.04$\pm$0.11& 4.13$\pm$0.27& 0.71$\pm$0.08& 0.16$\pm$0.01\\
7$_{7}$-6$_{6}$& 611.265&117.4& -6.26$\pm$0.15& 3.22$\pm$0.35& 0.47$\pm$0.09& 0.14$\pm$0.01\\
               &        &     &  0.82$\pm$0.15& 2.58$\pm$0.36&-0.33$\pm$0.08&-0.12$\pm$0.02\\
7$_{6}$-6$_{5}$& 611.328&117.4& -6.64$\pm$0.14& 2.86$\pm$0.33& 0.39$\pm$0.09& 0.14$\pm$0.02\\
8$_{8}$-7$_{7}$& 698.542&150.9& -6.89$\pm$0.19& 3.89$\pm$0.46& 0.35$\pm$0.08& 0.09$\pm$0.01\\ 
8$_{7}$-7$_{6}$& 698.604&150.9& -6.64$\pm$0.19& 3.62$\pm$0.45& 0.38$\pm$0.08& 0.09$\pm$0.01\\
\hline
\multicolumn{7}{c}{CH $^{2}\Pi_{3/2}$}\\
\hline
$3/2_{2+}-1/2_{1-}$& 532.724&25.7& -5.10$\pm$0.07& 4.84$\pm$0.19& 2.00$\pm$0.13& 0.39$\pm$0.01\\
                   &        &    & -0.32$\pm$0.14& 1.80$\pm$0.33&-0.24$\pm$0.07&-0.13$\pm$0.02\\
$3/2_{1+}-1/2_{0-}$& 532.793&25.7& -5.77$\pm$0.14& 4.83$\pm$0.34& 1.12$\pm$0.13& 0.22$\pm$0.01\\
$3/2_{2-}-1/2_{1+}$& 536.761&25.8& -5.40$\pm$0.08& 5.68$\pm$0.20& 2.04$\pm$0.12& 0.34$\pm$0.01\\
                   &        &    &  0.73$\pm$0.15& 2.57$\pm$0.36&-0.31$\pm$0.07&-0.11$\pm$0.01\\
$3/2_{1-}-1/2_{1+}$& 536.782&25.8& -4.93$\pm$0.25& 3.64$\pm$0.60& 1.98$\pm$0.10& 0.09$\pm$0.01\\ 
$3/2_{1-}-1/2_{0+}$& 536.796&25.8& -5.68$\pm$0.15& 4.92$\pm$0.38& 1.18$\pm$0.12& 0.19$\pm$0.01\\ 
\hline
\multicolumn{7}{c}{CH$^+$}\\
\hline
$1-0$& 835.138&40.1&  4.17$\pm$0.16& 12.42$\pm$0.43& -5.90$\pm$0.33 & -0.45$\pm$0.01\\
	 &        &    & -7.46$\pm$0.12&  3.50$\pm$0.29&  1.24$\pm$0.17 &  0.33$\pm$0.02\\
	 &        &    &-16.90$\pm$0.22&  9.24$\pm$0.59& -2.72$\pm$0.28 & -0.28$\pm$0.01\\
\hline
\multicolumn{7}{c}{OH $^{2}\Pi_{1/2}$}\\
\hline 
$J$=$3/2$--$1/2, F$=$2$-- $-1+$&1834.747&269.8&-4.30$\pm$0.18& 3.42$\pm$0.10& 6.8$\pm$0.3& 1.27$\pm$0.05\\
                               &        &     &-5.12$\pm$0.06&10.39$\pm$0.28& 9.5$\pm$0.6& 0.56$\pm$0.05\\
\hline
\multicolumn{7}{c}{OH$^+$, F$=1-0$}\\
\hline 
$J$=$2$--$1, F$=$3/2$--$1/2$& 971.805& 46.7& 3.46$\pm$0.28&11.87$\pm$0.69& -7.66$\pm$0.77& -0.61$\pm$0.03\\
$J$=$1$--$1, F$=$3/2$--$1/2$&1033.004& 49.6& 3.88$\pm$0.19& 4.60$\pm$0.47& -2.03$\pm$0.36& -0.42$\pm$0.04\\
$J$=$1$--$1, F$=$3/2$--$3/2$&1033.119& 49.6& 3.61$\pm$0.25&10.92$\pm$0.64& -5.78$\pm$0.59& -0.51$\pm$0.03\\
\hline
\multicolumn{7}{c}{CN}\\
\hline 
$3_{03}-2_{03}$$^*$ & 339.517& 32.6& -5.66$\pm$0.30& 3.19$\pm$0.80& 0.67$\pm$0.17& 0.20$\pm$0.02\\
$3_{03}-2_{02}$$^*$ & 340.008& 32.6& -5.47$\pm$0.18& 2.14$\pm$0.34& 0.83$\pm$0.35& 0.36$\pm$0.04\\
$3_{03}-2_{02}$$^*$ & 340.020& 32.6& -5.45$\pm$0.37& 3.14$\pm$0.50& 1.15$\pm$0.31& 0.34$\pm$0.05\\
$3_{03}-2_{02}$$^*$ & 340.032& 32.6& -5.47$\pm$0.19& 2.93$\pm$0.46& 5.77$\pm$1.01& 1.85$\pm$0.11\\
$3_{03}-2_{02}$$^*$ & 340.035& 32.6& -5.60$\pm$0.20& 3.17$\pm$0.57& 5.90$\pm$1.04& 1.75$\pm$0.12\\
%
%
$5_{05}-4_{04}$& 566.731& 81.6& -5.61$\pm$0.09& 3.56$\pm$0.22& 0.88$\pm$0.09& 0.23$\pm$0.01\\
$5_{06}-4_{05}$& 566.947& 81.7& -5.66$\pm$0.07& 3.52$\pm$0.16& 1.11$\pm$0.09& 0.30$\pm$0.01\\
$6_{06}-5_{05}$& 680.047&114.2& -5.38$\pm$0.09& 2.87$\pm$0.21& 0.47$\pm$0.06& 0.16$\pm$0.01\\
$6_{07}-5_{06}$& 680.264&114.3& -5.85$\pm$0.11& 3.80$\pm$0.25& 0.75$\pm$0.07& 0.17$\pm$0.01\\
\hline
\multicolumn{7}{c}{CS}\\
\hline
7-6$^*$& 342.883& 65.8& -5.80$\pm$0.02& 3.10$\pm$0.06&15.50$\pm$0.50& 4.70$\pm$0.30\\
     &        &     & -7.30$\pm$0.30& 6.40$\pm$0.30& 5.00$\pm$0.80& 0.73$\pm$0.29\\ 
10-9 & 489.751&129.3& -5.68$\pm$0.06& 3.69$\pm$0.14& 1.30$\pm$0.10& 0.33$\pm$0.01\\
11-10& 538.689&155.2& -5.71$\pm$0.07& 3.53$\pm$0.17& 1.01$\pm$0.20& 0.27$\pm$0.01\\
12-11& 587.616&183.4& -5.04$\pm$0.11& 4.52$\pm$0.25& 0.89$\pm$0.30& 0.18$\pm$0.01\\
13-12& 636.532&213.9& -5.24$\pm$0.12& 4.04$\pm$0.29& 0.61$\pm$0.30& 0.14$\pm$0.01\\
14-13& 685.435&246.8& -5.05$\pm$0.12& 3.93$\pm$0.28& 0.73$\pm$0.30& 0.18$\pm$0.01\\
15-14& 734.324&282.0& -5.66$\pm$0.16& 4.17$\pm$0.38& 0.70$\pm$0.40& 0.16$\pm$0.01\\ 
\hline
\multicolumn{7}{c}{$^{13}$CS}\\
\hline
8-7$^*$ & 369.909&79.9& -5.20$\pm$0.40& 3.20$\pm$0.80& 0.85$\pm$0.36& 0.25$\pm$0.05\\
\hline
\multicolumn{7}{c}{C$^{34}$S}\\
\hline
7-6$^*$ & 337.396&50.2& -5.00$\pm$0.30& 3.40$\pm$0.60& 1.52$\pm$0.46& 0.42$\pm$0.06\\
\hline
\multicolumn{7}{c}{OCS}\\
\hline
28-27$^*$  & 340.449&237.0& -4.91$\pm$0.21& 2.94$\pm$0.40& 0.77$\pm$0.11& 0.25$\pm$0.0.3\\
29-28$^*$  & 352.600&253.9& -4.85$\pm$0.18& 4.09$\pm$0.44& 1.12$\pm$0.11& 0.25$\pm$0.0.3\\
30-29$^*$  & 364.749&271.4& -5.37$\pm$0.24& 2.79$\pm$0.80& 0.80$\pm$0.16& 0.27$\pm$0.0.4\\
\hline
\multicolumn{7}{c}{o-H$_2$S}\\
\hline
$3_{2,1}-3_{1,2}$$^*$&369.102&154.5& -5.27$\pm$0.32& 3.30$\pm$0.58& 1.29$\pm$0.22& 0.37$\pm$0.04\\
$2_{21}-2_{12}$& 505.565& 79.4& -5.71$\pm$0.08& 3.10$\pm$0.20& 0.70$\pm$0.08& 0.22$\pm$0.01\\
$2_{12}-1_{01}$& 736.034& 55.1& -5.40$\pm$0.03& 3.39$\pm$0.07& 4.00$\pm$0.14& 1.11$\pm$0.02\\
               &        &     &  0.22$\pm$0.09& 0.97$\pm$0.22&-0.19$\pm$0.08&-0.19$\pm$0.04\\
$3_{03}-2_{12}$& 993.102&350.1& -6.98$\pm$0.16& 4.34$\pm$0.40& 2.66$\pm$0.42& 0.58$\pm$0.05\\
$2_{21}-1_{10}$&1072.837& 79.4& -5.59$\pm$0.19& 4.05$\pm$0.45& 1.73$\pm$0.30& 0.38$\pm$0.04\\
\hline
\multicolumn{7}{c}{p-H$_2$S}\\
\hline
$3_{31}-3_{22}$& 568.051&166.0& -4.96$\pm$0.19& 2.08$\pm$0.45& 0.19$\pm$0.07& 0.08$\pm$0.02\\
$2_{02}-1_{11}$& 687.304& 54.7& -5.77$\pm$0.05& 3.49$\pm$0.13& 1.43$\pm$0.09& 0.39$\pm$0.01\\   
\hline
\multicolumn{7}{c}{H$_2^{34}$S}\\
\hline   
$2_{12}-1_{01}$& 734.269& 55.0& -5.54$\pm$0.18& 1.94$\pm$0.42& 0.27$\pm$0.12& 0.15$\pm$0.03\\
\hline
\multicolumn{7}{c}{o-H$_2$CS}\\
\hline
$10_{1,10}-9_{1,9}$$^*$ & 338.083 & 102.4& -5.60$\pm$0.27& 3.09$\pm$1.07& 0.96$\pm$0.21& 0.30$\pm$0.03\\
$10_{1,9}-9_{1,8}$$^*$  & 348.534 & 105.2& -5.25$\pm$0.29& 4.61$\pm$0.86& 1.31$\pm$0.18& 0.27$\pm$0.03\\ 
$11_{1,11}-10_{1,10}$$^*$& 371.847 & 120.3& -5.62$\pm$0.33& 4.36$\pm$0.71& 0.98$\pm$0.15& 0.21$\pm$0.03\\ 
\hline
\multicolumn{7}{c}{HCl}\\
\hline   
$1_{3/2}-0_{3/2}$& 625.902& 30.0& -5.85$\pm$0.04& 3.63$\pm$0.08& 2.46$\pm$0.10& 0.64$\pm$0.01\\
$1_{5/2}-0_{3/2}$& 625.919& 30.0& -5.68$\pm$0.03& 3.82$\pm$0.08& 2.45$\pm$0.10& 0.73$\pm$0.01\\
$1_{1/2}-0_{3/2}$& 625.932& 30.0& -5.89$\pm$0.06& 3.89$\pm$0.14& 2.44$\pm$0.10& 0.42$\pm$0.01\\
$2_{5/2}-1_{5/2}$&1251.434& 90.1& -5.21$\pm$0.12& 2.22$\pm$0.37& 0.52$\pm$0.28& 0.30$\pm$0.08\\
$2_{7/2}-1_{5/2}$&1251.452& 90.1& -5.27$\pm$0.03& 4.51$\pm$0.08& 2.26$\pm$0.20& 0.47$\pm$0.04\\
\hline
\multicolumn{7}{c}{H$^{37}$Cl}\\
\hline  
$1_{3/2}-0_{3/2}$& 624.964& 30.0& -5.74$\pm$0.09& 4.07$\pm$0.23& 1.15$\pm$0.11& 0.27$\pm$0.01\\
$1_{5/2}-0_{3/2}$& 624.978& 30.0& -5.86$\pm$0.07& 3.19$\pm$0.18& 1.18$\pm$0.12& 0.42$\pm$0.02\\
$1_{1/2}-0_{3/2}$& 624.988& 30.0& -5.74$\pm$0.18& 3.89$\pm$0.45& 0.17$\pm$0.13& 0.17$\pm$0.01\\
\hline
\multicolumn{7}{c}{o-NH$_3$}\\
\hline    
$1_{0}-0_{0}$& 572.498& 27.5& -5.41$\pm$0.03& 4.13$\pm$0.07& 3.41$\pm$0.10& 0.78$\pm$0.01\\
               &        &     &  0.00$\pm$0.15& 1.30$\pm$0.36&-0.12$\pm$0.06&-0.09$\pm$0.02\\
$3_{0}-2_{0}$&1763.524&170.4& -5.30$\pm$0.13& 5.41$\pm$0.31& 2.54$\pm$0.25& 0.44$\pm$0.02\\
\hline
\multicolumn{7}{c}{p-NH$_3$}\\
\hline 
$2_{1}-1_{1}$&1168.452& 79.3& -5.68$\pm$0.11& 2.68$\pm$0.28& 2.29$\pm$0.41& 0.80$\pm$0.07\\
$3_{1}-2_{1}$&1763.601&165.1& -4.37$\pm$0.09& 4.18$\pm$0.22& 1.01$\pm$0.32& 0.23$\pm$0.07\\
$3_{2}-2_{2}$&1763.823&149.1& -4.55$\pm$0.15& 3.92$\pm$0.36& 1.67$\pm$0.25& 0.40$\pm$0.03\\
\hline
\multicolumn{7}{c}{N$_2$H$^+$}\\
\hline
4-3$^*$& 372.673& 44.7& -6.00$\pm$0.10& 2.90$\pm$0.10& 6.61$\pm$0.39& 2.14$\pm$0.06\\
6-5& 558.967& 93.9& -5.68$\pm$0.09& 2.89$\pm$0.21& 0.76$\pm$0.09& 0.25$\pm$0.02\\ 
7-6& 652.096&125.2& -6.01$\pm$0.09& 2.47$\pm$0.19& 0.43$\pm$0.06& 0.16$\pm$0.01\\ 
\hline
\multicolumn{7}{c}{NO}\\
\hline
$^2\Pi_{+1/2}$, {\tiny{J=7/2-5/2, F=7/2-5/2}}$^*$& 350.691& 36.1& -4.40$\pm$0.20& 4.90$\pm$0.30& 4.32$\pm$0.13& 0.80$\pm$0.10\\
$^2\Pi_{-1/2}$, {\tiny{J=7/2-5/2, F=7/2-5/2}}$^*$& 351.052& 36.1& -4.62$\pm$0.70& 9.98$\pm$1.50& 2.01$\pm$0.14& 0.17$\pm$0.10\\
%
$^2\Pi_{+1/2}$, {\tiny{J=13/2-11/2, F=11/2-9/2}}& 651.434&115.4& -5.39$\pm$0.17& 4.46$\pm$0.40& 0.48$\pm$0.08& 0.10$\pm$0.01\\
$^2\Pi_{-1/2}$, {\tiny{J=13/2-11/2, F=11/2-9/2}}& 651.773&115.5& -4.92$\pm$0.09& 3.78$\pm$0.22& 0.58$\pm$0.06& 0.15$\pm$0.01\\ 
\hline
\multicolumn{7}{c}{E-CH$_3$OH}\\
\hline
$7_{0,0}-6_{-1,0}$   & 495.173& 70.2& -5.62$\pm$0.20& 3.16$\pm$0.47& 0.35$\pm$0.10& 0.11$\pm$0.02\\
$7_{1,0}-6_{0,0}$    & 504.294& 78.2& -6.25$\pm$0.19& 4.10$\pm$0.46& 0.49$\pm$0.10& 0.11$\pm$0.01\\
$2_{-2,0}-1_{-1,0}$  & 520.179& 25.0& -5.33$\pm$0.13& 3.57$\pm$0.30& 0.57$\pm$0.09& 0.15$\pm$0.01\\
$13_{-4,0}-13_{-3,0}$& 524.269&291.2& -5.86$\pm$0.18& 3.61$\pm$0.44& 0.40$\pm$0.08& 0.10$\pm$0.01\\
$11_{2,0}-10_{1,0}$  & 558.345&167.6& -5.57$\pm$0.24& 3.10$\pm$0.57& 0.34$\pm$0.11& 0.10$\pm$0.02\\ 
$3_{-2,0}-2_{-1,0}$  & 568.566& 31.9& -5.87$\pm$0.17& 4.08$\pm$0.41& 0.58$\pm$0.10& 0.14$\pm$0.01\\
$12_{-2,0}-11_{-2,0}$& 581.092&199.2& -6.21$\pm$0.20& 2.55$\pm$0.46& 0.23$\pm$0.07& 0.09$\pm$0.01\\
$13_{1,0}-12_{2,0}$  & 601.849&224.4& -6.24$\pm$0.25& 3.44$\pm$0.60& 0.30$\pm$0.09& 0.08$\pm$0.01\\ 
$9_{1,0}-8_{0,0}$    & 602.233&117.6& -5.86$\pm$0.21& 2.86$\pm$0.49& 0.33$\pm$0.10& 0.11$\pm$0.02\\
$4_{-2,0}-3_{-1,0}$  & 616.980& 41.2& -5.63$\pm$0.11& 3.39$\pm$0.27& 0.62$\pm$0.09& 0.17$\pm$0.01\\
$13_{0,0}-12_{0,0}$  & 625.749&215.9& -6.31$\pm$0.31& 4.52$\pm$0.74& 0.38$\pm$0.11& 0.08$\pm$0.01\\
$13_{-2,0}-12_{-2,0}$& 629.652&229.4& -6.56$\pm$0.23& 3.21$\pm$0.55& 0.30$\pm$0.09& 0.09$\pm$0.01\\
$10_{0,0}-9_{-1,0}$  & 638.280&132.7& -5.53$\pm$0.18& 2.98$\pm$0.42& 0.31$\pm$0.08& 0.10$\pm$0.01\\
$14_{1,0}-13_{2,0}$  & 649.540&256.9& -5.84$\pm$0.17& 3.71$\pm$0.42& 0.27$\pm$0.05& 0.07$\pm$0.01\\
$10_{1,0}-9_{0,0}$   & 651.617&140.8& -5.79$\pm$0.12& 3.79$\pm$0.30& 0.40$\pm$0.06& 0.10$\pm$0.01\\
$5_{-2,0}-4_{-1,0}$  & 665.442& 52.8& -5.53$\pm$0.07& 2.87$\pm$0.16& 0.54$\pm$0.06& 0.20$\pm$0.01\\ 
$17_{-1,0}-16_{0,0}$ & 672.903&352.0& -5.76$\pm$0.14& 3.09$\pm$0.33& 0.35$\pm$0.06& 0.11$\pm$0.01\\
$3_{3,0}-2_{2,0}$    & 675.773& 53.7& -5.81$\pm$0.20& 4.17$\pm$0.49& 0.44$\pm$0.08& 0.09$\pm$0.01\\  
$11_{0,0}-10_{-1,0}$ & 685.505&158.2& -5.56$\pm$0.23& 4.30$\pm$0.56& 0.50$\pm$0.08& 0.09$\pm$0.01\\  
$14_{2,0}-13_{1,0}$  & 705.182&258.2& -5.11$\pm$0.16& 2.61$\pm$0.38& 0.23$\pm$0.06& 0.08$\pm$0.01\\ 
$6_{-2,0}-5_{-1,0}$  & 713.983& 66.8& -5.93$\pm$0.10& 3.47$\pm$0.22& 0.63$\pm$0.07& 0.17$\pm$0.01\\
$15_{0,0}-14_{0,0}$  & 721.011&282.9& -6.11$\pm$0.14& 2.64$\pm$0.33& 0.44$\pm$0.10& 0.16$\pm$0.02\\
$13_{1,0}-12_{0,0}$  & 802.241&224.4& -5.78$\pm$0.13& 2.14$\pm$0.31& 0.43$\pm$0.11& 0.19$\pm$0.02\\
$6_{-4,0}-5_{-3,0}$  & 815.071&128.8& -5.41$\pm$0.12& 2.63$\pm$0.28& 0.43$\pm$0.08& 0.15$\pm$0.01\\
$6_{3,0}-5_{2,0}$    & 820.762& 88.6& -5.83$\pm$0.22& 3.45$\pm$0.52& 0.46$\pm$0.12& 0.13$\pm$0.02\\
\hline
\multicolumn{7}{c}{A-CH$_3$OH}\\
\hline
$4_{1,+0}-3_{0,+0}$  & 492.279&	37.6& -5.75$\pm$0.11& 3.44$\pm$0.26& 0.80$\pm$0.11& 0.22$\pm$0.01\\
$5_{3,+0}-4_{2,+0}$  & 493.699& 84.6& -5.41$\pm$0.17& 2.67$\pm$0.41& 0.30$\pm$0.08& 0.11$\pm$0.01\\ 
$5_{1,+0}-4_{0,+0}$  & 538.571& 49.1& -5.76$\pm$0.07& 3.03$\pm$0.16& 0.72$\pm$0.07& 0.22$\pm$0.01\\
$6_{3,+0}-5_{2,+0}$  & 542.001&	98.6& -5.60$\pm$0.14& 1.88$\pm$0.33& 0.31$\pm$0.10& 0.16$\pm$0.02\\
$6_{3,-0}-5_{2,-0}$  & 542.082& 98.6& -6.69$\pm$0.19& 3.43$\pm$0.47& 0.60$\pm$0.14& 0.16$\pm$0.02\\
$2_{2,-0}-1_{1,-0}$  & 579.085& 44.7& -5.44$\pm$0.11& 3.01$\pm$0.27& 0.54$\pm$0.09& 0.17$\pm$0.01\\
$2_{2,+0}-1_{1,+0}$  & 579.921& 44.7& -6.39$\pm$0.21& 4.30$\pm$0.50& 0.60$\pm$0.13& 0.13$\pm$0.01\\
$12_{4,+0}-11_{4,+0}$& 580.196&261.4& -5.87$\pm$0.20& 2.57$\pm$0.49& 0.20$\pm$0.09& 0.08$\pm$0.02\\
$12_{3,-0}-11_{3,-0}$& 580.213&230.8& -5.51$\pm$0.26& 3.53$\pm$0.62& 0.31$\pm$0.10& 0.09$\pm$0.01\\ 
$6_{1,+0}-5_{0,+0}$  & 584.450& 62.9& -5.54$\pm$0.12& 3.07$\pm$0.27& 0.66$\pm$0.10& 0.20$\pm$0.02\\
$13_{1,+0}-12_{1,+0}$& 622.659&223.8& -6.25$\pm$0.18& 3.58$\pm$0.43& 0.40$\pm$0.09& 0.10$\pm$0.01\\
$3_{2,-0}-2_{1,-0}$  & 626.626& 51.6& -6.25$\pm$0.13& 3.67$\pm$0.31& 0.71$\pm$0.10& 0.18$\pm$0.01\\ 
$7_{1,+0}-6_{0,+0}$  & 629.921& 79.0& -5.32$\pm$0.11& 3.16$\pm$0.11& 0.65$\pm$0.10& 0.19$\pm$0.01\\
$13_{1,-0}-12_{1,-0}$& 633.423&227.5& -6.74$\pm$0.24& 4.14$\pm$0.59& 0.50$\pm$0.12& 0.11$\pm$0.01\\
$8_{3,-0}-7_{2,-0}$  & 638.818&133.4& -5.41$\pm$0.13& 3.13$\pm$0.31& 0.51$\pm$0.09& 0.15$\pm$0.01\\
$4_{2,-0}-3_{1,-0}$  & 673.746& 60.9& -5.70$\pm$0.09& 3.66$\pm$0.22& 0.69$\pm$0.07& 0.18$\pm$0.01\\
$8_{1,+0}-7_{0,+0}$  & 674.991& 97.4& -5.42$\pm$0.09& 3.18$\pm$0.22& 0.56$\pm$0.07& 0.17$\pm$0.01\\
$14_{4,+0}-13_{4,+0}$& 676.824&324.0& -5.48$\pm$0.15& 3.89$\pm$0.36& 0.47$\pm$0.07& 0.11$\pm$0.01\\
$4_{2,+0}-3_{1,+0}$  & 678.785& 60.9& -5.99$\pm$0.09& 4.13$\pm$0.20& 0.87$\pm$0.07& 0.20$\pm$0.01\\
$9_{3,+0}-8_{2,+0}$  & 686.732&154.3& -5.36$\pm$0.15& 4.20$\pm$0.37& 0.54$\pm$0.08& 0.12$\pm$0.01\\
$9_{3,-0}-8_{2,-0}$  & 687.225&154.3& -5.52$\pm$0.23& 2.70$\pm$0.55& 0.32$\pm$0.11& 0.11$\pm$0.02\\
$5_{2,+0}-4_{1,+0}$  & 728.863& 72.5& -5.46$\pm$0.18& 4.19$\pm$0.43& 0.95$\pm$0.17& 0.21$\pm$0.02\\
$16_{0,+0}-15_{0,+0}$& 771.576&315.2& -4.57$\pm$0.19& 3.21$\pm$0.44& 1.70$\pm$0.31& 0.41$\pm$0.04\\
$4_{4,-0}-3_{3,-0}$  & 829.891&103.6& -5.26$\pm$0.14& 3.40$\pm$0.36& 0.67$\pm$0.11& 0.18$\pm$0.02\\ 
$7_{2,+0}-6_{1,+0}$  & 830.349&102.7& -5.60$\pm$0.12& 2.67$\pm$0.29& 0.57$\pm$0.11& 0.20$\pm$0.02\\
$12_{3,-0}-11_{2,-0}$& 832.754&230.8& -5.45$\pm$0.16& 2.73$\pm$0.38& 0.49$\pm$0.12& 0.17$\pm$0.02\\
$13_{1,+0}-12_{0,+0}$& 894.614&223.8& -6.08$\pm$0.18& 3.59$\pm$0.43& 0.71$\pm$0.15& 0.19$\pm$0.02\\
$9_{2,-0}-8_{1,-0}$  & 902.935&142.2& -5.45$\pm$0.22& 3.12$\pm$0.52& 0.62$\pm$0.18& 0.19$\pm$0.03\\
$9_{4,+0}-8_{3,+0}$  &1071.514&184.8& -4.69$\pm$0.10& 1.95$\pm$0.23& 0.97$\pm$0.19& 0.47$\pm$0.05\\ 
\hline
\multicolumn{7}{c}{o-H$_2$CO}\\
\hline    
$5_{15}-4_{14}$$^*$&351.769& 31.7& -5.57$\pm$0.02&3.51$\pm$0.04&10.85$\pm$0.10 & 2.91$\pm$0.02\\
$5_{33}-4_{32}$$^*$&364.275&158.4& -5.24$\pm$0.16&4.78$\pm$0.35& 2.98$\pm$0.21 & 0.58$\pm$0.08\\
$5_{32}-4_{31}$$^*$&364.289&158.4& -5.65$\pm$0.26&4.21$\pm$0.63& 2.96$\pm$0.38 & 0.66$\pm$0.17\\
$7_{17}-6_{16}$ & 491.968&106.3& -5.75$\pm$0.09& 3.59$\pm$0.21& 0.98$\pm$0.10& 0.26$\pm$0.01\\
$7_{35}-6_{34}$ & 510.156&203.9& -5.21$\pm$0.56& 3.19$\pm$0.97& 0.33$\pm$0.15& 0.10$\pm$0.02\\
$7_{34}-6_{33}$ & 510.238&203.9& -5.43$\pm$0.25& 3.64$\pm$0.60& 0.31$\pm$0.09& 0.08$\pm$0.01\\
$7_{16}-6_{15}$ & 525.666&112.8& -5.56$\pm$0.09& 3.56$\pm$0.25& 0.77$\pm$0.09& 0.20$\pm$0.01\\
$8_{18}-7_{17}$ & 561.899&133.3& -4.69$\pm$0.12& 3.53$\pm$0.28& 0.80$\pm$0.11& 0.21$\pm$0.02\\
$8_{17}-7_{16}$ & 600.331&141.6& -5.20$\pm$0.15& 3.53$\pm$0.36& 0.61$\pm$0.11& 0.16$\pm$0.02\\ 
$9_{19}-8_{18}$ & 631.703&163.6& -5.09$\pm$0.12& 2.90$\pm$0.28& 0.53$\pm$0.09& 0.17$\pm$0.01\\
$9_{36}-8_{35}$ & 656.465&263.4& -4.84$\pm$0.22& 4.43$\pm$0.54& 0.36$\pm$0.05& 0.08$\pm$0.01\\
$9_{18}-8_{17}$ & 674.810&174.0& -5.18$\pm$0.22& 4.49$\pm$0.52& 0.49$\pm$0.05& 0.10$\pm$0.01\\ 
\hline
\multicolumn{7}{c}{p-H$_2$CO}\\
\hline
$5_{05}-4_{04}$$^*$& 362.736& 52.3& -5.58$\pm$0.04& 3.47$\pm$0.10 &7.27$\pm$0.16& 1.96$\pm$0.04\\
$5_{24}-4_{23}$$^*$& 363.946& 99.5& -5.57$\pm$0.21& 3.85$\pm$0.47 &2.48$\pm$0.27& 0.61$\pm$0.13\\
$5_{23}-4_{22}$$^*$& 365.363& 99.7& -5.69$\pm$0.10& 3.53$\pm$0.28 &1.98$\pm$0.12& 0.53$\pm$0.09\\
$7_{07}-6_{06}$& 505.834& 97.4& -5.53$\pm$0.09& 1.92$\pm$0.20& 0.36$\pm$0.07& 0.18$\pm$0.02\\
$7_{25}-6_{24}$& 513.076&145.4& -5.30$\pm$0.17& 2.29$\pm$0.39& 0.23$\pm$0.07& 0.09$\pm$0.01\\
\hline
\multicolumn{7}{c}{SO}\\ 
\hline
$3_3-3_2$$^*$    & 339.342& 25.5& -5.39$\pm$0.12& 3.38$\pm$0.32& 1.58$\pm$0.12& 0.44$\pm$0.05\\
$7_8-6_7$$^*$    & 340.714& 81.2& -5.78$\pm$0.07& 5.35$\pm$0.17& 7.99$\pm$0.20& 1.41$\pm$0.10\\
$8_8-7_7$$^*$    & 344.311& 87.5& -5.76$\pm$0.06& 5.12$\pm$0.13& 8.57$\pm$0.18& 1.57$\pm$0.10\\
$9_8-8_7$$^*$    & 346.529& 78.8& -5.56$\pm$0.05& 5.91$\pm$0.14&10.42$\pm$0.18& 1.66$\pm$0.10\\   
$12_{11}-11_{10}$& 514.853&167.6& -5.43$\pm$0.18& 4.51$\pm$0.44& 0.61$\pm$0.10& 0.13$\pm$0.01\\
$12_{12}-11_{11}$& 516.335&174.2& -4.97$\pm$0.23& 5.65$\pm$0.55& 0.72$\pm$0.12& 0.12$\pm$0.01\\
$12_{13}-11_{12}$& 517.354&165.8& -5.08$\pm$0.18& 5.15$\pm$0.43& 0.73$\pm$0.11& 0.13$\pm$0.01\\
$13_{12}-12_{11}$& 558.087&194.4& -5.83$\pm$0.18& 5.50$\pm$0.44& 0.95$\pm$0.13& 0.16$\pm$0.01\\
$13_{13}-12_{12}$& 559.319&201.1& -4.90$\pm$0.20& 5.73$\pm$0.48& 0.90$\pm$0.13& 0.15$\pm$0.01\\ 
$13_{14}-12_{13}$& 560.178&192.7& -6.30$\pm$0.15& 4.47$\pm$0.36& 0.75$\pm$0.11& 0.16$\pm$0.01\\
$14_{13}-13_{12}$& 601.258&223.2& -5.90$\pm$0.18& 5.08$\pm$0.44& 0.79$\pm$0.12& 0.15$\pm$0.01\\
$14_{14}-13_{13}$& 602.292&230.0& -4.87$\pm$0.27& 4.56$\pm$0.65& 0.53$\pm$0.13& 0.11$\pm$0.01\\ 
$14_{15}-13_{14}$& 603.021&221.6& -5.60$\pm$0.17& 5.04$\pm$0.40& 0.81$\pm$0.11& 0.15$\pm$0.01\\
$15_{14}-14_{13}$& 644.378&254.2& -5.47$\pm$0.13& 4.61$\pm$0.32& 0.80$\pm$0.10& 0.16$\pm$0.01\\
$15_{15}-14_{14}$& 645.254&260.9& -5.18$\pm$0.15& 5.14$\pm$0.37& 0.80$\pm$0.10& 0.14$\pm$0.01\\
$15_{16}-14_{15}$& 645.875&252.6& -5.51$\pm$0.12& 4.86$\pm$0.29& 0.86$\pm$0.09& 0.16$\pm$0.01\\
$16_{15}-15_{14}$& 687.456&287.2& -5.45$\pm$0.12& 4.91$\pm$0.29& 0.82$\pm$0.08& 0.15$\pm$0.01\\
$16_{16}-15_{15}$& 688.204&294.0& -5.24$\pm$0.10& 3.39$\pm$0.25& 0.62$\pm$0.09& 0.17$\pm$0.01\\
$16_{17}-15_{16}$& 688.735&285.7& -5.72$\pm$0.16& 5.36$\pm$0.38& 0.96$\pm$0.08& 0.14$\pm$0.01\\
$19_{18}-18_{17}$& 816.493&398.5& -6.06$\pm$0.19& 3.88$\pm$0.46& 0.68$\pm$0.09& 0.16$\pm$0.02\\
$19_{19}-18_{18}$& 816.971&405.4& -6.01$\pm$0.20& 4.64$\pm$0.48& 0.76$\pm$0.13& 0.15$\pm$0.01\\
$19_{20}-18_{19}$& 817.306&397.2& -5.32$\pm$0.19& 4.41$\pm$0.47& 0.76$\pm$0.14& 0.16$\pm$0.01\\ 
\hline
\multicolumn{7}{c}{$^{34}$SO}\\ 
\hline
$7_8-6_7$$^*$        & 333.902& 79.9& -5.12$\pm$0.18& 4.15$\pm$0.43& 1.74$\pm$0.16& 0.39$\pm$0.05\\
$8_8-7_7$$^*$        & 337.582& 86.1& -5.41$\pm$0.27& 4.05$\pm$0.62& 1.12$\pm$0.15& 0.26$\pm$0.04\\
$9_8-8_7$$^*$        & 339.858& 77.3& -5.18$\pm$0.13& 4.75$\pm$0.30& 2.06$\pm$0.12& 0.41$\pm$0.05\\
\hline
\multicolumn{7}{c}{SO$_2$}\\ 
\hline
$21_{2,20}-21_{1,21}$$^*$& 332.091& 219.5& -4.74$\pm$0.18& 5.00$\pm$0.43& 1.81$\pm$0.14& 0.34$\pm$0.04\\
$4_{3,1}-3_{2,2}$$^*$    & 332.505&  31.3& -5.41$\pm$0.14& 5.26$\pm$0.36& 2.80$\pm$0.16& 0.50$\pm$0.07\\
$8_{2,6}-7_{1,7}$$^*$    & 334.673&  43.2& -5.25$\pm$0.17& 5.43$\pm$0.37& 3.52$\pm$0.22& 0.61$\pm$0.08\\
$23_{3,21}-23_{2,22}$$^*$& 336.089& 276.0& -4.55$\pm$0.21& 4.44$\pm$0.46& 1.87$\pm$0.18& 0.40$\pm$0.06\\
$18_{4,14}-18_{3,15}$$^*$& 338.306& 196.8& -5.03$\pm$0.19& 5.29$\pm$0.41& 2.31$\pm$0.16& 0.41$\pm$0.06\\
$20_{1,19}-19_{2,18}$$^*$& 338.612& 198.9& -6.55$\pm$0.34& 6.04$\pm$0.50& 3.95$\pm$0.46& 0.61$\pm$0.09\\
$19_{1,19}-18_{0,18}$$^*$& 346.652& 168.1& -5.11$\pm$0.14& 5.99$\pm$0.33& 3.45$\pm$0.16& 0.54$\pm$0.07\\
$24_{2,22}-23_{3,21}$$^*$& 348.388& 292.7& -4.98$\pm$0.18& 5.06$\pm$0.40& 2.47$\pm$0.18& 0.46$\pm$0.06\\
$10_{6,4}-11_{5,7}$$^*$  & 350.863& 138.9& -5.41$\pm$0.20& 4.15$\pm$0.40& 0.82$\pm$0.10& 0.19$\pm$0.03\\
$5_{3,3}-4_{2,2}$$^*$    & 351.257&  35.9& -5.16$\pm$0.10& 5.16$\pm$0.21& 3.30$\pm$0.12& 0.60$\pm$0.07\\
$14_{4,10}-14_{3,11}$$^*$& 351.874& 135.9& -5.10$\pm$0.10& 4.90$\pm$0.22& 2.31$\pm$0.10& 0.44$\pm$0.05\\
$12_{4,8}-12_{3,9}$$^*$  & 355.046& 111.0& -5.02$\pm$0.10& 5.29$\pm$0.21& 3.84$\pm$0.13& 0.68$\pm$0.08\\
$13_{4,10}-13_{3,11}$$^*$& 357.165& 122.9& -5.12$\pm$0.16& 5.27$\pm$0.34& 2.70$\pm$0.16& 0.48$\pm$0.06\\
$15_{4,12}-15_{3,13}$$^*$& 357.241& 149.7& -5.25$\pm$0.11& 5.10$\pm$0.26& 2.54$\pm$0.12& 0.47$\pm$0.06\\
$11_{4,8}-11_{3,9}$$^*$  & 357.388& 100.0& -5.11$\pm$0.11& 5.12$\pm$0.22& 2.99$\pm$0.12& 0.55$\pm$0.06\\
$8_{4,4}-8_{3,5}$$^*$    & 357.581&  72.4& -5.01$\pm$0.13& 4.64$\pm$0.26& 2.58$\pm$0.14& 0.52$\pm$0.06\\ 
$9_{4,6}-9_{3,7}$$^*$    & 357.672&  80.6& -5.14$\pm$0.13& 5.23$\pm$0.32& 2.85$\pm$0.15& 0.51$\pm$0.06\\ 
$7_{4,4}-7_{3,5}$$^*$    & 357.892&  65.0& -4.85$\pm$0.10& 4.71$\pm$0.18& 3.02$\pm$0.10& 0.60$\pm$0.06\\ 
$6_{4,2}-6_{3,3}$$^*$    & 357.926&  58.6& -5.24$\pm$0.26& 5.65$\pm$0.63& 2.98$\pm$0.28& 0.49$\pm$0.06\\ 
$4_{4,0}-4_{3,1}$$^*$    & 358.038&  48.5& -5.32$\pm$0.12& 4.36$\pm$0.25& 2.07$\pm$0.11& 0.45$\pm$0.05\\ 
$20_{0,20}-19_{1,19}$$^*$& 358.216& 185.3& -4.97$\pm$0.10& 5.26$\pm$0.22& 3.78$\pm$0.13& 0.68$\pm$0.08\\ 
$21_{4,18}-21_{3,19}$$^*$& 363.159& 252.1& -5.39$\pm$0.20& 5.47$\pm$0.48& 2.62$\pm$0.20& 0.45$\pm$0.06\\ 
$15_{2,14}-14_{1,13}$$^*$& 366.215& 119.3& -5.18$\pm$0.10& 5.09$\pm$0.23& 3.17$\pm$0.13& 0.58$\pm$0.06\\ 
$6_{3,3}-5_{2,4}$$^*$    & 371.172&  41.1& -5.19$\pm$0.13& 4.86$\pm$0.29& 2.75$\pm$0.14& 0.53$\pm$0.06\\  
$28_{0,28}-27_{1,27}$& 501.108&354.3& -5.41$\pm$0.17& 3.12$\pm$0.42& 0.38$\pm$0.09& 0.11$\pm$0.01\\ 
$15_{3,13}-14_{2,12}$& 508.710&132.5& -4.63$\pm$0.19& 3.40$\pm$0.46& 0.41$\pm$0.09& 0.11$\pm$0.01\\
$18_{6,12}-18_{5,13}$& 559.882&245.5& -5.88$\pm$0.24& 3.06$\pm$0.57& 0.32$\pm$0.11& 0.10$\pm$0.02\\ 
$13_{6,8}-13_{5,9}$  & 561.266&171.9& -5.27$\pm$0.25& 4.81$\pm$0.61& 0.61$\pm$0.12& 0.11$\pm$0.01\\
$12_{6,6}-12_{5,7}$  & 561.393&160.0& -5.13$\pm$0.21& 2.96$\pm$0.49& 0.32$\pm$0.09& 0.10$\pm$0.02\\
$8_{5,3}-7_{4,4}$    & 613.076& 94.4& -4.86$\pm$0.19& 4.74$\pm$0.45& 0.53$\pm$0.09& 0.11$\pm$0.01\\
$14_{4,10}-13_{3,11}$& 626.087&135.9& -4.54$\pm$0.10& 2.23$\pm$0.24& 0.41$\pm$0.08& 0.17$\pm$0.02\\
$9_{5,5}-8_{4,4}$    & 632.193&102.7& -5.09$\pm$0.20& 4.61$\pm$0.48& 0.56$\pm$0.10& 0.11$\pm$0.01\\
$15_{4,12}-14_{3,11}$& 639.651&149.7& -5.16$\pm$0.20& 3.76$\pm$0.48& 0.37$\pm$0.08& 0.09$\pm$0.01\\
$10_{5,5}-9_{4,6}$   & 651.300&111.9& -5.06$\pm$0.15& 5.06$\pm$0.35& 0.62$\pm$0.07& 0.11$\pm$0.01\\
$18_{3,15}-17_{2,16}$& 653.110&180.6& -4.83$\pm$0.18& 5.01$\pm$0.44& 0.55$\pm$0.08& 0.10$\pm$0.01\\  
$22_{7,15}-22_{6,16}$& 660.918&352.8& -4.81$\pm$0.21& 3.69$\pm$0.49& 0.39$\pm$0.09& 0.10$\pm$0.01\\
$19_{7,13}-19_{6,14}$& 661.962&294.8& -5.00$\pm$0.18& 3.60$\pm$0.44& 0.35$\pm$0.08& 0.09$\pm$0.01\\
$17_{7,11}-17_{6,12}$& 662.404&260.8& -4.64$\pm$0.25& 4.50$\pm$0.60& 0.34$\pm$0.08& 0.07$\pm$0.01\\
$16_{7,9}-16_{6,10}$ & 662.567&245.1& -4.99$\pm$0.15& 3.91$\pm$0.36& 0.42$\pm$0.07& 0.10$\pm$0.01\\
$14_{7,7}-14_{6,8}$  & 662.799&216.6& -5.34$\pm$0.19& 4.12$\pm$0.46& 0.43$\pm$0.08& 0.10$\pm$0.01\\
$13_{7,7}-14_{6,8}$  & 662.877&203.8& -5.19$\pm$0.23& 4.73$\pm$0.57& 0.36$\pm$0.08& 0.07$\pm$0.01\\
$12_{7,5}-12_{6,6}$  & 662.934&191.8& -4.93$\pm$0.22& 3.92$\pm$0.53& 0.40$\pm$0.08& 0.08$\pm$0.01\\
$16_{4,12}-15_{3,13}$& 665.247&164.5& -4.75$\pm$0.21& 5.53$\pm$0.49& 0.59$\pm$0.08& 0.10$\pm$0.01\\
$11_{5,7}-10_{4,6}$  & 670.366&122.0& -5.16$\pm$0.16& 4.07$\pm$0.37& 0.51$\pm$0.08& 0.12$\pm$0.01\\ 
$7_{6,2}-6_{5,1}$    & 695.633&114.0& -4.89$\pm$0.13& 3.73$\pm$0.30& 0.51$\pm$0.07& 0.13$\pm$0.01\\
$19_{4,16}-18_{3,15}$& 702.104&214.3& -4.73$\pm$0.17& 4.19$\pm$0.39& 0.49$\pm$0.08& 0.11$\pm$0.01\\
$21_{4,18}-20_{3,17}$& 727.379&252.1& -5.61$\pm$0.11& 2.39$\pm$0.26& 0.71$\pm$0.15& 0.30$\pm$0.03\\ 
$19_{5,15}-18_{4,14}$& 820.150&236.2& -5.26$\pm$0.17& 5.27$\pm$0.41& 0.94$\pm$0.13& 0.17$\pm$0.01\\
$20_{5,15}-19_{4,16}$& 840.751&254.6& -4.79$\pm$0.20& 5.25$\pm$0.49& 0.70$\pm$0.11& 0.12$\pm$0.01\\
$15_{6,10}-14_{5,9}$ & 848.523&198.6& -4.85$\pm$0.18& 4.76$\pm$0.43& 0.78$\pm$0.12& 0.15$\pm$0.01\\
\hline
\multicolumn{7}{c}{$^{34}$SO$_2$}\\ 
\hline
$19_{1,19}-18_{0,18}$$^*$&344.581&167.7&-5.23$\pm$0.24&4.15$\pm$0.54&1.03$\pm$0.12&0.23$\pm$0.03\\
$13_{4,10}-13_{3,11}$$^*$&344.808&121.6&-5.18$\pm$0.28&3.49$\pm$0.55&0.65$\pm$0.10&0.17$\pm$0.03\\
$7_{4,4}-7_{3,5}$$^*$    &345.520& 63.7&-4.70$\pm$0.24&3.06$\pm$0.49&0.76$\pm$0.12&0.23$\pm$0.03\\
$20_{0,20}-19_{1,19}$$^*$&357.102&184.8&-5.27$\pm$0.20&3.04$\pm$0.48&0.72$\pm$0.10&0.23$\pm$0.03\\
$6_{3,3}-5_{2,4}$$^*$    &362.158& 40.7&-4.92$\pm$0.28&3.40$\pm$0.47&0.80$\pm$0.16&0.22$\pm$0.04\\
\hline
\multicolumn{7}{c}{o-H$_2$O}\\
\hline    
$1_{10}-1_{01}$& 556.936& 61.0& -3.63$\pm$0.10& 3.35$\pm$0.13& 3.82$\pm$0.44& 1.07$\pm$0.02\\
               &       &      & -0.52$\pm$0.17& 2.25$\pm$0.33&-0.66$\pm$0.42&-0.28$\pm$0.04\\
$3_{12}-3_{03}$&1097.365&249.4& -5.27$\pm$0.05& 2.85$\pm$0.13& 5.94$\pm$0.46& 1.96$\pm$0.08\\
               &        &     & -6.34$\pm$0.35&11.65$\pm$1.02& 6.38$\pm$0.11& 0.52$\pm$0.07\\
$3_{12}-2_{21}$&1153.127&249.4& -5.10$\pm$0.09& 2.16$\pm$0.25& 2.18$\pm$0.41& 0.95$\pm$0.09\\
               &        &     & -5.78$\pm$0.35&10.93$\pm$1.02& 5.97$\pm$1.11& 0.51$\pm$0.06\\ 
$3_{21}-3_{12}$&1162.912&305.3& -5.32$\pm$0.04& 2.89$\pm$0.10& 8.12$\pm$0.50& 2.65$\pm$0.08\\
               &        &     & -6.04$\pm$0.76&14.17$\pm$1.44& 6.79$\pm$1.41& 0.45$\pm$0.04\\
\hline
\multicolumn{7}{c}{p-H$_2$O}\\
\hline
$2_{11}-2_{02}$& 752.033&136.9& -5.56$\pm$0.03& 3.45$\pm$0.09&  8.11$\pm$0.36& 2.21$\pm$0.05\\
               &        &     & -6.20$\pm$0.28&12.60$\pm$0.93&  5.48$\pm$0.19& 0.41$\pm$0.05\\
$2_{02}-1_{11}$& 987.927&100.9& -5.33$\pm$0.04& 3.82$\pm$0.10& 11.12$\pm$0.45& 2.74$\pm$0.06\\
               &        &     & -5.81$\pm$0.40&14.78$\pm$1.30&  8.01$\pm$0.19& 0.54$\pm$0.05\\ 
$1_{11}-0_{00}$&1113.343& 53.4& -3.30$\pm$0.20& 3.26$\pm$0.34&  6.17$\pm$0.63& 1.78$\pm$0.11\\
               &        &     & -0.48$\pm$0.20& 2.60$\pm$0.27& -3.88$\pm$0.51&-1.40$\pm$0.17\\  
               &        &     &-11.98$\pm$0.34&13.75$\pm$0.84&-11.23$\pm$0.93&-0.77$\pm$0.03\\
$2_{20}-2_{11}$&1228.789&195.9& -5.24$\pm$0.10& 2.82$\pm$0.27&  6.23$\pm$0.64& 2.08$\pm$0.20\\ 
               &        &     & -5.98$\pm$0.91& 8.47$\pm$1.59&  5.66$\pm$1.80& 0.63$\pm$0.20\\
\hline
\multicolumn{7}{c}{HF}\\
\hline  
1-0&1232.476& 59.2& 13.01$\pm$0.13& 1.17$\pm$0.31& -1.16$\pm$0.52& -0.93$\pm$0.21\\
   &        &     & -0.05$\pm$0.13& 2.31$\pm$0.31& -2.81$\pm$0.76& -1.25$\pm$0.16\\
   &        &     & -3.88$\pm$0.15& 2.50$\pm$0.35&  3.08$\pm$0.79&  1.21$\pm$0.15\\
   &        &     &-12.58$\pm$0.28& 8.81$\pm$0.71&-10.02$\pm$1.39& -1.07$\pm$0.07\\
\hline
\end{longtable}
}

\end{appendix}

\end{document}